На правах рукописи

Моторина Галина Геннадьевна

# Динамика нагрева плазмы и энергетических распределений ускоренных электронов во время солнечных вспышек по данным рентгеновского и ультрафиолетового излучения

01.03.02 – Астрофизика и звездная астрономия

ДИССЕРТАЦИЯ

на соискание ученой степени

кандидата физико-математических наук

Научный руководитель

кандидат физико-математических наук

Кудрявцев Игорь Владимирович

Санкт-Петербург – 2017

# Оглавление

# Введение

Солнце — единственная звезда Солнечной системы, вокруг которой обращаются планеты и их спутники, астероиды, кометы и космическая пыль. Возникающие на Солнце активные процессы, выражающиеся в появлении солнечных пятен, факелов, протуберанцев, изменениях структуры солнечной короны, усилении солнечного ветра и др. имеют значительное влияние на атмосферу Земли. Следствиями этих изменений являются полярные сияния, геомагнитные бури, которые могут влиять на работу технических средств (помехи радиосвязи на коротких волнах, нарушение работы радионавигационных устройств, околоземных спутников и др., вплоть до отключения электричества (г. Квебек, Канада, 1989 г.)) и самочувствие людей. Данная проблема называется «космической погодой» и отмечена космическими агентствами NASA и ESA как одна из приоритетных проблем, требующих изучения. Вспышечные процессы являются наиболее мощными процессами космической погоды. Именно эти взрывные процессы определяют околоземную космическую погоду и могут оказывать заметное влияние на техносферу, биосферу и климат Земли. Поэтому исследование вспышечных процессов имеет не только большое фундаментальное, но и прикладное значение, а исследование солнечной активности является исследованием с высоким приоритетом.

Солнечные вспышки являются магнитными взрывными процессами, спонтанно происходящими в солнечной атмосфере, приводящие к эффективному ускорению частиц и нагреву плазмы. Эти явления охватывают все слои солнечной атмосферы: фотосферу, хромосферу и корону Солнца, и генерируют все виды электромагнитного излучения: от радиоволн до рентгеновских и гамма-лучей. Диагностика вспышечной плазмы, как правило, осуществляется изучением крайнего (жесткого или далекого) ультрафиолетового (КУФ) излучения, в то время как информацию о нетепловой компоненте плазмы, распределении высокоэнергичных ускоренных электронов, можно получить из данных рентгеновского излучения (РИ). В соответствии с современными представлениями, именно данный диапазон длин волн наиболее чувствителен к тепловым и нетепловым процессам, которые происходят в области первичного энерговыделения.

Солнечные вспышки отличаются друг от друга морфологией, длительностью, расположением на диске Солнца, интенсивностью излучения, поэтому существуют различные их классификации (см., напр., [1, 127]): по оптическим характеристикам, где вспышки разделяются на пять классов в зависимости от полной энергии, излучаемой в линии $H_\alpha$ (введена с 1964 г.), по типам радиовсплесков, по потоку высокоэнергичных частиц для событий с энергиями ≥10 МэВ (для протонных вспышек), по РИ, где классификация происходит по яркости в рентгеновских лучах в диапазоне длин волн от 1 до 8 Å, зарегистрированную за 1 минуту, основанную на патрульных измерениях серии искусственных спутников Земли (ИСЗ), главным образом Geostationary Operational Environmental Satellites (GOES) [138] (введена с 1969 г.). Приведенные выше классификации являются официально принятыми, хотя существует множество других признаков, по которым описывают данные явления. Наиболее широкое применение получила классификация по РИ, которая является статистическим индикатором геоэффективности солнечных вспышек в течение уже 30 лет, а также ценным ресурсом для изучения прошлого солнечной активности и прогноза космической погоды (напр., [35, 106, 28, 60]). Особый интерес представляют вспышки классов C ($10^{-6} \leq I < 10^{-5}$), M ($10^{-5} \leq I < 10^{-4}$) и X ($I \geq 10^{-4}$), которые могут оказывать значительное влияние на межпланетное пространство и магнитосферу Земли, где $I$ - максимум интенсивности (Вт/м$^2$) в интервале 1-8 Å. Каждый класс дополнительно делится на 9 подгрупп, например, C1.3, где индекс (множитель) показывает, во сколько раз вспышка сильнее минимальной величины вспышки класса C, т.е. $1.3 \times 10^{-6}$ Вт/м$^2$.

В настоящее время для интерпретации вспышечных явлений активно привлекается «стандартная» (двумерная) модель солнечной вспышки (CSHKP) [41, 124, 80, 64, 115, 116, 132], хотя и существует ряд других моделей (см. обзор, напр., [30]). В частности, считается (см., напр., [115]), что выделение первичной энергии происходит в результате магнитного пересоединения, что приводит впоследствии к выбросу плазмоида и ускорению заряженных частиц в области вершины вспышечной петли (см. рис. 1.1, построенный на основе модели из работы [116]). Часть ускоренных частиц, направляясь вверх, покидает солнечную корону через открытые силовые линии магнитного поля в виде солнечного ветра. Другая часть энергичных электронов, распространяясь вдоль магнитных силовых линий, высыпается в основаниях петли, обуславливая генерацию жесткого РИ и нагрев хромосферы [111]. «Испаряющаяся» горячая плазма с температурой $(5-30) \times 10^6$ K заполняет корональную часть магнитной петли и высвечивается в ультрафиолетовом и мягком рентгеновском диапазонах. Таким образом, в диссертации для диагностики ускоренных электронов и вспышечной плазмы основной упор сделан на наблюдения в этих энергетических диапазонах.

Оценки показывают, что, по крайней мере, для некоторых событий описанный выше сценарий хорошо согласуется с наблюдениями [136, 117, 59]. Стоит отметить, что довольно часто пики температуры горячей корональной плазмы могут опережать пики жесткого РИ [126]. Поскольку жесткое РИ генерируется ускоренными электронами, то это свидетельствует о важной роли тепловых механизмов энерговыделения в солнечных вспышках.

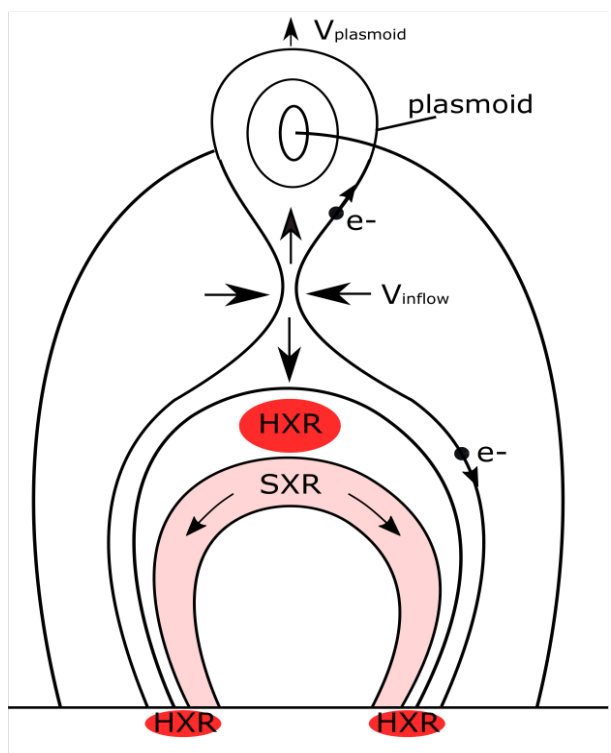

Рис. 1.1. «Стандартная» (двумерная) модель солнечной вспышки. На рисунке схематично представлены геометрия вспышечной области и характерные процессы, генерируемые во время вспышечного процесса.

В рамках «стандартной» модели появление горячей плазмы внутри вспышечных петель может происходить не только за счет "хромосферного испарения", связанного с высыпанием ускоренных электронов, но и другими способами: например, в окрестности пересоединяющегося токового слоя, расположенного в вершине магнитного каспа, а также на ударных волнах, связанных с истечением плазмы из токового слоя (см., напр., обзор [112]). Также нагрев плазмы может осуществляться за счет сокращения (*shrinkage*) магнитных трубок, вышедших из области пересоединения - Ферми и бетатронный механизмы (см., напр., [2, 121]), а также вследствие сильных электрических полей из-за неустойчивости Рэлея-Тейлора [141, 123]. Все эти (и другие) эффекты содержатся в «стандартной» модели CSHKP (в ее более современных модификациях).

Продолжительность солнечных вспышек в целом не превышает нескольких минут, однако в некоторых случаях может достигать нескольких часов. В различных диапазонах энергетического спектра эта величина может варьироваться. Как правило, солнечную вспышку подразделяют на три фазы: фазу роста, максимума (импульсная фаза) и спада интенсивности. В работе [7] на примере события 26 июля 2002 года, зарегистрированного спектрометром ИРИС спутника КОРОНАС-Ф [10], на временном профиле вспышки было выделено три интервала, для которых характерны разные зависимости энергетического спектра жесткого РИ: неустойчивый вид – на стадии роста, нестепенная и степенная зависимости – на стадии максимума и спада интенсивности соответственно. Отметим также, что иногда наблюдаются отдельные всплески жесткого РИ в кривых блеска вспышек (см., напр., [71]), а во время импульсной фазы вспышки кроме жесткого РИ могут возникать гамма-излучение и нейтроны, что свидетельствует о существовании протонов с энергиями больше 100 МэВ, а также радио всплески третьего и других типов (см., напр., [73]).

Область КУФ излучения соответствует фотонам с длинами волн <1000 Å, отвечающий за переход от хромосферы к короне Солнца, при этом в более коротковолновой области преобладает линейчатое излучение хромосферы и короны, хотя если говорить о вспышках, вклад в излучение свободно-связанных и связанно-связанных переходов может быть сравнимым (см., напр., [98, 99]). Тем не менее, изображения в различных линиях (*He*, *O*, *Ne*, *Mg*, *Si*, *Ca*, *Fe*, *Al*) (см., напр., [30]) позволяют изучать структурные особенности на разных высотах над фотосферой. Фотоны с длинами волн 100–0.1 Å (0.1–100 кэВ) относят к рентгеновскому диапазону спектра, которое можно примерно разделить на мягкое РИ с энергиями примерно <20 кэВ и жесткое РИ с энергиями >20 кэВ (см., напр., [11]). Источником излучения в рентгеновском диапазоне обычно является плазма с температурой порядка ~$10^6$ К и выше. Принято считать, что спектр электронов соответствует тепловому тормозному излучению горячей оптически тонкой плазмы в рамках «квазитепловой» модели. Этому излучению, обусловленному электронами, находящимися в равновесии со средой, соответствует мягкое РИ.

Жесткое РИ, генерируемое в солнечных вспышках при столкновении быстрых электронов с частицами солнечной плазмы, является нетепловым тормозным излучением высокоэнергичных электронов. По данным жесткого РИ можно получить информацию об инжектируемых электронах во вспышечном источнике, для этого, как правило, используется приближение моделями толстой и тонкой мишени [36, 24]. В последнее десятилетие были получены многочисленные результаты регистрации жесткого РИ во время вспышек на Солнце с высоким пространственным, временным и энергетическим разрешением (см. обзоры [29, 83, 78, 66]), позволяющие детально исследовать структуру жесткого РИ вспышек. Отсюда следует, что изучение наблюдений в смежных диапазонах электромагнитного спектра позволяет более детально исследовать структуру энергетических спектров ускоренных электронов.

В диссертации используются данные современных космических аппаратов (КА), а именно:

- изображающий рентгеновский спектрометр Reuven Ramaty High Energy Solar Spectroscopic Imager (RHESSI), который регистрирует РИ с высоким пространственным (~2.3″) и спектральным (~1–10 кэВ) разрешением, и таким образом позволяет определить источник и форму распределения электронов в энергетическом диапазоне от 3 кэВ до ~17 МэВ при помощи германиевых детекторов [92].

- инструмент Atmospheric Imaging Assembly (AIA) на борту космической станции Solar Dynamics Observatory (SDO), регистрирующий КУФ излучение от всего диска Солнца с пространственным (~1.5″) и временным (12с) разрешением в семи каналах, соответствующим линиям ионизированного железа: *Fe XVIII* (94 Å), *Fe VIII, XXI* (131 Å), *Fe IX* (171 Å), *Fe XII, XXIV* (193 Å), *Fe XIV* (211 Å), *He II* (304 Å), *Fe XVI* (335 Å). Также AIA регистрирует излучение в линии *C IV* (1600 Å) и континууме (1700, 4500 Å), соответствующие фотосфере и переходному слою. Однако так как в работе рассматриваются области на Солнце, соответствующие горячей вспышечной плазме, то используются AIA данные только в шести КУФ каналах 94, 131, 171, 193, 211, 335 Å, соответствующие температурному диапазону ~0.5-16 МК [88].

- спектрометр ИРИС (Исследование Рентгеновского Излучения Солнца) на борту орбитальной станции КОРОНАС–Ф (космическая программа «Комплексные орбитальные околоземные наблюдения активности Солнца») [10, 42, 7], который регистрировал РИ с временным разрешением 2.5с – в 12-и каналах в режиме "патруль" (2-250 кэВ); 1с – в 64-х каналах (2-150 кэВ) и 0.01 секунды – в 4-х энергетических каналах (24-180 кэВ) в режиме "всплеск". Высокая чувствительность прибора позволяла изучать динамику РИ на суб–секундных временных шкалах во время солнечных вспышек.

- спутник GOES [138], регистрирующий мягкое РИ в двух энергетических каналах: 0.5–4 и 1–8 Å, позволяющий определить класс вспышки, а также ее температуру и меру эмиссии.

Данная работа главным образом посвящена восстановлению энергетических распределений электронов, генерирующих крайнее ультрафиолетовое и рентгеновское излучение во время солнечных вспышек, для чего необходимо решать обратную задачу. **Для этого были последовательно поставлены и решены следующие задачи**:

1) Восстановление энергетических распределений жесткого РИ солнечной вспышки по данным регистрации с учетом приборной функции спектрометра ИРИС на борту КА КОРОНАС-Ф;

2) Реконструкция энергетических распределений ускоренных электронов по восстановленным энергетическим спектрам жесткого РИ из п.(1);

3) Восстановление энергетических распределений электронов, излучающих КУФ и мягкое РИ, зарегистрированное на КА SDO/AIA и RHESSI соответственно, в рамках много-температурной модели;

4) Оценка теплового баланса горячих вспышечных петель и интенсивности РИ из различных вспышечных областей по данным КА SDO/AIA и RHESSI соответственно.

## Актуальность исследования

Изучение солнечной активности становится все более актуальным в связи с резким улучшением пространственного, временного и спектрального разрешения современных отечественных и зарубежных солнечных телескопов наземного и космического базирования, что позволяет детально изучать такие явления, как солнечные вспышки (см. обзоры [29, 78, 30, 10]).

Принято считать, что солнечные вспышки являются результатом пересоединения магнитного поля в солнечной короне, где большая доля выделяемой энергии переходит в ускоренные частицы. Несмотря на то, что качественная картина вспышек, движимых импульсно освобождающейся энергией из корональных магнитных петель, является общепринятой, подробные процессы ускорения частиц и их распространения остаются нерешенной проблемой астрофизики и физики плазмы.

Данное исследование особенно актуально в связи с ближайшими космическими миссиями ESA Solar Orbiter и NASA Solar Probe Plus, которые планируется запустить в 2018 году, а также планируемый российский КА Интергелиозонд, где основной задачей будет изучение внутренней гелиосферы.

## Цель диссертационной работы

Целью исследования является восстановление энергетических распределений электронов, ускоренных во время солнечных вспышек, диагностика вспышечной плазмы на основе данных крайнего ультрафиолетового и РИ, а также оценка теплового баланса горячих вспышечных петель и интенсивности РИ из различных вспышечных областей в рамках «стандартной» модели. Предложенные методики позволяют в хорошей степени восстанавливать как спектры ускоренных электронов, так и основные параметры вспышечной плазмы, такие как температура, мера эмиссии, концентрация, энергия солнечной вспышки.

## Научная новизна

1. Разработана методика, позволяющая определять вначале спектр регистрируемого детекторами жесткого РИ спектрометра ИРИС на борту КА КОРОНАС-Ф (с учетом приборной функции), а затем по нему восстанавливать энергетические спектры излучающих электронов.

2. Разработана методика нахождения дифференциальной меры эмиссии (ДМЭ) на основе аппроксимации модельными функциями одновременно данных КА RHESSI и SDO/AIA. Впервые с помощью комбинированного анализа SDO/AIA и RHESSI данных найдено энергетическое распределение электронов для широкого диапазона энергий: 0.1 – 20 кэВ.

3. Предложена новая функциональная форма для описания ДМЭ, для которой выведено выражение для энергетического распределения электронов во вспышечной области.

4. Сделан вывод о каппа-распределении электронов во вспышечной области посредством ДМЭ. Произведено сравнение результатов данного подхода для одновременно SDO/AIA и RHESSI данных с использованием методики из п.(2) и результатов аппроксимации модельными функциями только данных КА RHESSI.

## Теоретическая и практическая значимость

Так как исследование главным образом направлено на изучение солнечной активности и улучшение существующих методов диагностики во вспышечной плазме, то результаты данной работы могут быть использованы, в первую очередь, для более детальной спектральной и пространственной диагностики процессов, наблюдаемых в активных областях Солнца, таких как, солнечные вспышки, протуберанцы, волокна, петлевые аркады. Также полученные знания помогут прояснить процессы, протекающие на звездах других галактик, а также более детально изучить процессы в лабораторной плазме, получаемой в токамаках.

Разработанные методики позволят решать некорректно поставленные обратные задачи сложных нестационарных нелинейных процессов, а также помогут улучшить диагностику процессов нагрева и ускорения частиц на Солнце, получать и анализировать полный спектр ускоренных электронов, что, в свою очередь, открывает новые возможности в исследовании физики Солнца.

## Положения, выносимые на защиту

1. Динамика реконструированных энергетических распределений высокоэнергичных электронов для солнечной вспышки 15 апреля 2002 г. по данным спектрометра ИРИС на борту КА КОРОНАС-Ф с учетом приборной функции. Методика основана на решении интегрального уравнения, описывающего трансформацию спектра рентгеновских квантов в процессе регистрации, и восстановлении спектров ускоренных электронов в источнике генерации тормозного излучения.

2. Результаты разработанной методики восстановления энергетических распределений электронов в солнечных вспышках с помощью аппроксимации модельными функциями ДМЭ одновременно данных крайнего ультрафиолетового и рентгеновского излучения по данным КА

SDO/AIA и RHESSI соответственно.

3. Функциональная форма ДМЭ, для которой выведено выражение для энергетического распределения электронов в солнечной вспышке, с помощью данной ДМЭ и разработанной методики из п.(2) автоматически можно получать основные параметры вспышечной плазмы (температуру и меру эмиссии).

4. Анализ теплового баланса и результаты сравнения потоков жесткого РИ из различных частей вспышечной петли в рамках «стандартной» модели солнечной вспышки.

## Степень достоверности и апробация результатов

Результаты диссертации опубликованы в 12 статьях, из которых 5 статей опубликовано в рецензируемых изданиях, рекомендованных ВАК, и 7 – в сборниках трудов конференций.

**Статьи в рецензируемых изданиях:**

1. **Моторина Г.Г.**, Кудрявцев И.В., Лазутков В.П., Матвеев Г.А., Савченко М.И., Скородумов Д.В., Чариков Ю.Е. К вопросу о реконструкции энергетического распределения электронов, ускоренных во время солнечных вспышек // Журнал технической физики. – 2012. – Т.82. – №12. – С.11-15. – doi: 10.1134/S1063784212120201.

2. **Motorina G.G.**, Kontar E.P. Differential emission measure and electron distribution function reconstructed from RHESSI and SDO observations // Geomagnetism and Aeronomy. – 2015. – V.55. – №7. – P.995-999. – doi: 10.1134/S0016793215070154.

3. Battaglia M., **Motorina G.**, Kontar E.P. Multi-thermal representation of the kappa-distribution of solar flare electrons and application to simultaneous X-ray and EUV observations // Astrophysical Journal. – 2015. – V.815. – №1. – Id.73. – 8 p. – doi: 10.1088/0004-637X/815/1/73.

4. **Моторина Г.Г.**, Кудрявцев И.В., Лазутков В.П., Савченко М.И.,

Скородумов Д.В., Чариков Ю.Е. Реконструкция энергетического спектра электронов, ускоренных во время солнечной вспышки 15 апреля 2002 года, на основе измерений рентгеновским спектрометром ИРИС // Журнал технической физики. – 2016. – Т.86. – №4. – С.47-52. – doi: 10.1134/S1063784216040186.

5. Tsap Yu.T., **Motorina G.G.**, Kopylova Yu.G. Flare coronal loop heating and hard X-ray emission from solar flares of August 23, 2005, and November 9, 2013 // Geomagnetism and Aeronomy. – 2016. – V.56. – №8. – P.1104-1109. – doi: 10.1134/S0016793216080235.

**Статьи в сборниках трудов конференций:**

1. **Нахатова Г.Г.**, Кудрявцев И.В. К вопросу о реконструкции энергетических спектров ускоренных во время солнечных вспышек электронов, на основе данных по тормозному рентгеновскому излучению // Сборник трудов XIV ежегодной конференции по физике Солнца «Солнечная и солнечно-земная физика-2010». – 2010. – ГАО РАН, Санкт-Петербург. – С.287-290.

2. **Нахатова Г.Г.**, Кудрявцев И.В. К вопросу о реконструкции энергетических спектров ускоренных во время солнечных вспышек электронов // Труды XII конференции молодых ученых «Взаимодействие полей и излучения с веществом». – 2011. – Иркутск. – С.90-92.

3. **Моторина Г.Г.**, Кудрявцев И.В., Лазутков В.П., Матвеев Г.А., Савченко М.И., Скородумов Д.В., Чариков Ю.Е. Восстановление энергетического распределения электронов, ускоренных во время солнечной вспышки 26 июля 2002 года, по данным жесткого рентгеновского излучения // Сборник трудов XV ежегодной конференции по физике Солнца «Солнечная и солнечно-земная физика-2011». – 2011. – ГАО РАН, Санкт-Петербург. – С.171-174.

4. **Моторина Г.Г.**, Кудрявцев И.В., Лазутков В.П., Матвеев Г.А., Савченко М.И., Скородумов Д.В., Чариков Ю.Е. Реконструкция

энергетического спектра электронов, ускоренных в солнечной вспышке 15 апреля 2002 года // Сборник трудов XVI ежегодной конференции по физике Солнца «Солнечная и солнечно-земная физика-2012». – 2012. – ГАО РАН, Санкт-Петербург. – С.301-304.

5. **Моторина Г.Г.**, Кудрявцев И.В., Лазутков В.П., Савченко М.И., Скородумов Д.В., Чариков Ю.Е. Эволюция энергетических спектров жесткого рентгеновского излучения солнечной вспышки 15 апреля 2002 года // Сборник трудов XVII ежегодной конференции по физике Солнца «Солнечная и солнечно-земная физика-2013». – 2013. – ГАО РАН, Санкт-Петербург. – С.161-164.

6. **Моторина Г.Г.**, Контарь Э.П. Дифференциальная мера эмиссии, полученная в результате комбинирования RHESSI, SDO/AIA наблюдений // Сборник трудов XVIII ежегодной конференции по физике Солнца «Солнечная и солнечно-земная физика-2014». – ГАО РАН, Санкт-Петербург. – С.307-310.

7. **Моторина Г.Г.**, Контарь Э.П. Временная эволюция энергетического распределения электронов в солнечных вспышках на основе RHESSI и SDO/AIA наблюдений // Сборник трудов XIX ежегодной конференции по физике Солнца «Солнечная и солнечно-земная физика-2015». – 2015. – ГАО РАН, Санкт-Петербург. – С.289-292.

**Основные результаты диссертации докладывались на следующих конференциях:**

1. Всероссийская ежегодная конференция по физике Солнца «Солнечная и солнечно-земная физика-2010», 3-9 октября 2010 г., Санкт-Петербург, Россия.

2. Всероссийская ежегодная конференция по физике Солнца «Солнечная и солнечно-земная физика-2011», 2-8 октября 2011 г., Санкт-Петербург, Россия.

3. Международная байкальская молодежная научная школа по

фундаментальной физике, 19-24 сентября 2011 г., Иркутск, Россия.

4. Всероссийская ежегодная конференция по физике Солнца «Солнечная и солнечно-земная физика-2012», 24-28 сентября 2012 г., Санкт-Петербург, Россия.

5. Российская молодёжная конференция по физике и астрономии «Физика СПб», 24-25 октября 2012 г., Санкт-Петербург, Россия.

6. Восьмая ежегодная конференция «Физика плазмы в солнечной системе», 4-8 февраля 2013 г., Москва, Россия.

7. Всероссийская астрономическая конференция «Многоликая Вселенная» (ВАК-2013), 23-27 сентября 2013 г., Санкт-Петербург, Россия.

8. Российская молодёжная конференция по физике и астрономии «Физика СПб», 23-24 октября 2013 г., Санкт-Петербург, Россия.

9. The 40th COSPAR scientific assembly, 2-10 August, 2014, Moscow, Russia.

10. RADIOSUN Workshop on solar flares and energetic particles at Pulkovo Observatory, 11-14 August, 2014, Saint-Petersburg, Russia.

11. Всероссийская ежегодная конференция по физике Солнца «Солнечная и солнечно-земная физика-2014», 20-24 октября 2014 г., Санкт-Петербург, Россия.

12. Glasgow-Cambridge Mini Flare Workshop, 15-16 April, 2015, Glasgow, UK.

13. The XIII Russian-Finnish Symposium on Radio Astronomy «Multi-Wavelength Study of Stellar Flares and the Properties of Active Galactic Nuclei», 25-29 May, 2015, Saint-Petersburg, Russia.

14. 14th RHESSI Workshop, 11-15 August, 2015, Newark, New Jersey, USA.

15. First Joint Solar Probe Plus-Solar Orbiter Workshop «The Origins of the Heliosphere», 2-4 September, 2015, Florence, Italy.

16. Всероссийская ежегодная конференция по физике Солнца

«Солнечная и солнечно-земная физика-2015», 5–9 октября 2015 г., Санкт-Петербург, Россия.

17. Всероссийская ежегодная конференция по физике Солнца «Солнечная и солнечно-земная физика-2016», 10–14 октября 2016 г., Санкт-Петербург, Россия.

Кроме того, результаты были представлены и обсуждались на научных семинарах в ГАО РАН (Санкт-Петербург, Россия), ИСЗФ РАН (Иркутск, Россия), Университете Глазго (Глазго, Великобритания), Обсерватории Онджеева (Онджеев, Чехия).

Апробацией результатов является участие в научных проектах РФФИ №14-02-00924а «Радио- и рентгеновская диагностика ускоренных электронов в солнечных вспышках»; №15-02-03835а «Исследование энерговыделения в активных областях с помощью многоволновых наблюдательных данных и современного трехмерного моделирования»; №16-32-000535мол_а «Новые наблюдения и диагностика миллиметрового излучения солнечных вспышек»; №16-32-50055мол_нр «Вспышечное энерговыделение в солнечных корональных петлях по данным рентгеновских наблюдений», ФЦП «Кадры» №8524, Marie Curie International Research Staff Exchange Scheme «Radiosun» (PEOPLE–2011–IRSES–295272), РНФ №16-12-10448, программах ПРАН П-7 и НШ-7241.2016.2, а также получение Гранта для студентов ВУЗов, расположенных на территории Санкт-Петербурга, аспирантов ВУЗов, отраслевых и академических институтов, расположенных на территории Санкт-Петербурга в 2013, 2015г.г. и Стипендии Президента Российской Федерации для обучения за рубежом студентов и аспирантов российских ВУЗов в 2013/2014 учебном году, которая позволила автору пройти стажировку в аспирантуре Университета Глазго (научный руководитель – к.ф.-м.н. Э.П. Контарь).

## Личный вклад автора

Результаты исследований отражены в работах [12-20, 33, 100, 131]. Автор принимала участие в постановке задач, проведении теоретических расчетов, обработке, анализе и интерпретации результатов наблюдений с космических аппаратов (КОРОНАС-Ф, SDO/AIA, RHESSI, GOES). Автором были созданы расчетные программы для реконструкции энергетических распределений электронов по наблюдательным данным в крайнем ультрафиолетовом и рентгеновском диапазонах. В работе [100] автором была предложена функциональная форма для дифференциальной меры эмиссии, и выведено выражение для энергетического распределения электронов во вспышечном источнике. Определение задач исследования, обсуждение полученных результатов и подготовка статей к публикации проводилось совместно с научным руководителем и соавторами.

## Структура и объем диссертации

Диссертационная работа состоит из введения, трех глав, заключения, списка литературы. Общий объем диссертации составляет 120 страниц, включая 36 рисунков и 2 таблицы. Список литературы включает 141 наименование.

## Краткое содержание диссертации

Во **введении** отражены актуальность исследования, цель работы, научная новизна, теоретическая и практическая значимость, положения, выносимые на защиту, степень достоверности и апробация результатов, личный вклад автора и краткое содержание диссертации.

**Глава 1** посвящена реконструкции энергетических спектров ускоренных в солнечных вспышках электронов по данным жесткого РИ. Во введении к главе 1 описаны цели главы 1 и используемые методы обработки данных. Результаты главы 1 легли в основу работ [12-15, 18-20]. В разделе 1.1.1 описаны элементарные процессы (механизмы) в плазме, ответственные за излучение в рентгеновском и крайнем ультрафиолетовом диапазонах во время солнечных вспышек. Приведено сравнение данных механизмов на примере конкретных событий. Раздел 1.1.2 посвящен описанию модельного подхода к реконструкции спектров ускоренных электронов по данным жесткого РИ. Приведены модели столкновительной тонкой и толстой мишени [36, 24], модель «теплой» толстой мишени (*warm thick-target*) [79] для описания нетепловой части энергетических распределений электронов. Рассмотрена модель «квазитепловой» плазмы для описания тепловой части энергетических спектров ускоренных электронов, показана связь нетепловой и тепловой частей и необходимость определения нижней энергетической границы $E_c$. Раздел 1.1.3 описывает немодельный подход к реконструкции спектров ускоренных электронов по данным жесткого РИ, который главным образом основан на методах регуляризации, в частности, рассмотрен метод регуляризации Тихонова [25]. Данный метод позволяет решать напрямую некорректную (вследствие наличия погрешностей в наблюдательных данных) обратную задачу путем добавления параметра регуляризации. Представлено сравнение модельного и немодельного подходов для одного события, сделаны выводы об их достоинствах и недостатках. Раздел 1.2 описывает методику обработки данных спектрометра ИРИС на борту КА КОРОНАС-Ф на примере вспышки 15 апреля 2002 г., для которой были зарегистрированы все фазы эволюции вспышки. В разделе 1.3 рассмотрен метод восстановления энергетического спектра жесткого РИ по данным регистрации с учетом приборной функции спектрометра ИРИС с помощью метода случайного поиска в комбинации с методом наименьших квадратов. Обсуждаются особенности восстановленных спектров жесткого РИ. В разделе 1.4 проведена реконструкция энергетических распределений ускоренных электронов по восстановленным спектрам жесткого РИ с использованием метода регуляризации Тихонова нулевого порядка для различных значений параметра регуляризации *α*. Рассмотрена подробная эволюция спектров жесткого РИ и спектров ускоренных электронов. Для данного события выявлена особенность в спектрах электронов, связанная с наличием локального минимума в области энергий 40−60 кэВ, которая не может быть выявлена прямыми методами. Аналогично, для вспышки 26 июля 2002 наблюдалась подобная особенность в спектре в диапазоне 40-50 кэВ [13]. В разделе 1.5 приведено заключение к главе 1 и основные выводы.

**Глава 2** посвящена восстановлению энергетических распределений электронов на основе одновременного анализа КУФ и мягкого РИ. Результаты главы 2 легли в основу работ [16, 17, 33, 100]. Раздел 2.1 описывает основные вопросы, связанные с получением информации по данным КУФ и мягкого РИ, а также цели и задачи главы 2. В разделе 2.2 введена дифференциальная мера эмиссии (ДМЭ, *differential emission measure*) и описана ее связь с энергетическим распределением надтепловых электронов. В разделе 2.2.1 предложена функциональная форма ДМЭ, выведен аналитический вид для энергетического распределения электронов. Раздел 2.2.2 описывает каппа-распределение через ДМЭ, а также вывод энергетического распределения электронов из ДМЭ в виде каппа-распределения. Раздел 2.3 посвящен комбинированному анализу данных КУФ и мягкого РИ, зарегистрированных спутниками SDO/AIA и RHESSI соответственно. В разделе 2.3.1 рассмотрен метод одновременной аппроксимации модельными функциями (*forward fitting method*) ДМЭ одновременно SDO/AIA и RHESSI данных, описана единая матрица температурного отклика, включающая температурный отклик КА SDO/AIA и температурный отклик КА RHESSI. В разделе 2.4 представлен анализ событий 14 августа 2010 г. и 8 мая 2015 г. с помощью разработанной методики анализа одновременно SDO/AIA и RHESSI данных. В разделе 2.4.1 для вспышки 14 августа 2010 г. произведено восстановление ДМЭ тремя различными способами: 1) метод регуляризации только SDO/AIA данных, используется программа *data2dem_reg.pro*, разработанная авторами [63]; 2) аппроксимация много-температурной ДМЭ функцией только данных КА RHESSI, используется *f_multi_therm_2pow.pro* (доступна в среде *OSPEX*); 3) аппроксимация предложенной ДМЭ и моделью тонкой мишени данных КА SDO/AIA и RHESSI одновременно. Производится сравнение новой методики аппроксимации модельными функциями данных КА SDO/AIA и RHESSI одновременно, показано, что результаты хорошо согласуются с наблюдательными данными (критерий $\chi^2$=0.83). По найденной ДМЭ построено энергетическое распределение электронов $\langle nVF(E) \rangle$ для диапазона энергий: 0.1–20 кэВ, а также представлены параметры вспышечной плазмы: мера эмиссии $EM$=4.64×10$^{46}$ см$^{-3}$, максимальная температура $T_{max}$=0.58 кэВ, спектральный индекс $\delta$=2.9, низкоэнергетическая граница $E_c$=7.78 кэВ. Раздел 2.4.2 описывает временную эволюцию параметров *EM*, *T*, концентрации *n* для события 8 мая 2015г., полученных с помощью рассматриваемого метода. Полученные результаты показывают присутствие горячей вспышечной плазмы и холодной фоновой плазмы вдоль луча зрения во вспышечном источнике. В разделе 2.5 рассматривается энергетическое распределение электронов в рамках каппа-распределения. Раздел 2.5.1 описывает применение аппроксимации данных КА SDO/AIA и RHESSI одновременно двумя функциями ДМЭ в виде каппа-распределения $\xi_\kappa^{cold}(T)$ и $\xi_\kappa^{hot}(T)$, соответствующие холодной (*cold*) и горячей (*hot*) компонентам вспышечной плазмы, к солнечной вспышке 14 августа 2010 г. Производится сравнение полученного энергетического распределения $\langle nVF(E) \rangle$ по SDO/AIA и RHESSI данным и $\langle nVF(E) \rangle$, вычисленного только из RHESSI данных. В разделе 2.5.2 обсуждаются результаты аппроксимации одновременно SDO/AIA и RHESSI данных модельными функциями $\xi_\kappa^{cold}(T)$ и $\xi_\kappa^{hot}(T)$. В разделе 2.5.3 производится вывод полной электронной концентрации и энергии солнечной вспышки, приведены оценки для события 14.08.2010. Полученные результаты показывают, что для типичной вспышки, число электронов, найденное с использованием новой методики ($n=0.45\times10^{10}$ см$^{-3}$), значительно меньше, чем с помощью стандартного каппа-распределения на основе модели тонкой мишени *thin_kappa* ($n=14\times10^{10}$ см$^{-3}$) [72], которое учитывает только электрон-ионное тормозное излучение, или одной ДМЭ функции $\xi_\kappa(T)$ ($n=4.5\times10^{10}$ см$^{-3}$), вычисленные только из RHESSI данных. В то же время учет свободно-связанного излучения от надтепловых электронов может давать значительный вклад в РИ [47, 48, 40, 56]. Это означает, что гораздо меньшее число надтепловых электронов, чем считалось ранее, присутствует в корональных источниках. Совместный анализ RHESSI и SDO/AIA данных подтверждает результат аппроксимации модельными функциями RHESSI данных отдельно, но необходим ввод дополнительной ДМЭ компоненты для учета низкотемпературной части излучения, регистрируемого КА SDO/AIA, к которому КА RHESSI малочувствителен, что свидетельствует о дополнительном вкладе корональной 1-2 МК плазмы. В разделе 2.6 приведены основные результаты.

**Глава 3** посвящена оценке теплового баланса горячих вспышечных петель на основе размерностных соотношений по данным РИ, зарегистрированного на КА RHESSI, на примере двух событий 23.08.2005 г. и 09.11.2013 г. Результаты главы 3 легли в основу работы [131]. В разделе 3.1 дано описание применимости «стандартной» CSHKP модели солнечной вспышки [41, 124, 80, 64, 115, 116, 132] и эффекта Нойперта [101] как основного критерия. В разделе 3.2 рассматривается скорость энергетических потерь петли, которая в первую очередь определяется электронной теплопроводностью. Так как рассматривается горячая вспышечная плазма с температурой $T \geq 10^7$ К, радиационными потерями можно пренебречь [105, 45]. Производится оценка порогового значения энергии $E_{loop}$, ниже которой ускоренные электроны термализуются в короне [136, 118]. Для значений $nL = 6\times10^{19}$ см$^{-2}$ и $\mu = 0.5$, где $L$ — длина корональной петли, $n$ — концентрация электронов, $\mu$ - косинус питч-угла ускоренного электрона, $E_{loop}$ = 25 кэВ. Выводится зависимость отношения потоков жесткого РИ в основаниях петли и корональной части от относительной энергии фотонов при различных значениях показателя спектра $\delta$. В разделе 3.3 производится анализ событий 23.08.2005 г., 09.11.2013 г. Для события 23 августа 2005 г. рассмотрена временная эволюция меры эмиссии и температуры, показателя спектра, интегрального потока и нижнего предела энергии ускоренных электронов, зависимости отношения характерных времен нагрева $\tau_h$ и охлаждения $\tau_{cond}$ плазмы. Обсуждаются результаты и сделан вывод, что ускоренные электроны не могут обеспечить нагрев корональной плазмы. Для события 9 ноября 2013 г. было произведено сравнение полученных оценок с картами КУФ и рентгеновскими контурами в диапазоне энергий 23–27.5 кэВ, где из оценок следует, что интенсивность в вершине петли должна преобладать по сравнению с интенсивностью в основаниях, что не соответствует источникам жесткого РИ. Полученные результаты представлены в разделе 3.4.

В **заключении** сделаны выводы по проведенной в диссертации работе.

# Глава 1
# Реконструкция энергетических распределений электронов, ускоренных во время солнечных вспышек, на основе жесткого рентгеновского излучения

## 1.1 Введение к главе 1

В данной главе рассмотрен вопрос реконструкции спектра ускоренных электронов непосредственно в области источника жесткого РИ. Глава 1 посвящена решению обратной задачи с использованием последовательно метода случайного поиска и метода регуляризации Тихонова для решения интегральных уравнений, с помощью которых производится реконструкция энергетических распределений ускоренных во время солнечных вспышек электронов для различных моментов времени на основе спектров жесткого РИ излучения, а также исследованию эволюции процесса ускорения электронов. С помощью метода случайного поиска в комбинации с методом наименьших квадратов производилось восстановление исходного («истинного») спектра тормозного РИ на основе аппаратурных спектров. Это необходимо делать для учета погрешности измерений, полученных спектрометром при регистрации жесткого РИ. Таким образом, получив «истинный» спектр, были реконструированы спектры распределений быстрых электронов в солнечной вспышке. Для этого была разработана методика, которая позволяет находить спектры распределений высокоэнергичных электронов не в первом приближении, а непосредственно по регистрируемым спектрам фотонов получать истинные спектры распределения высокоэнергичных электронов.

**Цель главы** – восстановление по полученным спектрам жесткого РИ энергетических распределений электронов с учетом энергетического разрешения спектрометров, с помощью которых производится регистрация жесткого РИ солнечных вспышек, изучение полученных особенностей реконструированных спектров ускоренных во вспышках электронов с целью выявления физических механизмов, связанных с ускорением заряженных частиц во время солнечных вспышек.

В процессе работы были апробированы метод регуляризации Тихонова и метод случайного поиска (оптимизации) и созданы расчетные программы на языке Си для поиска квазиоптимального решения и дальнейшей реконструкции энергетических распределений электронов, ускоренных в солнечных вспышках. С помощью разработанной методики комбинации этих двух методов обработано конкретное вспышечное событие (12.04.2002). Результаты главы 1 опубликованы в работах [12-15, 18-20].

## 1.1.1 Элементарные процессы в плазме, ответственные за излучение в рентгеновском и крайнем ультрафиолетовом диапазонах

Диагностика РИ, которое происходит непосредственно в оптически тонкой среде, является одним из наиболее прямых методов, с помощью которых проводится исследование энергичных электронов в солнечных вспышках. Основными механизмами, ответственными за излучение плазмы в рентгеновской области спектра, являются свободно-свободные (тормозное излучение, или *bremsstrahlung*), свободно-связанные (рекомбинационное излучение) и связанно-связанные переходы (линейчатое излучение).

Под свободно-свободными переходами понимается торможение ускоренных электронов на частицах солнечной плазмы, сопровождаемое излучением квантов, где доминирующими взаимодействиями в основном являются электрон-ионные (для энергий выше ~10-20 кэВ). Вклад электрон-электронного тормозного излучения в РИ, как правило, игнорируется, хотя для электронов (и фотонов) с энергиями выше ~300 кэВ такое упущение не оправдано, и его учет в целом делает спектр электронов, необходимых для генерации заданного жесткого рентгеновского спектра, более крутым на высоких энергиях [76]. Таким образом, тормозное излучение наиболее эффективно для очень горячей плазмы с температурой $T>10^6$ К и высокой степенью ионизации, и главным образом ответственно за непрерывное излучение солнечной короны (см., напр., [30, 21]). На рис. 1.1 [138] представлено сравнение вклада трех механизмов излучения (тормозное, рекомбинационное и двухфотонное) в континуум для изотермических спектров на основе численных расчетов атомной базы данных *CHIANTI 4.2* [54, 85] для температуры 10 МК. Более подробно база данных *CHIANTI* будет описана в главе 2.

Свободно-связанными переходами является процесс захвата свободного электрона ионом или атомом с последующим испусканием кванта, однако второй вариант менее вероятен [6]. Для горячих корональных источников с температурой T~10-30 МК рекомбинационное излучение может быть значительным (см. рис. 1.1) и давать существенный вклад в РИ [47, 48, 23, 40, 56]. Для корональных обилий элементов оба вклада (свободно-свободный и свободно-связанный) в континуум примерно одинаковы при 20 МК [138]. Что касается двухфотонного излучения, которое заключается в излучении двух фотонов при запрещенных переходах атомов и ионов водорода и гелиеподобных ионов, то, из рис. 1.1 видно, что для рентгеновской плазмы с $T$~10 МК его вклад является незначительным. Рис. 1.1 взят из работы [138].

Вследствие связанно-связанных переходов при переходах электронов в атомах между энергетическими уровнями происходит поглощение и рассеяние в линиях, причем электроны до и после взаимодействия находятся в связанном состоянии. Данный вид излучения возникает в хромосфере/переходной области/короне и проявляется в оптическом, ультрафиолетовом и мягком рентгеновском солнечном спектре.

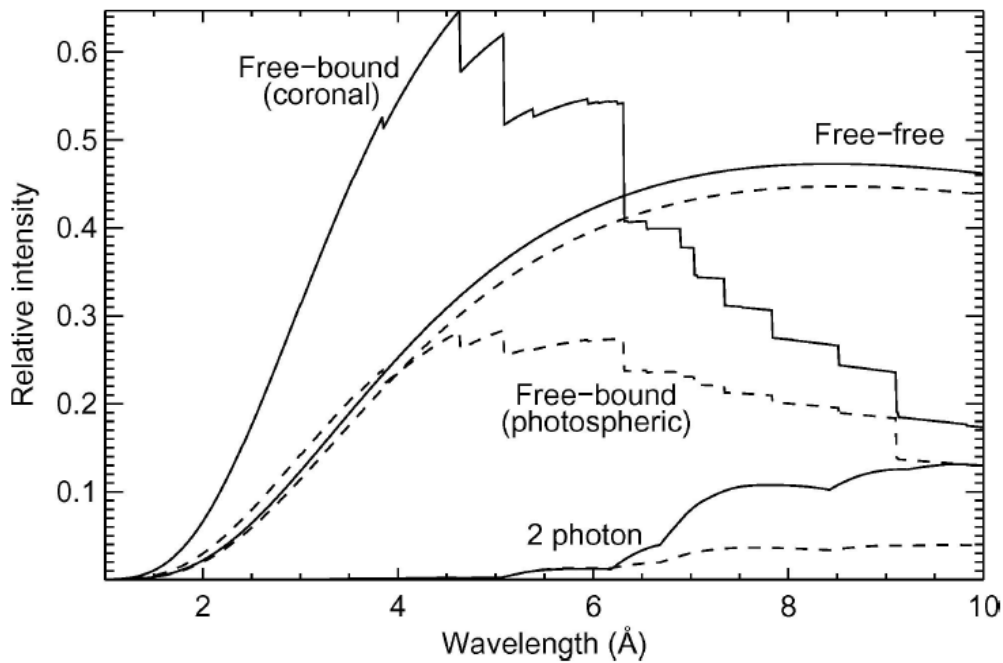

Рис. 1.1. Сравнение вклада трех механизмов в континуум для изотермических спектров, рассчитанных с помощью *CHIANTI 4.2* при 10 МК. Спектры при свободно-связанных (*free-bound*), свободно-свободных (*free-free*) и двухфотонных (*2 photon*) переходах для корональных (сплошные линии) и фотосферных (пунктирные линии) обилий элементов. Кривые нормированы к пиковой интенсивности суммарного излучения для корональных обилий. Острые края в спектрах свободно-связанных переходов представляют переходы ионов в различные состояния.

Для спектра Солнца в крайнем ультрафиолетовом (КУФ) диапазоне, от 10 до 1200 Å, доминирует излучение спектральных линий водорода (*H*), гелия (*He*), кислорода (*O*), натрия (*Na*), магния (*Mg*), кремния (*Si*) и железа (*Fe*). На рис. 1.2 представлен пример спектра спокойного Солнца, зарегистрированного изображающим КУФ спектрометром, Extreme ultraviolet Imaging Spectrometer (EIS), на борту спутника Hinode [49]. Рис. 1.2 взят из работы [108]. Спектр включает в себя резонансные линии ионов железа: от *Fe VIII* (формируется при чуть менее 1 МК) до *Fe XIV* (~2 МК) [108].

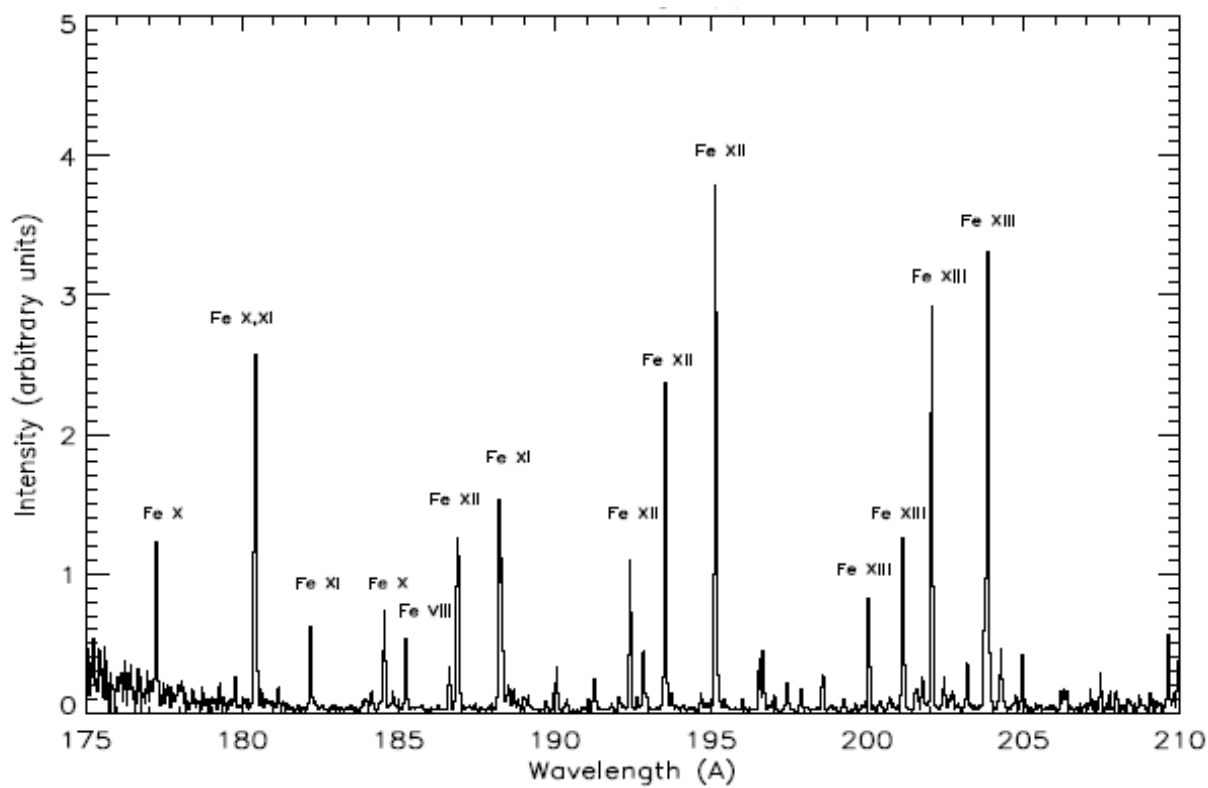

Рис. 1.2. Спектр спокойного Солнца, зарегистрированный спектрометром EIS на борту спутника Hinode в диапазоне 175-210 Å, включающий в себя линии ионов *Fe*, образовавшихся при корональных температурах.

Следует также отметить, что во время солнечных вспышек, в зависимости от их особенностей, вклад в КУФ диапазон от свободно-свободных, свободно-связанных и связанно-связанных переходов может быть значительным (см., напр., [98, 99]). Возрастание и убывание потока КУФ излучения (10–1030 Å) может коррелировать с жестким РИ в диапазоне энергий 10-20 кэВ [55, 57] вследствие взаимодействия солнечной плазмы со свободными электронами, освобожденными нетепловыми частицами, ускоренными во время импульсной фазы вспышки (т.е. свободно-свободные, свободно-связанные и связанно-связанные переходы).

На рис. 1.3 (гистограмма) приведен пример спектра потока фотонов солнечной вспышки 26 апреля 2006 года, зарегистрированного КА RHESSI, и сравнение с синтетическим спектром изотермической плазмы с $T = 14$ МК, рассчитанного с использованием кода *CHIANTI* [54]. При данной температуре можно видеть, что особенность в спектре РИ при *E*=6.7 кэВ в основном состоит из линий сильно ионизированного *Fe*, а более слабая особенность при *E*=8 кэВ – *Fe/Ni* линий [107]. Рис. 1.3. взят из работы [108].

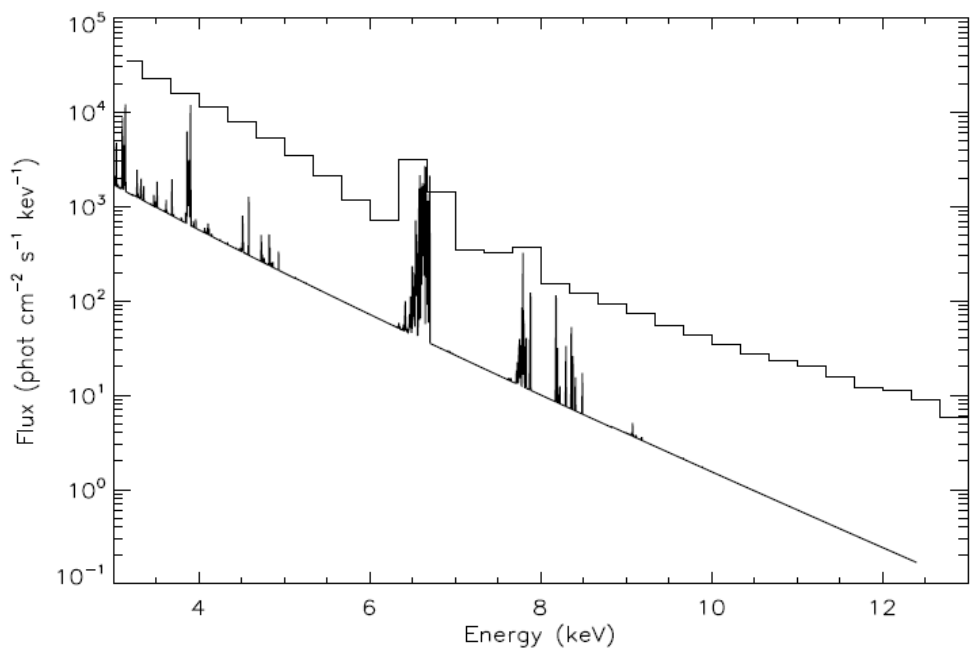

Рис. 1.3. Спектр фотонов солнечной вспышки 26 апреля 2006 года, зарегистрированный с помощью КА RHESSI, (гистограмма) и синтетический спектр фотонов, рассчитанный для изотермической плазмы с *T*=14 МК, используя код *CHIANTI*.

Суммируя все вышесказанное, можно сделать вывод, что жесткое (нетепловое) РИ в солнечных вспышках, как правило, определяется тормозным механизмом, возникающим в результате излучения фотонов при рассеянии (торможении) электронов в электрическом кулоновском поле ионов окружающей среды. Данное излучение энергичных электронов является более эффективным, чем обратное комптоновское рассеяние, возникающее в результате увеличения энергии фотонов при рассеянии на свободных электронах, или синхротронное излучение из той же популяции электронов, которое определяется движением релятивистских электронов в магнитном поле [81, 82]. В свою очередь, мягкое (тепловое) РИ и КУФ излучение, также тепловое излучение, в основном определяются тормозным, рекомбинационным и линейчатым излучением.

## 1.1.2 Реконструкция распределения ускоренных электронов по данным жесткого рентгеновского излучения (модельный подход)

Как отмечалось выше, жесткое РИ, генерируемое во время солнечных вспышек, является тормозным излучением высокоэнергичных электронов. Параметры жесткого РИ - спектр, поток, направленность и поляризация несут непосредственную информацию о функции распределения ускоренных электронов в области источника излучения, знание которой с учетом процессов переноса в плазме магнитной петли необходимо для решения вопроса о механизме ускорения частиц (см., напр., [43, 96]).

Для определения спектра инжектируемых электронов по наблюдаемому спектру рентгеновских квантов впервые Брауном [36] была предложена модель толстой мишени, в которой рассмотрен случай степенной зависимости спектра фотонов $\sim \varepsilon^{-\delta}$, для которого было найдено решение для спектра инжектируемых электронов $\sim E^{-(\delta+1)}$. Данная задача восстановления спектра ускоренных электронов традиционно решалась в приближении модели толстой–тонкой мишени [36, 24, 90, 68, 50, 23, 9] (прямая задача) при условии, что угловой зависимостью тормозного сечения, направленностью, коллективными потерями, влиянием магнитного поля и временными характеристиками распределения электронов можно пренебречь. Для данных моделей используется нерелятивистское сечение тормозного излучения [74], что позволяет определить спектр электронов, ускоренных в солнечной вспышке, по спектру наблюдаемых рентгеновских квантов, а сами модели представляют собой два предельных случая. Рассмотрим работу [9], в которой обсуждаются оба случая. Чтобы связать энергетическое распределение электронов в излучающем объеме и в источнике, рассматривается уравнение непрерывности [89, 5, 24]

$$\frac{\partial N(E,t)}{\partial t} = f(E,t) - \frac{N(E,t)}{\tau_e} - \frac{\partial}{\partial E}\left( N(E,t)\frac{dE}{dt} \right) \qquad (1.1.1),$$

где *N* - усредненный спектр излучающих электронов, $\tau_e$ – характерное время выхода из излучающей области, *f* – инжектируемый спектр электронов [электроны/(кэВ с)]. Предполагая независимость от времени *t* (условие квазистационарности), степенную зависимость $N(E)=KE^{-(\delta-0.5)}$, *K*-нормировочный коэффициент (определяется количеством излучаемых квантов (см. [9])), учет только кулоновских потерь $dE/dt \approx -5.8 \cdot 10^{-9} nE^{-1/2}$ [кэВ/с], выражение (1.1.1) примет вид

$$f(E) = N(E)\left(\frac{1}{\tau_e} + \frac{1}{\tau_c}\right) \quad (1.1.2),$$

где $\tau_c = \dfrac{E^{3/2}}{5.8 \cdot 10^{-9} n \delta}$.

Если время жизни $\tau_c$ в кулоновских соударениях пучка электронов с частицами плазмы много меньше времени выхода ($\tau_c \ll \tau_e$), то есть ускоренные электроны сталкиваются с тепловой плазмой и теряют всю свою энергию в источнике жесткого РИ (модель толстой мишени)

$$f(E) = \frac{N(E)}{\tau_c} \approx 5.8 \cdot 10^{-9} \delta n E^{-3/2} N(E) \approx 5.8 \cdot 10^{-9} \delta n K E^{-(\delta+1)} \quad (1.1.3).$$

Если время жизни в кулоновских соударениях много больше времени выхода ($\tau_c \gg \tau_e$), то есть ускоренные электроны, теряя малую долю своей энергии, покидают мишень (модель тонкой мишени)

$$f(E) = \frac{N(E)}{\tau_e} \approx E^{1/2} K E^{-(\delta-1/2)} \approx K E^{-(\delta-1)} \quad (1.1.4).$$

Таким образом для нерелятивистского сечения тормозного излучения показатели степенного спектра жесткого РИ γ и ускоренных электронов δ в этих моделях однозначно связаны: γ=δ-1 (толстая мишень) и γ=δ+1 (тонкая мишень). Отсюда следует, что спектр электронов источника зависит от характерного времени выхода из излучающей области $\tau_e$. В работе [8] на основе решения кинетического уравнения для функции распределения электронов с учетом углового распределения в модели толстой мишени и для степенного спектра инжектируемых электронов были получены параметры жесткого РИ (спектр, степень линейной поляризации и направленность излучения). При этом модель толстой мишени подразумевает пренебрежение выходом излучающих электронов из области генерации излучения.

Что касается наблюдательного подтверждения, то в случае модели толстой мишени предполагается, что ускоренные электроны тормозятся в плазме и остаются в ней, что соответствует источникам жесткого РИ, наблюдаемым на уровне холодной, столкновительно плотной хромосферы [36, 67]. Для случая тонкой мишени ускоренные электроны проходят сквозь мишень, не оставаясь в ней, и жесткий спектр РИ почти идентичен спектру ускоренных или инжектируемых электронов [90, 68], т.е. излучение генерируется электронами, уходящими в верхние слои солнечной плазмы с относительно малой плотностью и при этом их функция распределения не успевает релаксировать в результате кулоновских столкновений. В залимбовых вспышках, где предполагаемые источники жесткого РИ в основаниях вспышечной петли закрыты диском Солнца, протяженные источники жесткого РИ, наблюдаемые на значительной высоте в короне, объясняются моделью тонкой мишени (см., напр., [50, 129]). Так как корональные источники жесткого РИ обычно гораздо слабее хромосферных источников, в основаниях предполагается, что тормозное излучение в виде модели толстой мишени является доминирующим. На рис. 1.4 (левая панель) представлена типичная вспышка с источником жесткого РИ в основаниях петли и источником мягкого РИ в вершине петли для солнечной вспышки 19 января 2005 года [83]. Рис. 1.4 взят из работы [83].

Однако, наблюдения на КА RHESSI выявили новый класс событий, в которых жесткое РИ исходит преимущественно из корональной части петли, и лишь незначительная его часть из оснований [136, 125, 61]. В этих событиях сама корона выступает в роли толстой мишени для инжектируемого пучка электронов, и, таким образом, большинство электронов никогда не достигнет хромосферы и не сгенерирует там жесткое РИ (см. рис. 1.4, правая панель).

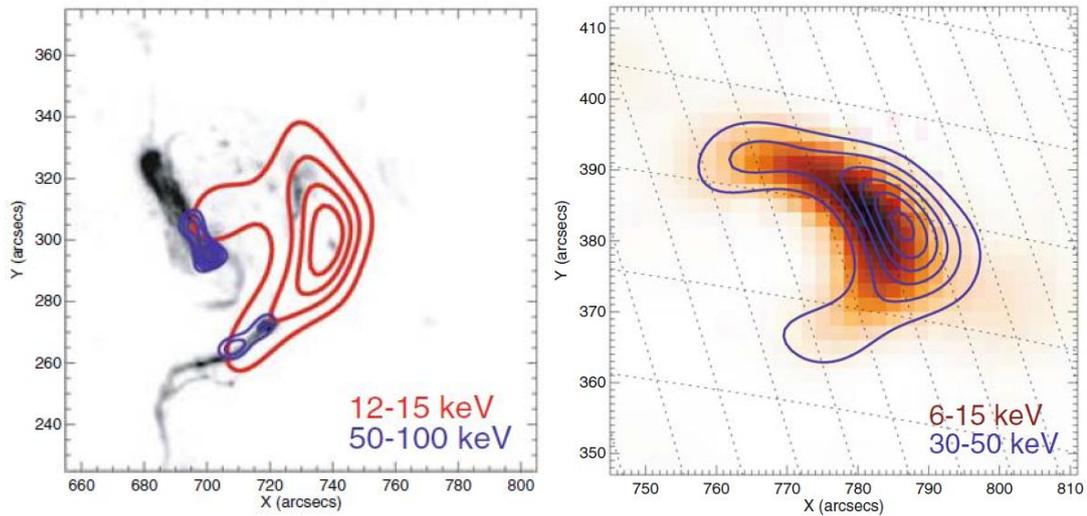

Рис. 1.4. Слева: карта, полученная КА TRACE для длины волны 1600 Å, с наложенными RHESSI контурами для энергий 12-15 кэВ (красные линии) и 50-100 кэВ (синие линии) для солнечной вспышки 19 января 2005 года за промежуток времени 08:11:40-08:13:41 UT. Справа: RHESSI изображение, реконструированное с использованием алгоритма MEM (метод максимальной энтропии) [69], для энергий 6-15 кэВ, для события 14/15 апреля 2002 года в момент времени 00:02-00:12 UT, с наложенными контурами (синие линии) для энергий 30-50 кэВ.

Следует также отметить, что данные модели определяют верхнюю границу спектра ускоренных электронов, тогда как нижнюю границу $E_c$ невозможно определить без учета тепловой компоненты излучения. На рис. 1.5 представлен типичный спектр жесткого РИ, аппроксимируемый тепловой и нетепловой составляющими. Рис. 1.5 взят из работы [65]. Решение проблемы определения $E_c$ было представлено в работе [79], где была предложена модель «теплой» толстой мишени (*warm thick-target*) тормозного излучения, включающая эффекты столкновительной диффузии и термализации быстрых электронов. Режим «теплой» толстой мишени соответствует корональной и нагретой хромосферной плазме, а дополнительные параметры в модели необходимы для учета пространственных характеристик излучающей области, в частности, размеры мишени по сравнению с общей областью энерговыделения. Контарь и др. [79] представили аналитический подход и численные расчеты, в которых распределение инжектируемых электронов имеет степенной спектр на высоких энергиях, распределение Максвелла на низких энергиях и переход между этими двумя предельными случаями.

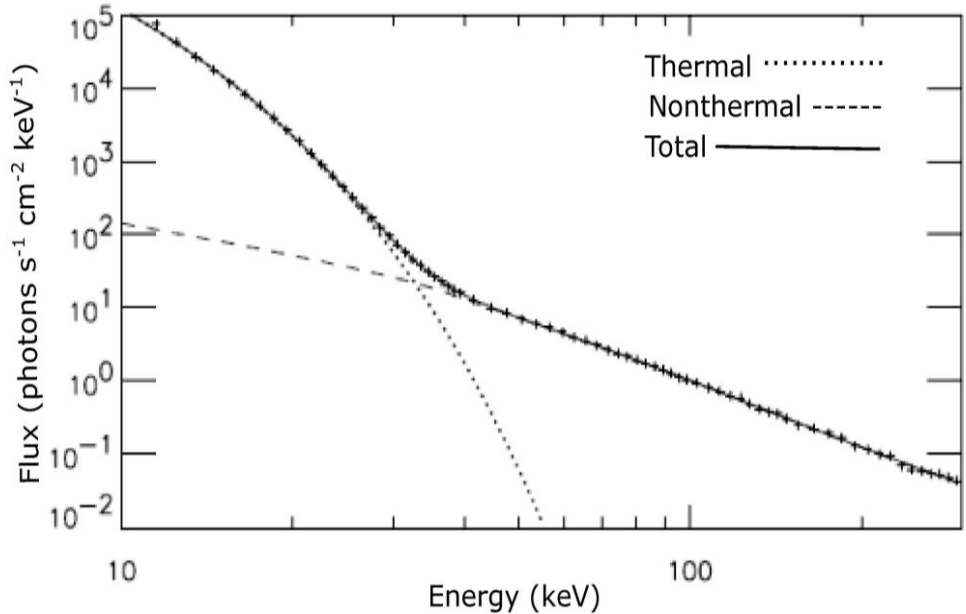

Рис. 1.5. Интегральный спектр фотонов для вспышки 23 июля 2002 (*total*, сплошная линия), аппроксимируемый тепловой и нетепловой моделями: изотермической *(thermal)* и двойной степенной функцией *(nonthermal)* в диапазоне энергий 10-300 кэВ за промежуток времени 00:30:00-00:30:20 UT.

Так как интерпретация жесткого РИ тесно связана с мягким РИ, то следует отметить модели, которые его описывают. Как правило, для описания теплового тормозного излучения рассматривается «квазитепловая» (однотемпературная) модель, предполагается тормозной механизм от оптически тонкой тепловой плазмы. Тогда интенсивность РИ оптически тонкой тепловой плазмы во вспышечной области объемом *V* на расстоянии в одну астрономическую единицу определяется [133, 46]

$$J(\varepsilon) = 1.07 \times 10^3 \frac{\sum_i EM_i Z_i^2}{\varepsilon \sqrt{T}} g_{ff}(\varepsilon, T) \exp(-\varepsilon/T) \qquad (1.1.7),$$

где *J(ε)* – интенсивность РИ, $Z_i$ и $EM_i = n_e n_i V$ - заряд и мера эмиссии ионов, $n_e$ и $n_i$ – концентрация электронов и ионов, $g_{ff}(E,T) \approx (T/E)^{0.4}$ – Гаунт-фактор для $T \leq \varepsilon$, *T* – температура [кэВ]. Выражение (1.1.7) можно упростить, введя дополнительные приближения, для сильно ионизированной плазмы $\sum_i EM_i Z_i^2 \approx 1.2 EM$, где $EM = n_e^2 V$ – мера эмиссии [см$^{-3}$], тогда
$$J(\varepsilon) \approx 1.07 \times 10^3 \left(1.2 EM / \varepsilon \sqrt{T}\right) g_{ff}(\varepsilon, T) \exp(-\varepsilon/T).$$

Таким образом, энергетический спектр высокоэнергичных электронов определяется на основе вида спектра РИ вспышек, который с учетом ошибок измерений аппроксимируется либо степенной функцией от энергии, либо законом квазитеплового излучения с заданными значениями температуры и меры эмиссии. Однотемпературное (изотермическое или *isothermal*) предположение часто является хорошим приближением для потока плазмы в трубке. Однако для интегральной интенсивности РИ вдоль луча зрения будет присутствовать вклад от многих трубок с различными температурами плазмы, поэтому, как правило, рассматривается локальное максвелловское распределение с локальными параметрами температуры *T*(**r**), концентрации *n*(**r**) и положения **r**. В главе 2 рассмотрена проблема восстановления энергетического спектра электронов в предположении много-температурной вспышечной плазмы.

Рассматривая модельный подход (аппроксимация модельными функциями или фитирование, *forward fitting method*), предполагается, что форма энергетического распределения электронов заведомо известна, и идет поиск параметров, которые определяют эту форму (см. рис. 1.5). С другой стороны, найденные параметры имеют физический смысл (спектральный индекс, полное число ускоренных электронов, полная энергия), что позволяет диагностировать источник излучения, рассматривать динамику вспышечных процессов и описывать определенные процессы во вспышечной плазме.

Для проведения подобных расчетов для главы 2 и главы 3 использовалось программное обеспечение *Solar Soft Ware* (*SSW*), являющееся набором интегрированных программных библиотек, баз данных и системных утилит (на языке программирования IDL), которое обеспечивает анализ данных. Частью *SSW* является объектно-ориентированный интерфейс *Object Spectral Executive* (*OSPEX*) [114] (см. http://hesperia.gsfc.nasa.gov/rhessi3/software/spectroscopy/spectral-analysis-oftware/index.html), использовавшийся для рентгеновского спектрального анализа данных. Список модельных функций и их параметров с описанием для аппроксимации рентгеновских спектров в *OSPEX* доступен в интернете (https://hesperia.gsfc.nasa.gov/ssw/packages/spex/idl/object_spex/fit_model_components.txt). Аппроксимация наблюдательных данных модельной функцией заключается в подборе параметров функции, для которых находится минимальное значение хи-квадрата ($\chi^2$) (см., напр., [110]).

## 1.1.3 Реконструкция распределения ускоренных электронов по данным жесткого рентгеновского излучения (немодельный подход)

Как отмечалось в разделе 1.1.2, для нахождения энергетического распределения электронов, излучающих жесткое РИ во время солнечных вспышек, необходимо решать обратную задачу относительно функции распределения ускоренных электронов. Данное интегральное уравнение можно решать различными методами, например, методом квадратурных формул: формулы прямоугольников и трапеций (формулы замкнутого типа), которые обычно используются для вычисления интегралов на больших промежутках. Однако, выбор квадратурной формулы должен быть согласован со свойствами ядра.

Следует отметить, что характерной чертой решаемой задачи является неустойчивость ее решения, т.е. малые изменения исходных данных могут привести к большим изменениям решения: погрешность в исходных данных может играть значительную роль. При наличии погрешности в данных эта неустойчивость может привести к «расходящимся» результатам численного вычисления. Поэтому, чтобы избежать такой проблемы, используются методы устойчивого решения некорректных задач. Одним из подходов решения задач такого рода являются методы регуляризации (методы регуляризации Тихонова и другие методы регуляризации). В основе данных методов лежит понятие регуляризирующего алгоритма. Согласно методу регуляризации Тихонова [25], для обеспечения устойчивости решения вводится условие минимума сглаживающего функционала, где вводится параметр регуляризации, который необходимо определить.

Преимущество метода регуляризации Тихонова перед модельными подходами заключается в том, что он наиболее точно воспроизводит форму исходного излучения, не предполагая никакой аналитической формы. В то же время, метод регуляризации Тихонова не может дать значения физических параметров плазмы, и это является его недостатком. В работах [109, 78] представлено сравнение модельного (рис. 1.5) и немодельного подхода (рис. 1.6) для события 23 июля 2002 года. Рис. 1.6 взят из работы [65]. Видно, что для энергии ~50 кэВ на спектре наблюдается особенность – небольшой провал, который с помощью методов аппроксимации модельными функциями не удалось бы обнаружить. Причина такого локального минимума на энергетическом спектре в настоящее время остается не до конца выясненной. Кроме этого нельзя исключать, что данный минимум может быть связан и погрешностью измерения, так как результаты решения интегральных уравнений очень чувствительны к изменению формы аппаратурных спектров.

Таким образом, при реконструкции распределения ускоренных электронов по данным жесткого РИ используются как модельный, так и немодельный подходы. Приближение толстой-тонкой мишени является более упрощенной процедурой, возможной для жесткой части спектра (или даже отдельных поддиапазонов энергии), в целом для спектра в широкой области энергий от 10 кэВ до 10 МэВ такое приближение невозможно.

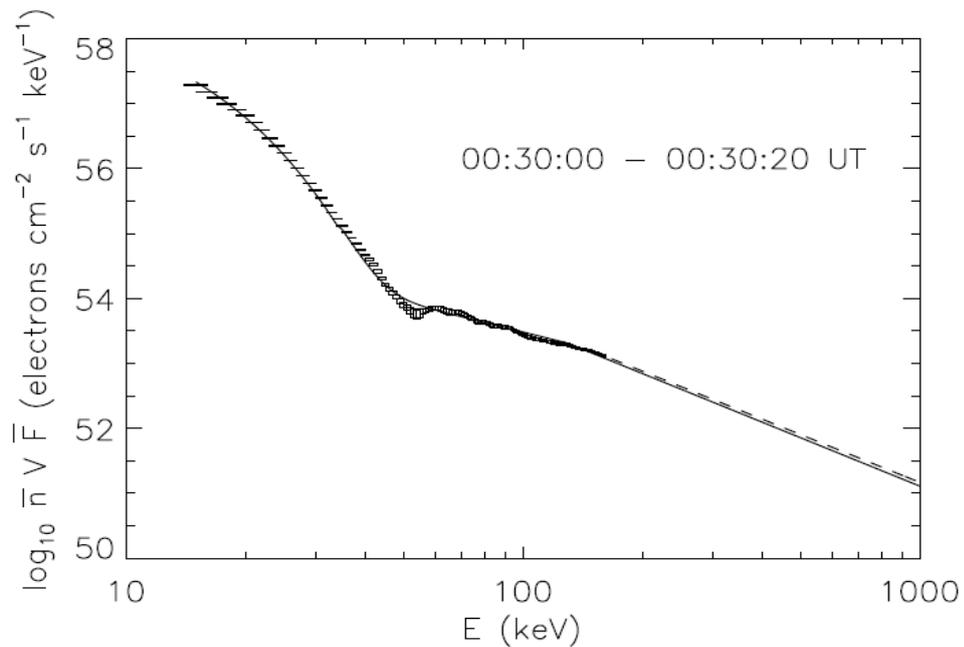

Рис. 1.6. Регуляризированный спектр электронов $\overline{n}V\overline{F}$ с продлением спектра в виде степенной зависимости до 1000 кэВ (пунктирная линия); спектр электронов $\overline{n}V\overline{F}$, полученный методом аппроксимации данных РИ тепловой и нетепловой моделями (сплошная линия) [65] для вспышки 23 июля 2002 года. Вертикальные бары для диапазона энергий 15-160 кэВ отражают ошибки равные 3σ вследствие статистического шума для наблюдаемого спектра фотонов КА RHESSI.

Более корректным, но и более трудоемким, является метод решения обратной задачи, связанный с решением интегрального уравнения и не требующий задания модели излучения. Аналитическое решение интегральных уравнений практически невозможно из-за сложной зависимости ядер уравнений от энергии электрона и рентгеновского кванта. Поэтому в обратной задаче используется численный метод на основе метода регуляризации Тихонова. Как показывают результаты подобных реконструкций [109, 38], в переходной области энергетического спектра возможны его уплощения и даже возрастания, которые невозможно получить прямым методом. Кроме того, как показывают последние измерения [83, 78, 66], жесткое РИ во время вспышек неоднородно распределено вдоль вспышечных петель. Спектры жесткого РИ из оснований и вершины также различны. Поэтому анализ интегрального РИ со всей петли может выявить особенности в спектре (если они имеют место), которые проявляются даже для интегральной интенсивности жесткого РИ. На первом этапе следует восстановить спектр РИ, трансформируемый в процессе регистрации спектрометром. Далее на основе «истинного» спектра жесткого РИ необходимо решить задачу реконструкции спектров ускоренных электронов.

## 1.2 Объекты и методики исследования

В качестве объектов исследования была выбрана вспышка 15 апреля 2002 года, которая началась в рентгеновском диапазоне в 23:05 UT, относится к рентгеновскому классу M1.2 и является одной из самых мощных, зарегистрированных спектрометром ИРИС во время полета спутника КОРОНАС–Ф [42, 7].

На рис. 1.7 приведен временной профиль жесткого РИ вспышки 15 апреля 2002 года [18]. Временная структура жесткого РИ вспышки представляет довольно сложную структуру - она состоит из многочисленных импульсов излучения секундной длительности (рис. 1.7), что свидетельствует о возможном импульсном характере ускорения электронов. Следует отметить, что спектрометр ИРИС начал регистрацию жесткого РИ данной вспышки с момента времени 23:09 UT, и излучение начальной фазы вспышки и фоновое излучение перед вспышкой не регистрировалось. Значения скорости счета фонового РИ были взяты из данных жесткого РИ на предыдущем витке в подобных магнитосферных условиях в отсутствии вспышек.

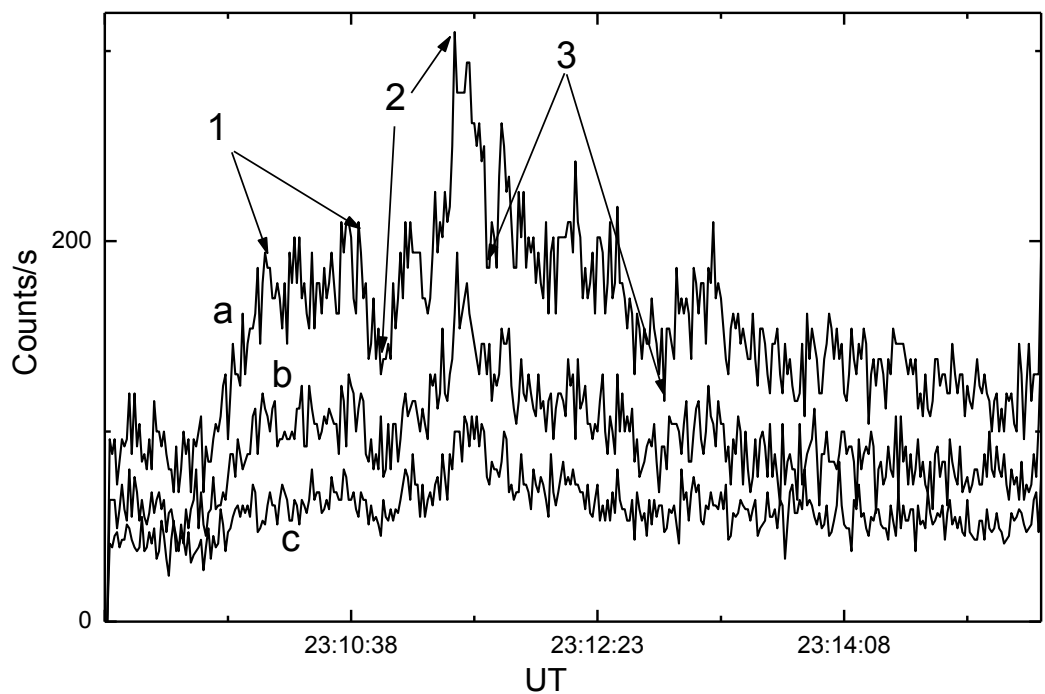

Рис. 1.7. Временной ход жесткого РИ вспышки 15 апреля 2002г. в 11-м (a), 12-м (b) и 13-м (c) энергетических каналах со средней энергией регистрируемых квантов 28.5, 32.9, 37.3 кэВ соответственно. Стрелками выделены временные интервалы, для которых проводилась реконструкция спектров: 1) 23:10:01-23:10:41 UT; 2) 23:10:52-23:11:22 UT; 3) 23:11:37-23:12:51 UT.

Для данного вспышечного события характерны несколько локальных максимумов интенсивности излучения. Таким образом, было выделено несколько групп временных интервалов, разделенных между собой локальными максимумами (минимумами). Для дальнейшего анализа энергетических спектров жесткого РИ и ускоренных во вспышках электронов для отмеченных на рис. 1.7 временных интервалов на первом этапе было проведено восстановление спектров, регистрируемых аппаратурой излучения, которое искажается при регистрации. После восстановления данных спектров производили реконструкцию энергетического распределения высокоэнергичных электронов, ускоренных во время солнечных вспышек.

# 1.3 Восстановление энергетического спектра жесткого рентгеновского излучения вспышки с учетом приборной функции спектрометра

Как известно, энергетический спектр излучения, регистрируемый детектором, искажается в результате конечного энергетического разрешения детекторов и зависимости эффективной площади детекторов от энергии квантов. Количество отсчётов $\Delta N_i$ за интервал времени $\Delta t = t'' - t'$ в энергетическом канале с номером $i$ определяется выражением

$$\Delta N_i = \int_{t'}^{t''} dt \int_{\varepsilon_{\min}}^{\varepsilon_{\max}} \int_{A_i}^{A_{i+1}} p(a,\varepsilon) J(\varepsilon,t) \, da \, d\varepsilon \qquad (1.3.1),$$

где $\varepsilon_{min}=6.4$ кэВ - минимальная регистрируемая энергия квантов, $\varepsilon_{max}$ – максимальная энергия жесткого РИ, $J(\varepsilon,t)$ - интенсивность падающего на детектор излучения, измеряемая в фотон/(см$^2$ кэВ с). Для данной вспышки $\varepsilon_{max}=152$ кэВ, что совпадает со средней энергией последнего энергетического канала; $A_i$ и $A_{i+1}$ – амплитуды сигналов детекторов, соответствующие нижней и верхней границам каналов с номером $i$, $i=8,...,39$. В нашем случае $A_i=i$ и $A_{i+1}=i+1$. Приборная функция $p(a,\varepsilon)$ предоставлена разработчиками спектрометра ИРИС и соавторами статьи [18] Лазутковым В.П., Савченко М.И. и Скородумовым Д.В.. Выражение для приборной функции имеет вид

$$p(a,\varepsilon) = \frac{Sw(\varepsilon)}{\sqrt{2\pi}\sigma(\varepsilon)} \exp\left( \frac{-(a - \tilde{A}(\varepsilon))^2}{2\sigma^2(\varepsilon)} \right) \qquad (1.3.2),$$

$$\sigma(\varepsilon) = 0.37\sqrt{\varepsilon} \qquad [\text{канал.}] \qquad (1.3.3),$$

где энергия кванта $\varepsilon$ выражается в кэВ, $a$ - амплитуда импульсов [канал.], $\tilde{A}(\varepsilon)=0.227(\varepsilon+20)$ [канал.], $\sigma(\varepsilon)$ - среднеквадратичный разброс по каналам [канал.], $S=100.5$ см$^2$ – геометрическая площадь детекторов и $w$ – вероятность поглощения РИ в кристаллах сцинтилляционных счетчиков (рис. 1.8).

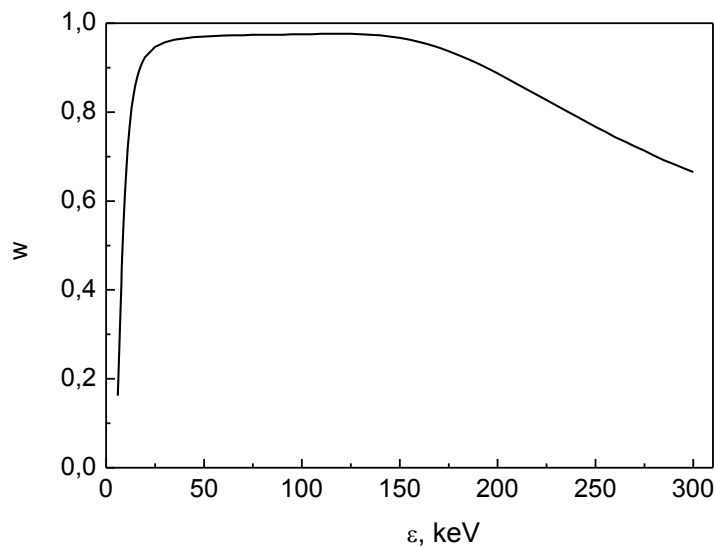

Рис. 1.8. Зависимость вероятности *w* от энергии кванта для спектрометра ИРИС КА КОРОНАС-Ф.

Величина *w* определяется пропусканием РИ защитной пленкой, входным бериллиевым окном, пенопластовой прокладкой между кристаллом и бериллиевым окном. Спад эффективности при больших энергиях обусловлен увеличением прозрачности кристалла для квантов. Средняя энергия регистрируемых квантов в канале *i* определяется выражением $\varepsilon_i=(i/0.227 - 20)$ кэВ, например, для $i=8$ $\varepsilon_i=15.24$ кэВ, для $i=39$ $\varepsilon_i=151.8$ кэВ.

Таким образом, задача определения «истинного» спектра жесткого РИ *J(ε,t)* сводится к решению интегрального уравнения Фредгольма 1-го рода (1.3.1). Однако, при решении уравнения (1.3.1) методом квадратурных формул, может получаться ложное, знакопеременное решение даже для положительно определенной функции *J(ε,t)* [3, 4]. Для устранения данной проблемы используют методы регуляризации, например, метод регуляризации Тихонова [25]. Однако в этом случае решение «регуляризированного» уравнения будет отличаться от истинного, к тому же отсутствует однозначный критерий выбора параметра регуляризации, что приводит к различным решениям исходного уравнения. Поэтому для нахождения решения уравнения (1.3.1) будем использовать метод случайного поиска (см., напр., [27]) в комбинации с методом наименьших квадратов. При этом будем искать положительную и не возрастающую функцию $J(\varepsilon,t)$.

На рис. 1.9 приведен «аппаратурный» спектр для промежутка времени 23:09:41-23:10:01 UT (отмечен цифрой 1 в соответствии с рис. 1.7), а на рис. 1.10 приведен «аппаратурный» спектр после вычитания фона, усредненный за рассматриваемый промежуток времени. Начиная с 23-го канала, как видно из рис. 1.10, спектр перестает быть спадающим, появляется разброс значений, а ошибки становятся сравнимы со значениями сигнала, и происходит регистрация только фонового излучения.

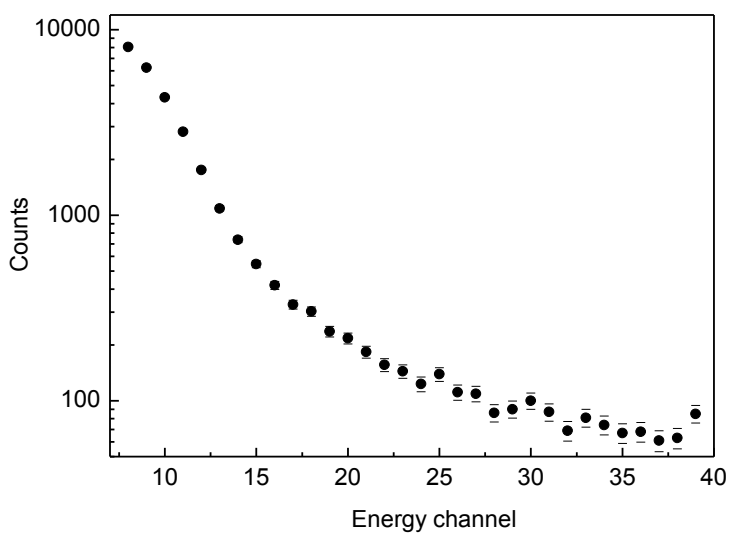

Рис. 1.9. Количество отсчетов за промежуток времени 23:09:41-23:10:01 UT для регистрируемого жесткого РИ с учетом ошибок измерений в различных каналах регистрации. Средняя энергия для $i$-го канала $\varepsilon_i=(i/0.227 - 20)$ кэВ.

Далее на рис. 1.11 приведены спектры жесткого РИ $J(\varepsilon)$ вспышки, которые были восстановлены в результате решения уравнения (1.3.1), для рассматриваемых временных интервалов (в соответствии с рис. 1.7). Как видно из рис. 1.11a, для рассматриваемой вспышки тепловое РИ переходит в жесткое РИ при энергии около 10-20 кэВ. При этом по мере роста интенсивности жесткого РИ нетепловая часть спектра появляется более отчетливо, и появляется излом спектра при энергиях ~20-25 кэВ (рис. 1.11b).

Для всех временных интервалов присутствует обрыв, который может объясняться либо отсутствием излучения больших энергий, либо тем, что интенсивность излучения становится меньше фонового уровня.

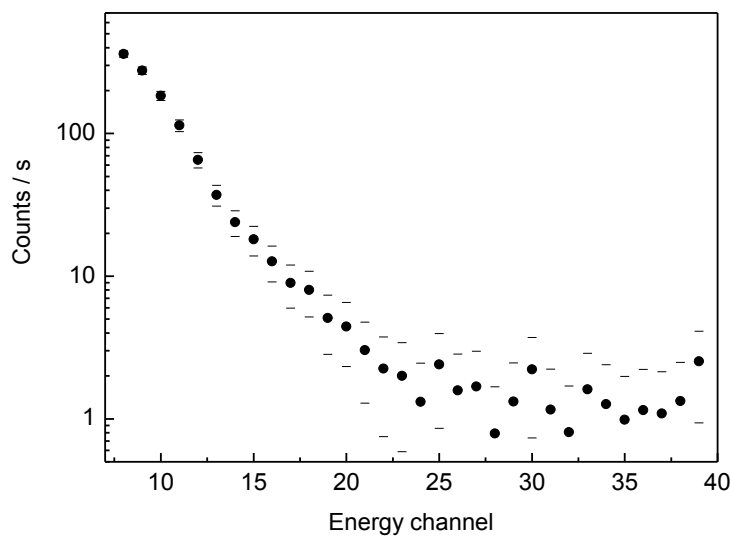

Рис. 1.10. Скорости счета в различных каналах детектирования для жесткого РИ после вычитания фона за промежуток времени 23:09:41-23:10:01 UT с учетом ошибок измерений.

В качестве примера опишем процедуру нахождения решения уравнения (1.3.1) применительно к временному интервалу 2 (23:10:52-23:11:22 UT, рис. 1.7), соответствующему абсолютному максимуму интенсивности жестко РИ во время вспышки. На рис. 1.12 приведены найденные данным методом функции $I(\varepsilon)$, где $I(\varepsilon) = \int_{t'}^{t''} J(\varepsilon,t)dt$, и рассчитанные согласно (1.3.1) $\Delta N_i$ для момента времени 23:10:52-23:11:22 UT. Будем искать решение для функции $I(\varepsilon)$ методом последовательных приближений. На первом шаге находим (кривая 1, рис. 1.12а) методом случайного поиска приближенное решение (1.3.1) в интервале энергий больше 15.2 кэВ при фиксированном отношении интенсивностей излучения с энергиями 6.4 кэВ и 10.8 кэВ к излучению с энергией 15.2 кэВ. Эти соотношения мы получаем на основе данных мягкого РИ, полученных спектрометром ИРИС в диапазоне 2.9–14.3 кэВ.

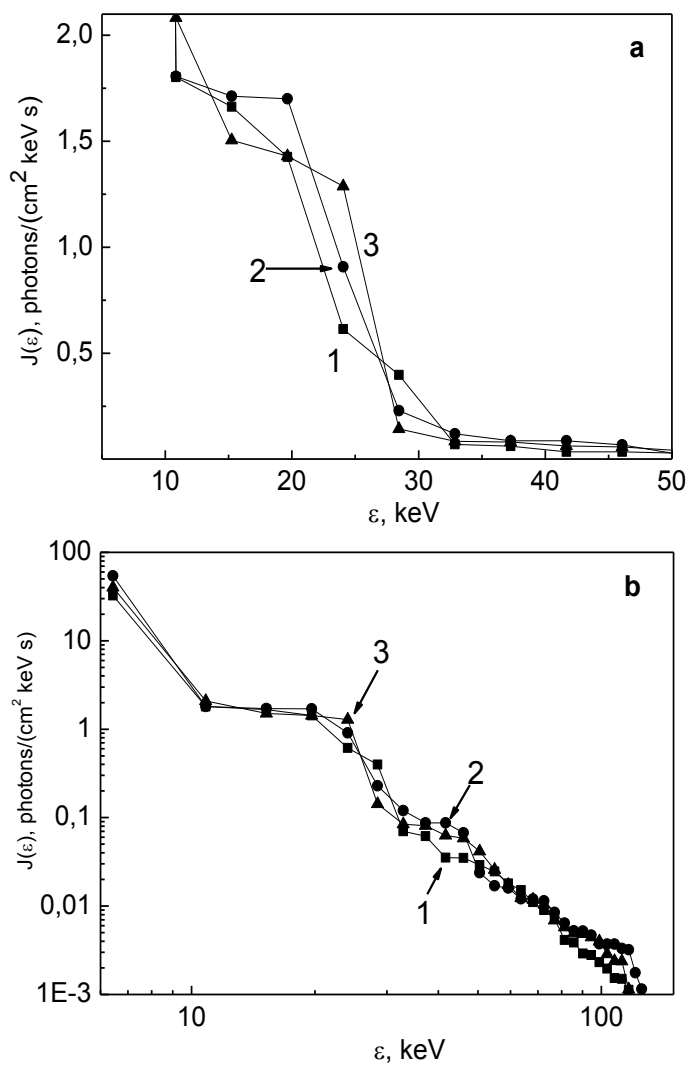

Рис. 1.11. Восстановленные энергетические спектры жесткого РИ *J(ε)* вспышки для временных диапазонов: 1) 23:10:01-23:10:41 UT; 2) 23:10:52-23:11:22 UT; 3) 23:11:37-23:12:51 UT в линейном (a) и логарифмическом масштабе (b).

На втором шаге для "улучшения" спектра в высокоэнергичной части спектра мы фиксируем значения *I(ε)* на низких энергиях (11 первых значений), решаем уравнение (1.3.1) и находим спектр в целом (кривая 2, рис. 1.12а). Фиксация найденных значений на малых энергиях возможна, так как вклад квантов высоких энергий в данный диапазон достаточно мал при спадающей функции *I(ε)*. На третьем этапе все значения *I(ε)* варьировались в интервале от 0.5×*I(ε)* до 1.5×*I(ε)* для нахождения приближенного решения с ещё меньшим среднеквадратичным отклонением (кривая 3, рис. 1.12а).

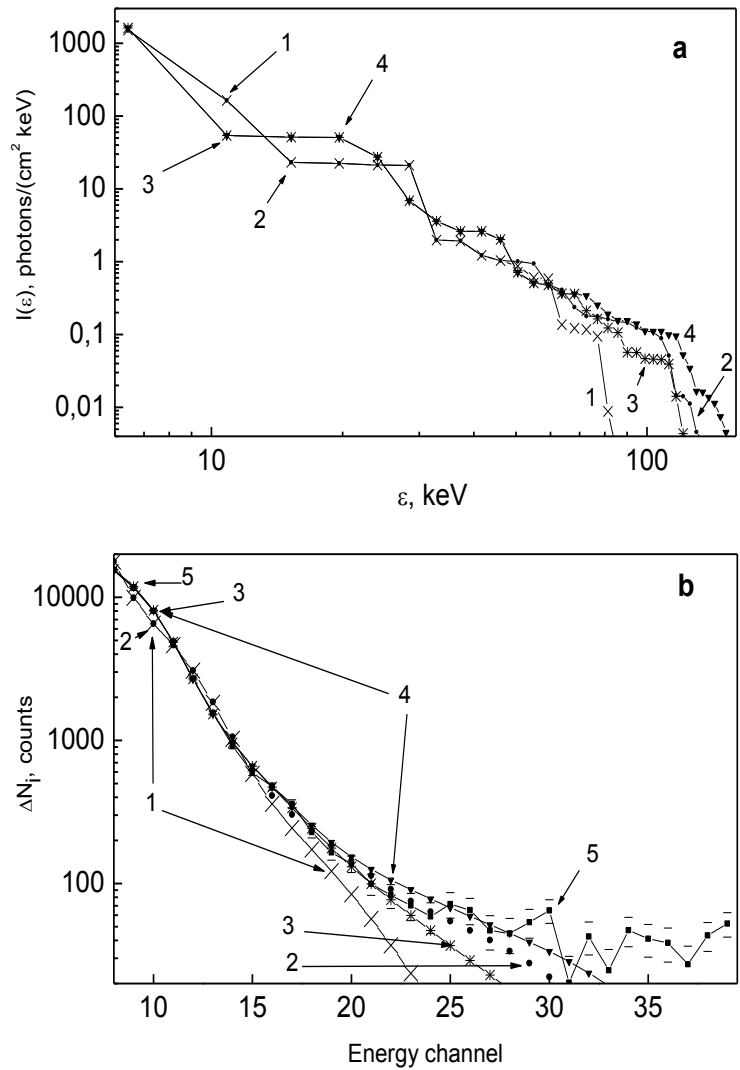

Рис. 1.12. a - Спектр жесткого РИ, реконструированный для временного интервала 2 (23:10:52-23:11:22 UT); b - рассчитанные (1-4) и исходный после вычитания фона (5) спектры. $I(\varepsilon)$ - интегральный по времени спектр жесткого РИ для данного временного интервала.

Нам удается еще уменьшить среднеквадратичное отклонение, зафиксировав первые 15 точек и разыграв оставшиеся точки (кривая 4, рис. 1.12а). Дальнейшее продолжение данной процедуры не приводит к повышению точности решения. Как видно из рис. 1.12а, восстановленный спектр РИ с энергиями, превышающими 20 кэВ, не может быть описан степенным законом.

Характерной особенностью данной реконструкции является наличие уплощения спектра в диапазоне энергий 10–20 кэВ. Данная особенность в спектре, возможно, связана с двумя причинами. Во–первых, в этой области энергии производится, как было отмечено выше, "сшивка" данных различных детекторов жесткого и мягкого РИ. И, во-вторых, в этой области энергий возможна суперпозиция теплового излучения горячей плазмы с нетепловым тормозным излучением быстрых электронов.

На рис. 1.12b приведены «аппаратурные» спектры жесткого РИ - рассчитанные согласно (1.3.1) с функциями $I(\varepsilon)$, представленными на кривых 1-4 (рис. 1.12a) соответственно, и экспериментальный после вычитания фона (кривая 5). Как видно из рис. 1.12b, рассчитанный спектр (кривая 4) согласуется с экспериментальным за исключением высокоэнергичной области (>25 канала), которая не может быть описана как тормозное излучение. Форма спектра этой высокоэнергичной области искажена фоновым излучением, изменяющимся во времени. Перейдем к рассмотрению задачи реконструкции спектра ускоренных электронов.

## 1.4 Реконструкция энергетического спектра ускоренных электронов

Как известно, интенсивность тормозного излучения $J(\varepsilon,t)$ [фотоны / (с кэВ см$^2$)] от элементарного источника с концентрацией частиц плазмы $n(\mathbf{r})$, находящемся в положении $\mathbf{r}$ на Солнце на расстоянии до места регистрации излучения $R$, возникающая при взаимодействии с потоком электронов $F(E, t)$ [электроны / (с кэВ см$^2$)] в момент времени $t$ можно представить в виде [38]

$$J(\varepsilon,t) = \frac{\overline{n} V}{4\pi R^2} \int_\varepsilon^\infty \overline{F}(E,t) Q(E,\varepsilon)\, dE \qquad (1.4.1),$$

где $J(\varepsilon,t)$ - поток рентгеновских квантов [фотоны / (с кэВ см$^2$)], $\overline{n} = \int n(\mathbf{r}) dV / V$ - среднее значение концентрации частиц плазмы в источнике, $V$ – объем излучающей области, $n(\mathbf{r})$ – концентрация плазмы излучающей области; $Q(E,\varepsilon)$ – сечение тормозного излучения кванта энергии $\varepsilon$ электроном с энергией $E$; $R$ – расстояние от Солнца до места регистрации излучения; $\overline{F}(E,t) = \int F(E,\mathbf{r},t)n(\mathbf{r})dV / \int n(\mathbf{r})dV$; $F$ – плотность потока быстрых электронов (измеряется в см$^{-2}$ сек$^{-1}$ кэВ$^{-1}$), которая связана с функцией распределения электронов $f(v,\mathbf{r},t)$ выражением $F(E,\mathbf{r},t)dE = v f(v,\mathbf{r},t)dv$, $v$ и $\mathbf{r}$ – скорость и координаты электрона [38]. В расчетах использовали тормозное сечение в борновском приближении [74, 78], рассматривая нерелятивистскую область энергий

$$Q(E,\varepsilon) = C \cdot \sigma(E,\varepsilon) \; ; \sigma(E,\varepsilon) = \frac{1}{E\varepsilon} \ln\left(\left(1+\sqrt{1-\frac{\varepsilon}{E}}\right) \middle/ \left(1-\sqrt{1-\frac{\varepsilon}{E}}\right)\right) \qquad (1.4.2),$$

где $C = 7.9 \times 10^{-25}$ см$^2$ кэВ для водородной плазмы (см., напр., [78]). Уравнение (1.4.1) далее можно представить в виде, более удобном для решения задачи

$$J(\varepsilon,t) = \int_{\varepsilon}^{\infty} F^*(E,t)\sigma(E,\varepsilon)dE \qquad (1.4.3),$$

где $F^*(\varepsilon,t) = \frac{\overline{n}VC}{4\pi R^2}\overline{F}(\varepsilon,t)$.

Таким образом, для нахождения функции распределения электронов $F^*$, излучающих жесткое РИ во время солнечных вспышек, необходимо решать интегральное уравнение (1.4.3). Следует учесть следующие возможные инструментальные ограничения рентгеновского спектрометра: в мощных вспышках энергетический спектр жесткого РИ может превышать энергетический диапазон прибора; измерения в высокоэнергичных каналах могут быть не достоверными по причине малой статистики квантов и, следовательно, спектр жесткого РИ может быть измерен только в ограниченном диапазоне энергий. Был рассмотрен случай, когда измерения производятся в диапазоне от $\varepsilon$ до $\varepsilon_{max}$ с интервалом $\Delta\varepsilon$.

Учитывая инструментальные ограничения, уравнение (1.4.3) можно представить в виде суммы двух интегралов

$$J(\varepsilon,t)=\int_{\varepsilon}^{\infty}F^{*}(E,t)\sigma(E,\varepsilon)dE= \int_{\varepsilon}^{\varepsilon_{\max}}F^{*}(E,t)\sigma(E,\varepsilon)dE + \int_{\varepsilon_{\max}}^{\infty}F^{*}(E,t)\sigma(E,\varepsilon)dE \quad (1.4.4).$$

Для энергий электронов больших верхней границы инструментальных измерений рентгеновских квантов $\varepsilon_{max}$ во втором интеграле функцию $F^*$ будем описывать степенной зависимостью

$$F^{*}(E \geq \varepsilon_{\max},t) = k_1(t) E^{-\beta(t)} \quad (1.4.5),$$

после чего уравнение (1.4.4) примет вид

$$\int_{\varepsilon}^{\varepsilon_{\max}} F^{*}(E,t)\sigma(E,\varepsilon)dE = J(\varepsilon,t) - k_1(t)\int_{\varepsilon_{\max}}^{\infty} E^{-\beta(t)}\sigma(E,\varepsilon)dE \quad (1.4.6).$$

При $\varepsilon=\varepsilon_{max}$ получаем

$$J(\varepsilon_{\max},t) = k_1(t)\int_{\varepsilon_{\max}}^{\infty}\sigma(E,\varepsilon_{\max})E^{-\beta(t)}dE \quad (1.4.7)$$

или

$$k_1(t) = J(\varepsilon_{\max},t) / \int_{\varepsilon_{\max}}^{\infty}\sigma(E,\varepsilon_{\max})E^{-\beta(t)}dE \quad (1.4.8).$$

Если спектр жесткого РИ измеряется во всем энергетическом диапазоне спектрометра с достаточной статистикой и $J(\varepsilon_{max})=0$, то в этом случае $k_1=0$. Для нахождения функции $F^*$ интегрального уравнения (1.4.6) использовали метод регуляризации Тихонова нулевого порядка [25]. Для этого исходное уравнение (1.4.6) было преобразовано в «регуляризованное»:

$$\alpha F_{\alpha}^{*}(\varepsilon) + \int_{\varepsilon_{\min}}^{\varepsilon_{\max}} R(E,\varepsilon)F_{\alpha}^{*}(E)dE = \hat{F}(\varepsilon) \quad (1.4.9),$$

где $\alpha$ – параметр регуляризации,

$$R(E,\varepsilon) = \int_{\varepsilon_{\min}}^{\varepsilon_{\max}}\sigma(t,E)\sigma(t,\varepsilon)dt \text{ и } \hat{F}(\varepsilon) = \int_{\varepsilon_{\min}}^{\varepsilon_{\max}}\sigma(x,\varepsilon)\left(J(x) - k_1\int_{\varepsilon_{\max}}^{\infty}\sigma(x,E)E^{-\beta}dE\right)dx.$$

Из рис. 1.12b следует, что величина фонового излучения с энергией 90 кэВ (25 канал) становится сравнимой с излучением вспышки. Это указывает на то, что тормозное излучение в данной области либо отсутствует, либо поток квантов этих энергий чрезвычайно мал. Поэтому будем считать, что для данной вспышки $\varepsilon_{max}$=152 кэВ, что соответствует последнему 39-му каналу регистрации. При решении уравнения Вольтерра 1-го рода (1.4.1) необходимо учитывать, что функция $\overline{F}(E,t)$ уже не обязательно должна быть монотонно спадающей. Поэтому для нахождения приближенного решения этого уравнения удобней использовать метод регуляризации Тихонова [4, 25] и рассматривать решения для различных параметров регуляризации. Решая преобразованное уравнение (1.4.9), получили реконструированные энергетические спектры электронов, генерирующих жесткое РИ, которые приведены на рис. 1.13 для моментов времени 1 – 3 согласно рис. 1.7.

Как было отмечено выше, не существует однозначного критерия выбора параметра регуляризации $\alpha$. Выбор параметра регуляризации – это результат компромисса между желанием получить сглаженную функцию и точностью решения. Увеличение этого параметра ведет к получению более гладкой функции в ущерб точности решения. Уменьшение же α позволяет выявлять особенности решения, но может приводить к появлению отрицательных значений искомой функции. Поэтому, для выявления особенностей спектра электронов мы будем приводить результаты решения уравнения (1.4.9) для различных значений параметра $\alpha$. Возможные отрицательные значения искомой функции при некоторых значениях энергии $E$ будем считать нулевыми.

Рассмотрим реконструированные энергетические распределения электронов $\langle nVF \rangle$ (*mean electron flux spectrum*) с энергиями большими 30 кэВ, которые генерируют жесткое РИ на различных стадиях развития вспышки (рис. 1.13). Этот диапазон энергий позволяет нам отступить от переходной области 10 – 20 кэВ между компонентами квазитеплового и нетеплового РИ. На рис. 1.13 (a, c, e) приведены результаты реконструкции энергетических распределений электронов для временных интервалов 1-3 (рис. 1.7), соответственно, и рассчитанные на их основе аппаратурные спектры (рис. 1.13, b, d, f) для различных параметров регуляризации *α*.

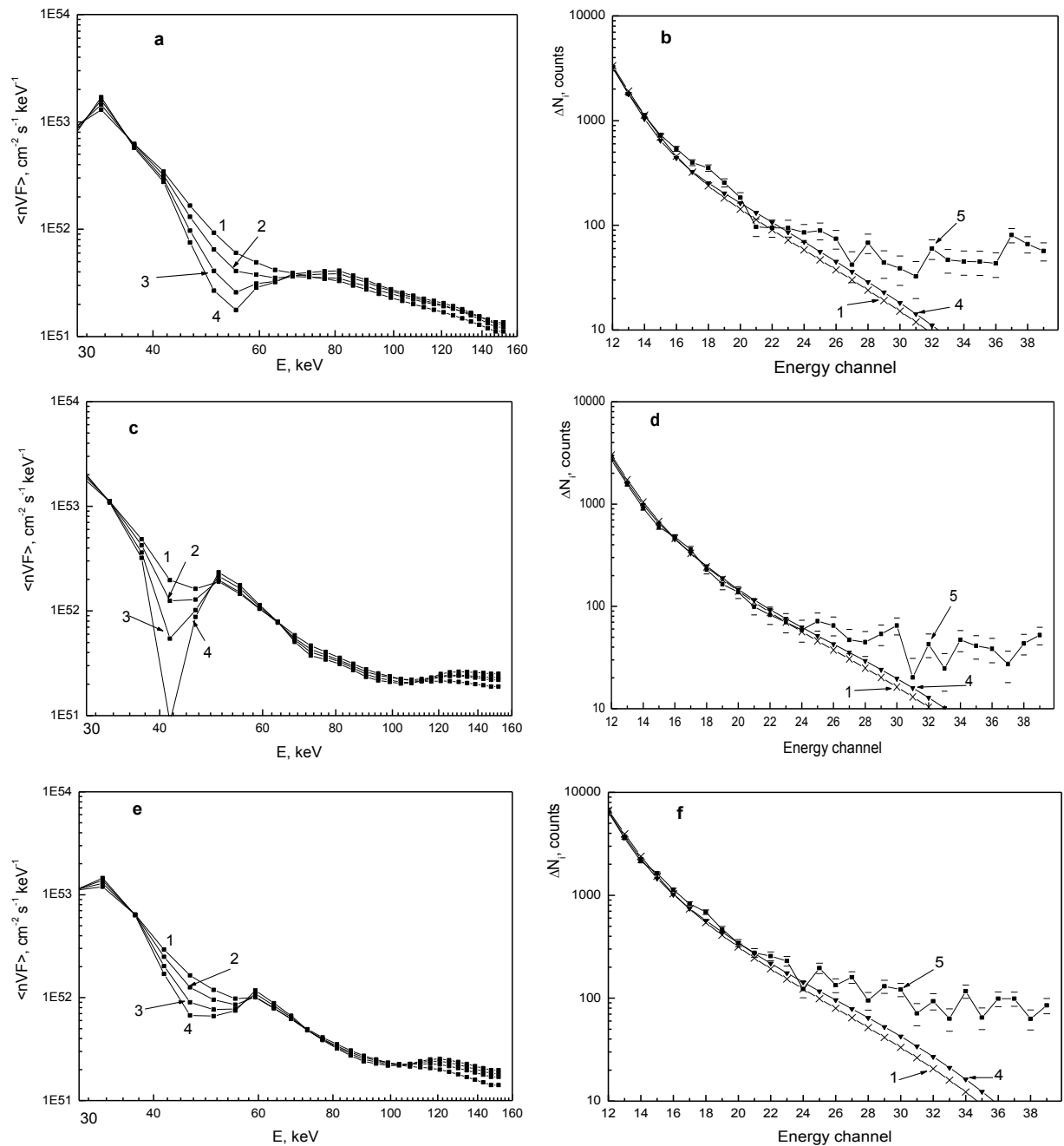

Рис. 1.13. Восстановленные энергетические спектры электронов (a, c, e) для различных параметров регуляризации *α*, расчетные аппаратурные (b, d, f, кривые 1 и 4) и измеренные с учетом вычитания фона (кривая 5) спектры жесткого РИ для рассматриваемых временных интервалов. 1 - α=1·10$^{-5}$; 2 - α =7·10$^{-6}$; 3 - α =5·10$^{-6}$; 4 - α=4·10$^{-6}$, $\langle nVF \rangle$ - среднее значение $\overline{n}\overline{VF}$ по времени для рассматриваемых временных интервалов.

Полученные реконструкции позволяют сделать несколько выводов. Во-первых, в области энергии, превышающей 100 кэВ, уменьшение параметра регуляризации *α* приводит к большим различиям в спектре, что говорит об искажении в этой области спектра жесткого РИ фоновым излучением, которое не удается полностью исключить. Во-вторых, спектр электронов в области 30 – 100 кэВ явно отличается от степенного. В-третьих, при уменьшении параметра регуляризации *α* наблюдается инверсия спектра: на первом временном интервале с меньшей интенсивностью жесткого РИ инверсия в энергетическом распределении электронов приходится на область энергий 50 – 60 кэВ.

На следующем временном интервале потоки электронов возрастают и область инверсии спектра смещается к меньшим энергиям: 40 – 50 кэВ. Отметим, что при уменьшении интенсивности жесткого РИ эта особенность смещается в область больших энергий. Подобная инверсия в реконструированном спектре электронов ранее была отмечена в работе [109] при анализе вспышки 23 июля 2002 года. Для её описания в работе [38] было сделано предположение, что нетепловая составляющая функции распределения электронов для энергий, меньших 40 кэВ, может быть $\sim E^7$, для энергии больше 40 кэВ - спадающей по степенному закону. Такая сильная зависимость от энергии в данной переходной области и приводит к различию графиков при различных значениях параметра регуляризации, которые можно видеть на рис.1.13. Следует также отметить, что подобная инверсия в спектре распределения электронов наблюдалась для вспышки 26 июля 2002 года (рис. 1.14), для которой была произведена реконструкция энергетического распределения электронов с помощью квадратурных формул и метода регуляризации Тихонова нулевого порядка без учета приборной функции по данным спектрометра ИРИС КА КОРОНАС-Ф (см. рис. 1.15) [13].

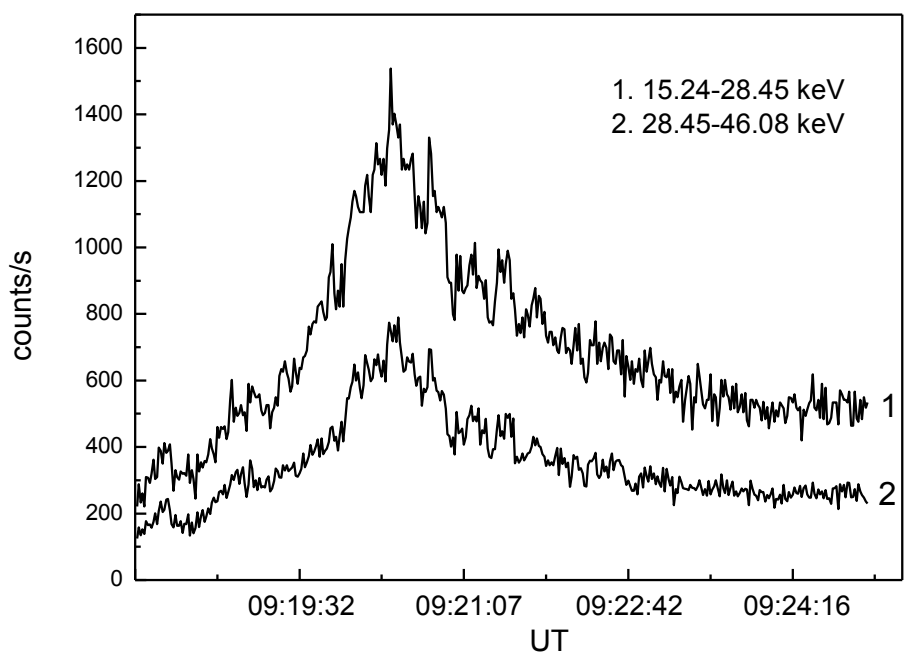

Рис. 1.14. Временной ход интенсивности жесткого РИ солнечной вспышки 26 июля 2002 года, зарегистрированной спектрометром ИРИС в энергетическом диапазоне 15.24 – 28.45 кэВ (кривая 1) и 28.45 – 46.08 кэВ (кривая 2).

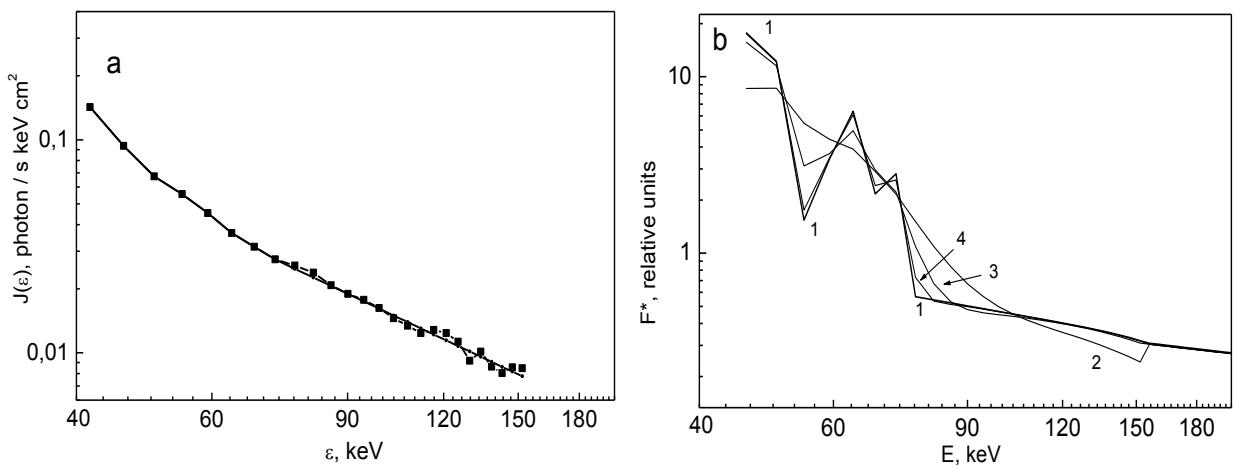

Рис. 1.15. a - Зарегистрированный спектр жесткого РИ вспышки 26 июля 2002 года в интервале времени 9:19:47–9:20:18 UT (квадраты) и расчётный спектр жесткого РИ (черная линия); b - реконструированная методом квадратурных формул функция $F^*$, соответствующая приведенному слева спектру жесткого РИ (кривая 1) и функция $F^*$, реконструированная методом регуляризации Тихонова для различных параметров регуляризации $\alpha$: 2 – $\alpha=10^{-6}$; 3 – $\alpha=10^{-7}$; 4 – $\alpha=10^{-8}$, где $F^* = \langle nVF \rangle / D$ [электроны/ см² с], множитель $D=4\pi R^2/C \approx 3.6 \times 10^{51}$ [1/кэВ].

Инверсия энергетического спектра электронов отмечалась ранее в работах [109, 14]. Возможные причины этого эффекта обсуждаются в работе [78]. Одной из них является формирование функции распределения надтепловых электронов при их распространении во вспышечной петле с учетом кулоновских столкновения с частицами окружающей плазмы. В этом случае на функции распределения формируется растущий с энергией участок [58] в результате зависимости частоты кулоновских столкновений от энергии частиц. При этом энергия, соответствующая локальному максимуму энергетического распределения электронов, может составлять десятки кэВ.

Другой возможной причиной появления минимума на энергетическом распределении электронов может быть то, что в спектрометр попадает не только первичное жесткое РИ, но и часть отраженного от нижних слоев солнечной плазмы излучения [78]. Рассмотрение данного эффекта для ряда вспышек показало, в этом случае положение локального минимума приходится на диапазон энергий 13 – 19 кэВ [77]. Однако, для рассматриваемой нами вспышки 15 апреля 2002 года, соответствующая локальному минимуму энергия находится в интервале между 40 кэВ и 60 кэВ. Также данная вспышка является лимбовой, где эффект вторичного отраженного излучения в жесткое РИ имеет малый вклад. В пользу первого механизма свидетельствует то, что динамика электронов с энергией 40 – 60 кэВ отличается от динамики электронов более низких энергий. Так, на рис. 1.13 видно, что на втором временном интервале количество электронов с энергиями 50 – 60 кэВ в несколько раз больше чем на первом участке, а для электронов с энергиями 30 – 40 кэВ различие гораздо меньше. Это говорит о наличии разных популяций электронов: электроны тепловой окружающей плазмы и ускоренные во вспышке электроны.

Тем не менее, однозначного ответа о причине такого локального минимума на энергетическом спектре электронов в настоящее время нет. В связи с этим, необходимо проведение дальнейших исследований с привлечением большого количества вспышек различных рентгеновских классов, теоретическое изучение процессов ускорения и распространения частиц во вспышечной плазме с целью выяснения, является ли данная особенность спектра результатом того, что имеется нижняя граница энергии ускоренных во вспышках электронов (что носит фундаментальный характер), либо она связана с процессом распространения ускоренных электронов во вспышечной плазме, когда в область излучения проникают только частицы больших энергий.

## 1.5. Заключение к главе 1

В работе проведена реконструкция энергетических спектров электронов, ускоренных во время вспышки 15 апреля 2002 года на различных этапах развития вспышки, на основе данных, полученных рентгеновским спектрометром ИРИС на космической станции КОРОНАС-Ф. Для этого последовательно решались два интегральных уравнения, описывающих искажение спектра рентгеновских квантов во время их регистрации спектрометром и процесс генерации тормозного излучения быстрыми электронами. Полученные спектры электронов характеризуются локальным минимумом в области энергий 40 – 60 кэВ, при этом положение этого минимума не остается постоянным во время развития вспышки. Данный минимум может быть связан с тем, что существует нижняя граница ускоренных во вспышке электронов или с особенностями распространения быстрых электронов во вспышечной плазме. Однако выяснение конкретных причин и механизмов, приводящих к формированию этой особенности спектра, требует дальнейших экспериментальных и теоретических исследований.

Таким образом, в результате проведения данной исследовательской работы на примере одного вспышечного события:

1. Разработана методика, которая позволяет определять на первой стадии спектр приходящего на аппаратуру спектра жесткого РИ с учетом приборной функции, а затем по нему восстанавливать энергетические спектры быстрых электронов, ускоренных в солнечных вспышках.

2. Для восстановленных спектров жесткого РИ показано, что тепловое РИ для рассматриваемой вспышки переходит в жесткое РИ при энергии около 10-20 кэВ. При этом по мере роста интенсивности жесткого РИ нетепловая часть спектра проявляется более отчетливо, чем на первом рассматриваемом интервале, и появляется излом спектра при энергии 20-25 кэВ. Данная особенность в спектре, может быть связана с тем, что в этой области энергий происходит суперпозиция теплового излучения горячей плазмы с нетепловым тормозным излучением ускоренных электронов.

3. Показано, что энергетические спектры жесткого РИ и спектры быстрых электронов имеют тенденцию к эволюции в течение вспышечного события.

4. Показано, что энергетические спектры быстрых электронов имеют особенность в спектрах (локальный минимум для энергий 50–60 кэВ для первого участка, который в дальнейшем сдвигается в область 40–50 кэВ), которая не может быть описана модельными аппроксимациями.

Все вышесказанное подтверждает возможность получать истинные спектры тормозного излучения от солнечных вспышек и по ним реконструировать спектры высокоэнергичных электронов, что показано на примере спектрометра ИРИС космического аппарата КОРОНАС-Ф.

# Глава 2
# Реконструкция энергетических распределений электронов на основе одновременных наблюдений солнечных вспышек в крайнем ультрафиолетовом и мягком рентгеновском диапазонах

## 2.1 Введение к главе 2

Одним из наиболее ярких проявлений солнечных вспышек является рентгеновское и радиоизлучение, которое свидетельствует о наличии нетепловых частиц и горячей плазмы. Наблюдения в этих диапазонах дают информацию о распределении электронов, и тем самым помогают найти ключевые параметры вспышечной плазмы, такие как полное число ускоренных электронов и энергию вспышки. Для вывода распределения ускоренных электронов, как правило, используется метод аппроксимации модельными функциями, предполагающий модельную зависимость, или методы инверсии спектра рентгеновского излучения (РИ) с помощью регуляризации (см. для обзора [78]), данный метод был использован в главе 1. Типичные спектры РИ, наблюдаемые КА RHESSI [92], согласуются с квази-изотермическим максвелловским распределением для энергий до ~20 кэВ и нетепловой частью в виде степенной зависимости для более высоких энергий (см., напр., [66], в качестве обзора), однако относительная значимость этих двух компонент может отличаться от вспышки к вспышке и во время различных ее фаз. Наиболее простая модель для функции распределения электронов содержит распределение Максвелла $F(E) \sim E \times \exp(-E/k_B T)$ и степенную функцию $F(E) \sim E^{-\delta}$. Основным недостатком данной модели является то, что она требует введения границы на низкие энергии $E_c$ для нетеплового спектра, что трудно определить наблюдательно (см., напр., [65, 113, 77]). В то же время, в связи с уменьшением энергетического спектра электронов с энергией (степенной индекс $\delta$, как правило, равен 3-6) основная часть энергии электронов передается надтепловыми электронами с энергией порядка нескольких $k_BT$. Вследствие этого, детальное изучение спектров электронов в диапазоне от нескольких кэВ до нескольких десятков кэВ является ключевым не только для понимания физики ускорения электронов, но и для более точных оценок полной энергии вспышек. Частично из-за этого в последние годы наблюдается растущий интерес к различным аналитическим распределениям, например, к каппа-распределению, близкому к распределению Максвелла на низких энергиях и имеющему степенную зависимость на высоких энергиях $F(E) \propto E(1 + E/k_B T_\kappa (\kappa - 1.5))^{-(\kappa+1)}$. Для энергий гораздо больше энергии $k_BT$ функция $F(E)$ имеет вид степенного спектра, а для индекса $\kappa \to \infty$ приближается к распределению Максвелла. Также следует отметить, что каппа-распределение поддерживается теоретическими соображениями об ускорении частиц в столкновительной плазме. В работе [34] показано, что распределение, близкое к каппа-распределению, может быть сформировано в сценарии стохастического ускорения, по крайней мере в диапазоне надтепловых частиц, в то время как убегание электронов может изменить высокоэнергетическое степенное распределение. В самом деле, некоторые рентгеновские спектры, наблюдаемые с помощью КА RHESSI во время солнечных вспышек, могут быть аппроксимированы модельной зависимостью с помощью каппа-распределений [72, 102, 103], это утверждение подходит в большей степени для событий, у которых присутствует один источник на всех энергиях. Тем не менее, аппроксимация модельными функциями, предложенная в работе [72], предполагает РИ на основе модели тонкой мишени и учитывает только электрон-ионное тормозное излучение. В то же время, надтепловые электроны посредством рекомбинаций испускают свободно-связанное излучение [47, 48, 40, 56], вклад которого в полный рентгеновский спектр может быть значительным. Учет свободно-связанного излучения приводит к увеличению полного потока РИ от того же числа электронов, тем самым уменьшая количество электронов, необходимое для генерации наблюдаемого излучения.

КА RHESSI регистрирует РИ в энергетическом диапазоне от нескольких кэВ до ~17 МэВ, однако нижняя граница чувствительности КА RHESSI находится около 3 кэВ (или даже выше, в зависимости от используемого аттенюатора, где аттенюаторы представляют собой два набора алюминиевых дисков, которые используются для предотвращения насыщения во время крупных вспышек). Вследствие этого, распределение электронов на более низких энергиях слабо ограничено. С другой стороны, объединение рентгеновских наблюдений с наблюдениями в КУФ диапазоне с помощью КА SDO/AIA [88] позволяет определять распределения электронов в гораздо более широком диапазоне энергий от ~0.1 кэВ до нескольких десятков кэВ. Баттаглиа и Контарь [32] впервые показали распределения электронов, найденные с помощью аппроксимации модельной зависимостью рентгеновских спектров по данным КА RHESSI, и распределения электронов, выведенные из дифференциальной меры эмиссии (ДМЭ) на основе данных КА SDO/AIA, где наблюдения обоих инструментов рассматривались независимо друг от друга. Улучшением по сравнению с вышесказанным было бы одновременная аппроксимация модельными функциями данных КА RHESSI и SDO/AIA. В работе [70] проведена подобная совместная аппроксимация для низкоэнергетической (тепловой) компоненты микровспышек, т.е. солнечных вспышек малой мощности, но для высокоэнергетической компоненты РИ (в случае ее наличия) аппроксимация модельными функциями производилась отдельно.

Таким образом, **целью главы 2** является реконструкция ДМЭ во вспышечной плазме по данным КА SDO/AIA и RHESSI одновременно,

восстановление по полученным значениям ДМЭ энергетических распределений электронов в солнечных вспышках.

Данная глава направлена на решение следующих задач:

(а) Разработать две модели с целью аппроксимации данных КУФ и мягкого РИ, которые представляют собой распределение модифицированной функции Бесселя второго рода, умноженное на степенную функцию, и каппа-распределение. Обе модели включают полный набор соответствующих механизмов излучения (свободно-свободное, свободно-связанное, двухфотонное излучение и излучение в линиях) на основе кода *CHIANTI 7.1* [54, 87]. Рассматриваемые в настоящей главе модели описывают распределение электронов в виде суммы распределений Максвелла, взвешенных по коэффициенту, пропорциональному ДМЭ, таким образом, что полное излучающее, усредненное по объему распределение представляет собой распределение модифицированной функции Бесселя второго рода, умноженное на степенную функцию, и каппа-распределение.

(б) Представить метод, который позволяет производить одновременную аппроксимацию модельными функциями всего спектра РИ, зарегистрированного с помощью КА RHESSI, и SDO/AIA данных в КУФ диапазоне с помощью новых моделей из пункта (а), объединяя функции температурного отклика КА SDO/AIA с матрицей температурного отклика КА RHESSI в единую матрицу отклика.

(в) Протестировать аппроксимацию RHESSI данных одновременно с SDO/AIA данными новыми моделями (из пункта (а)) на примере солнечной вспышки, имеющей источник в вершине петли.

Для проведения расчетов для главы 2 и главы 3 используется программное обеспечение *SSW*, интерфейс *OSPEX* для анализа данных. Также был задействован код *CHIANTI 7.1* [54, 87] (см. http://hesperia.gsfc.nasa.gov/ssw/packages/chianti/doc/cug.html), т.е. атомная база данных для спектроскопической диагностики астрофизической плазмы, состоящая из двух частей: базы данных и набора компьютерных программ на языках программирования Python и IDL. Вместе они позволяют рассчитать оптически тонкие спектры астрофизических объектов и обеспечивают спектроскопическую диагностику плазмы. База данных включает в себя энергетические уровни атомов, длины волн, вероятности радиационных переходов, коэффициенты скорости столкновительного возбуждения, ионизацию и коэффициенты скорости рекомбинации, а также данные для расчета свободно-свободного (тормозного), свободно-связанного и двухфотонного излучения в континууме (см. раздел 1.1.1). Коды для расчета излучения плазмы используются для изучения ультрафиолетового и РИ от солнечной или звездных атмосфер. Сравнение интенсивностей от теоретических данных с наблюдаемыми интенсивностями позволяет определить физические параметры плазмы [51]. Версия *CHIANTI 7.1* включает в себя дополненные данные для нескольких ионов, скоростей рекомбинации и обилий примесных элементов (*element abundances*). В частности, существенно расширился диапазон моделей *CHIANTI* для ионов железа от *Fe VIII* до *Fe XIV* для улучшения предсказанного излучения в диапазоне длин волн 50-170 Å. В настоящее время существует более новая версия *CHIANTI 8* [52], однако основные изменения в ней коснулись низкотемпературной плазмы или излучения в линиях, доминирующих для спокойного Солнца, и не повлияли на результаты данной работы.

Таким образом, в главе 2 были разработаны расчетные программы на языке IDL с использованием среды *OSPEX,* атомной базы данных *CHIANTI 7.1* и программных библиотек *SSW* для одновременной аппроксимации функциями ДМЭ данных с КА SDO/AIA и RHESSI и реконструкции энергетических распределений электронов в солнечных вспышках. Результаты главы 2 опубликованы в работах [16, 17, 100, 33].

## 2.2 Связь дифференциальной меры эмиссии и энергетического распределения электронов

Для теплового источника мягкого РИ в солнечных вспышках в предположении однотемпературной модели количество горячей плазмы можно представить в виде меры эмиссии (см., напр., [95])

$$EM = \int n^2 dV \qquad (2.2.1),$$

где $n$ – концентрация плазмы, а $V$ – объем излучающей области. Однако, из наблюдений следует, что во вспышечном источнике всегда присутствует плазма с разной температурой, поэтому распределение неоднородной плазмы по температуре $T$ во всем излучающем объеме $V$ может быть представлено с помощью ДМЭ $\xi(T)$ [см$^{-3}$К$^{-1}$] [95]

$$EM = \int \frac{d(n^2 V)}{dT} dT = \int \xi(T) dT \qquad (2.2.2),$$

где $\xi(T) = n^2 dV/dT$. Эта величина может быть измерена в широком диапазоне КУФ и мягких рентгеновских линий для любой области на Солнце (вдоль луча зрения).

### 2.2.1 Описание энергетического распределения электронов через дифференциальную меру эмиссии

Предполагая локальное распределение Максвелла с локальной температурой $T(\mathbf{r})$ и концентрацией $n(\mathbf{r})$ для пространственной координаты $\mathbf{r}$, меру эмиссии можно представить в виде (2.2.2). Распределение Максвелла для температуры $T(\mathbf{r})$ имеет вид (см., напр., [32])

$$F(E, r) = \frac{2^{3/2}}{\sqrt{\pi m_e}} \frac{n(\mathbf{r}) E}{(k_B T(\mathbf{r}))^{3/2}} \exp\left(-\frac{E}{k_B T(\mathbf{r})}\right) \qquad (2.2.3),$$

где $E$ – кинетическая энергия электронов, $m_e$ – масса электрона, $k_B$ – постоянная Больцмана. Отметим, что усредненное энергетическое распределение электронов $\langle nVF(E) \rangle = \int_V n(r) F(E, r) dV$ [электроны кэВ$^{-1}$с$^{-1}$см$^{-}$

²] из объема излучающей области $V$ – единственное может быть выведено непосредственно из рентгеновских данных без предположений о концентрации или излучающем объеме. Оно может быть связано с ДМЭ $\xi(T)$ с помощью максвелловского распределения электронов при температуре $T$ [37, 32]

$$\langle nVF(E)\rangle = \int_T n(r) \frac{2^{3/2}}{\sqrt{\pi m_e}} \frac{n(r)E}{(k_B T(r))^{3/2}} \exp\left(-\frac{E}{k_B T(r)}\right) \frac{dV}{dT} dT \qquad (2.2.4),$$

Отсюда, функция $\xi(T)$ может быть найдена из выражения

$$\langle nVF(E)\rangle = \frac{2^{3/2} E}{\sqrt{\pi m_e}} \int_0^\infty \frac{\xi(T)}{(k_B T)^{3/2}} \exp\left(-\frac{E}{k_B T}\right) dT \qquad (2.2.5).$$

Используя замену переменных $t = 1/T$ для (2.2.5)

$$\langle nVF(E)\rangle = \frac{2^{3/2} E}{\sqrt{\pi m_e} k_B^{3/2}} \int_0^\infty \frac{\xi(T(t))}{t^{1/2}} \exp(-Et/k_B) dt \qquad (2.2.6).$$

Исходя из уравнения (2.2.6), форма ДМЭ выбиралась таким образом, чтобы выражение $\xi(T(t))/t^{1/2}$ имело аналитический вид преобразования Лапласа для дальнейшего счета $\langle nVF(E)\rangle$, а ее поведение на низких и высоких энергиях было схожим с ДМЭ, которые были получены отдельно для КА SDO/AIA и RHESSI с помощью методов регуляризации и аппроксимации модельной зависимостью соответственно (см., напр., [32])

$$\xi(T) \propto T^\alpha \exp\left(-\frac{T}{T_0}\right) \qquad (2.2.7),$$

используя нормировку $\int_0^\infty \xi(T)dT = \int_0^\infty CT^\alpha \exp(-T/T_0)dT = EM$, константа $C = EM/(T_0^{\alpha+1}\Gamma(\alpha+1))$. Подставляя полученное значение $C$ в выражение (2.2.7)

$$\xi(T) = \frac{EM}{T_0 \Gamma(\alpha+1)} \left(\frac{T}{T_0}\right)^\alpha \exp\left(-\frac{T}{T_0}\right) \qquad (2.2.8),$$

где $\Gamma(x)$ – гамма функция, $\alpha > 0.5$: $\Gamma(x) = \int_0^\infty y^{x-1} \exp(-y)dy$. Представив выражение (2.2.7) как

$$\xi(T) = EM\left[\frac{1}{\Gamma(\alpha+1)}\left(\frac{1}{T_0}\right)^{\alpha+1} T^\alpha \exp\left(-\frac{T}{T_0}\right)\right] \qquad (2.2.9),$$

можно видеть, что в скобках стоит плотность гамма-распределения $\Gamma_{1/T_0,\alpha+1}(T)$ с параметрами $1/T_0$ и $\alpha+1$ для $T>0$, поэтому полученную ДМЭ в дальнейшем будем обозначать $\xi_\Gamma(T)$. Данная ДМЭ возрастает для $T<T_0$ вследствие $\xi_\Gamma(T) \propto (T/T_0)^\alpha$ и убывает с ростом $T$ для $T>>T_0$, так как $\xi_\Gamma(T) \propto \exp(-T/T_0)$. ДМЭ имеет один максимум $d\xi_\Gamma(T)/dT=0$ при температуре $T_{max}=\alpha T_0$. На рис. 2.1 (левая панель) представлен пример ДМЭ для $EM=10^{49}$ см$^{-3}$, $T_{max}=10$ МК, $\alpha=9$.

В результате подстановки уравнения (2.2.8) в (2.2.6) $\langle nVF(E)\rangle$ примет вид

$$\langle nVF(E)\rangle = \frac{2^{3/2} E}{\sqrt{\pi m_e} k_B^{3/2}} \frac{2EM(T_0 E/k_B)^{0.5\alpha-0.25}}{T_0^{\alpha+1}\Gamma(\alpha+1)} K_{\alpha-0.5}\left(2\sqrt{\frac{E/k_B}{T_0}}\right) \qquad (2.2.10),$$

где $K_n(z)$ – модифицированная функция Бесселя второго рода [44]. На рис. 2.1 (правая панель) показан пример $\langle nVF(E)\rangle$ для $EM=10^{49}$ см$^{-3}$, $T_{max}=10$ МК, $\alpha=9$.

То есть распределение электронов в виде (2.2.10) есть распределение модифицированной функции Бесселя второго рода, умноженное на степенную функцию, оно описывает распределение электронов в виде суммы распределений Максвелла, в дальнейшем будем обозначать его как $\langle nVF(E)\rangle_K$. Оно близко к распределению Максвелла на низких энергиях и убывает на высоких энергиях вследствие $K_{\alpha-0.5}(2\sqrt{E/k_B T_0})$, а для индекса $\alpha\to\infty$ $\langle nVF(E)\rangle_K$ приближается к распределению Максвелла. Таким образом, был предложен аналитический вид распределения надтепловых электронов, имеющий прямую связь с ДМЭ, для нахождения которой необходимо варьировать три параметра: $EM$, $T_0$ и $\alpha$. С помощью параметра $T_0$ можно найти максимальную температуру $T_{max} = \alpha T_0$ и среднюю температуру $\langle T \rangle = EM^{-1} \int_0^\infty T \xi_\Gamma(T) dT$, равную для данного случая $\langle T \rangle = T_0(\alpha + 1)$.

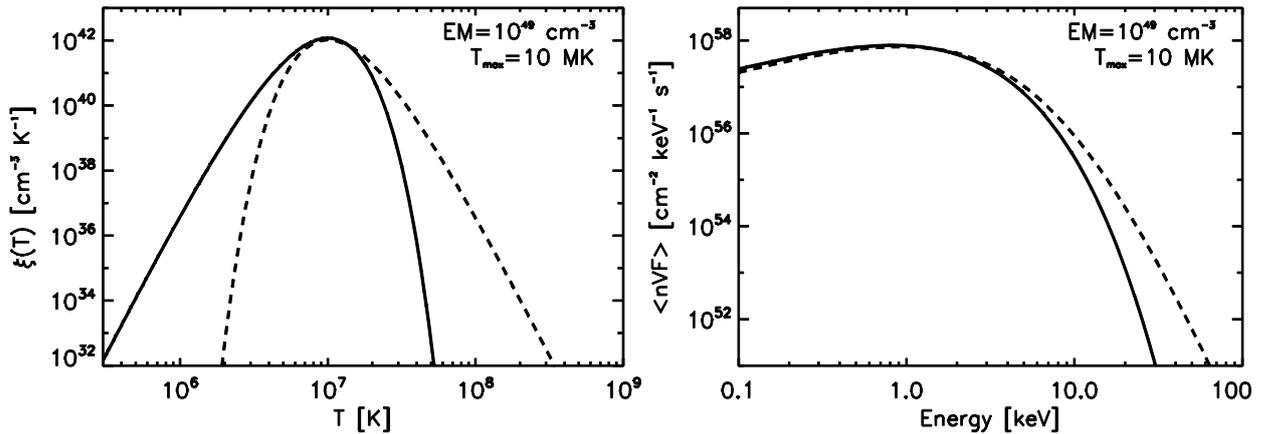

Рис. 2.1. Слева: ДМЭ, $\xi_\Gamma(T)$ в виде выражения (2.2.7) (сплошная линия) и $\xi_\kappa(T)$ в виде выражения (2.2.16) (пунктирная линия) для набора параметров: $EM=10^{49}$ см$^{-3}$, $T_{max}=10$ МК, $\alpha=9$. Справа: распределение электронов $\langle nVF(E) \rangle$ в виде выражения (2.2.9) (сплошная линия) и $\langle nVF(E) \rangle$ в виде выражения (2.2.18) (пунктирная линия) для того же набора параметров.

Из всего вышесказанного можно заключить, что используя эту модель для аппроксимации SDO/AIA данных и данных КА RHESSI в среде *OSPEX*, можно получить не только вид функций $\xi(T)$ и $\langle nVF(E) \rangle$, но также автоматически ключевые параметры плазмы, которые дают возможность диагностировать плазменные процессы. В разделе 2.4 приведен пример аппроксимации данных КА SDO/AIA и RHESSI функцией $\xi_\Gamma(T)$ и моделью тонкой мишени для солнечной вспышки 14.08.2010, где показано, что гамма-распределение хорошо описывает фоновую плазму вдоль луча зрения.

### 2.2.2 Описание каппа-распределения через дифференциальную меру эмиссии

Как было описано в разделе 2.1, в некоторых случаях распределения электронов, имеющие тепловое ядро (*thermal core*) и степенную зависимость, могут быть описаны с помощью каппа-распределения (напр., [134, 72])

$$f_\kappa(E) = n_\kappa \frac{2\sqrt{E}}{\sqrt{\pi(k_B T_\kappa)^3}} \frac{\Gamma(\kappa+1)}{\Gamma(\kappa-0.5)(\kappa-1.5)^{3/2}} \left(1 + \frac{E}{(\kappa-1.5)k_B T_\kappa}\right)^{-(\kappa+1)} \quad (2.2.11),$$

здесь $T_\kappa$ – средняя кинетическая температура электронов, определяемая через их среднюю энергию $\langle E \rangle = (3/2)k_B T_\kappa$. Предполагая изотропное распределение, используя замену переменных $f_\kappa(E)dE = 4\pi v^2 f_\kappa(v)dv$, можно получить распределение потока электронов по скоростям, которое представлено в работе [34]

$$f_\kappa(v) = n_\kappa \left(\frac{1}{\pi\kappa\theta^2}\right)^{3/2} \frac{\Gamma(\kappa+1)}{\Gamma(\kappa-1/2)}\left(1 + \frac{v^2}{\kappa\theta^2}\right)^{-\kappa-1} \quad (2.2.12),$$

где $\theta^2 = (2k_B T_\kappa / m_e)(\kappa-1.5)/\kappa$ – характеристическая скорость и $n_\kappa = \int f_\kappa(v) d^3v$ – концентрация, связанная с распределением ускоренных электронов. Параметры каппа-распределения имеют четкий физический смысл. В сценарии стохастического ускорения в столкновительной плазме в работе [34] характеристическая скорость связана с тепловой скоростью максвелловского распределения электронов, на которое действует ускорение, приводя к образованию степенного распределения, чей спектральный индекс $\kappa$ определяется через баланс между диффузным ускорением и столкновениями. Зная, что распределение электронов в излучающем объеме связано с ДМЭ выражением (2.2.6), введем ДМЭ в следующем виде:

$$\xi(T) \propto T^{-(\kappa+0.5)} \exp\left(-\frac{T_\kappa}{T}(\kappa-1.5)\right) \quad (2.2.13).$$

Данная ДМЭ убывает с ростом $T$, так как $\xi(T) \propto T^{-(\kappa+0.5)}$ для $T \gg T_\kappa$, и быстро растет для $T < T_\kappa$ из-за $\exp(-T_\kappa/T)$. ДМЭ имеет один максимум $d\xi(T)/dT = 0$ при температуре $T_{max} = T_\kappa(\kappa-1.5)/(\kappa+0.5)$. Более того, данная ДМЭ представляет собой каппа-распределение, описанное в уравнениях (2.2.11) и (2.2.12). Зная, что полную меру эмиссии *EM* можно определить в виде интеграла от $\xi(T)$ по всем температурам в плазме

$$EM = \int_0^\infty \xi(T)dT \propto \int_0^\infty T^{-(\kappa+0.5)} \exp\left(-\frac{T_\kappa}{T}(\kappa-1.5)\right)dT \qquad (2.2.14),$$

интеграл может быть решен с помощью гамма функции, в результате чего

$$EM = \frac{\Gamma(\kappa-0.5)}{(\kappa-1.5)^{(\kappa-0.5)}} T_\kappa^{-(\kappa-0.5)} \qquad (2.2.15).$$

Следовательно, $\xi(T)$ можно представить в виде

$$\xi(T) = \frac{EM(\kappa-1.5)^{(\kappa-0.5)}}{\Gamma(\kappa-0.5)T_\kappa}\left(\frac{T_\kappa}{T}\right)^{\kappa+0.5} \exp\left(-\frac{T_\kappa}{T}(\kappa-1.5)\right) \qquad (2.2.16),$$

т.е. константой пропорциональности в выражении (2.2.13) является величина $EM \cdot T_\kappa^{(\kappa-0.5)}(\kappa-1.5)^{(\kappa-0.5)}/\Gamma(\kappa-0.5)$. Для сравнения с ДМЭ из раздела 2.2.1 сделаем замену переменных: $\alpha=\kappa+0.5$, $T_0 = T_\kappa(\kappa-1.5)$, тогда

$$\xi(T) = \frac{EM}{T_0\Gamma(\alpha-1)}\left(\frac{T_0}{T}\right)^\alpha \exp\left(-\frac{T_0}{T}\right) \qquad (2.2.17),$$

где $\Gamma(x)$ – гамма функция, $\alpha>1$. На рис. 2.1 (левая панель) представлен пример ДМЭ для $EM=10^{49}$ см$^{-3}$, $T_{max} = 10$ МК, $\alpha=9$. Для однородной плазмы $n = n_e = n_i$ мера эмиссии может быть записана в виде $EM=n^2V$. Подставляя выражение (2.2.16) в (2.2.6), можно найти

$$\langle nVF(E)\rangle = n^2V \frac{2^{3/2}}{(\pi m_e)^{1/2}(k_BT_\kappa)^{1/2}} \frac{\Gamma(\kappa+1)}{(\kappa-1.5)^{1.5}\Gamma(\kappa-1/2)} \times$$

$$\times \frac{E/(k_BT_\kappa)}{(1+E/(k_BT_\kappa(\kappa-1.5)))^{\kappa+1}} \qquad (2.2.18).$$

Для сравнения с $\langle nVF(E)\rangle_K$ из раздела 2.2.1 аналогично используя замену переменных: $\alpha=\kappa+0.5$, $T_0 = T_\kappa(\kappa-1.5)$

$$\langle nVF(E)\rangle = EM \frac{2^{3/2}}{(\pi m_e)^{1/2}(k_BT_0)^{3/2}} \frac{\Gamma(\alpha+1/2)}{\Gamma(\alpha-1)} \frac{E}{(1+E/(k_BT_0))^{\alpha+1/2}} \qquad (2.2.19).$$

На рис. 2.1 (правая панель) показан пример $\langle nVF(E)\rangle$ для тех же параметров: $EM=10^{49}$ см$^{-3}$, $T_{max}=10$ МК, $\alpha=9$, в дальнейшем будем обозначать как $\langle nVF(E)\rangle_\kappa$. Если ввести изотропную функцию распределения электронов $\langle f(v)\rangle$, $n=\int\langle f(v)\rangle d^3v$, так что $F(E)dE=v\langle f(v)\rangle d^3v$ и $d^3v=4\pi v^2 dv$, можно записать

$$\langle f(v)\rangle = \frac{m_e n}{4\pi v^2}\frac{2^{3/2}}{(\pi m_e k_B T_\kappa)^{1/2}}\frac{\Gamma(\kappa+1)}{(\kappa-1.5)^{3/2}\Gamma(\kappa-1/2)}\frac{m_e v^2/(2k_B T_\kappa)}{[1+m_e v^2/(2k_B T_\kappa(\kappa-1.5))]^{\kappa+1}} \quad (2.2.20),$$

где $E=m_e v^2/2$, или упрощая

$$\langle f(v)\rangle = n\left(\frac{m_e}{2\pi k_B T_\kappa(\kappa-1.5)}\right)^{3/2}\frac{\Gamma(\kappa+1)}{\Gamma(\kappa-1/2)}\left(1+\frac{m_e v^2}{(\kappa-1.5)2k_B T_\kappa}\right)^{-\kappa-1} \quad (2.2.21).$$

Это идентично выражению (2.2.12), когда $T_\kappa$ выражается через характеристическую скорость. Таким образом, ДМЭ в выражении (2.2.16) представляет собой каппа-распределение, и определяется тремя параметрами (*EM*, $T_\kappa$ и $\kappa$), которые могут быть найдены путем аппроксимации модельными функциями рентгеновского и КУФ спектров. Она реализована в среде *OSPEX* (*f_multi_therm_pow_exp.pro*) и далее упоминается как $\xi_\kappa(T)$. Стоит отметить, что функция, реализованная в *OSPEX*, не дает напрямую параметры $T_\kappa$ и $\kappa$, но их можно вывести через соотношения $\alpha=\kappa+0.5$ и $T_{max}=T_\kappa(\kappa-1.5)/(\kappa+0.5)$ в диапазоне температур 0.086-8.6 кэВ.

## 2.3 Комбинирование RHESSI и SDO/AIA наблюдений

КА RHESSI чувствителен к плазме при температурах выше ~8 МК, в то время как совокупность шести КУФ каналов КА AIA (94, 131, 193, 171, 211, 335 Å), охватывает диапазон ~0.5-16 МК [88]. Таким образом, одновременный анализ КА RHESSI и SDO/AIA данных улучшает измерение низкотемпературной составляющей заданного распределения электронов и позволяет расширить доступный диапазон энергий вплоть до ~0.1 кэВ. В данной части главы представлен разработанный метод, с помощью которого данные КА RHESSI и SDO/AIA могут быть аппроксимированы модельными функциями одновременно путем создания единой матрицы температурного отклика, которая состоит из температурного отклика КА SDO/AIA и температурного отклика КА RHESSI. В разделах 2.4 и 2.5 разработанная методика применяется к двум солнечным вспышкам (14.08.2010, 08.05.2015), используя функции *ξ(T)* в виде выражений (2.2.7) и (2.2.15).

### 2.3.1 Аппроксимация модельными функциями одновременно RHESSI и SDO/AIA данных

Сигнал $g_i$, зарегистрированный в заданном энергетическом диапазоне РИ или КУФ каналах можно представить в виде произведения ДМЭ источника на функции отклика детекторов и температурного отклика

$$g_i = R_{ij} \xi_j dT_j \qquad (2.3.1),$$

где **g** = (**g**$^{AIA}$, **g**$^{RHESSI}$). В выражении (2.3.1) **g**$^{AIA}$ является вектором, содержащим КУФ данные (*DN/s*) от всей вспышечной области в шести КУФ каналах, **g**$^{RHESSI}$ – скорость счета, наблюдаемая с помощью КА RHESSI (*counts/s*). **R** является комбинированной матрицей отклика температуры, включая температурный отклик КА SDO/AIA (для *i=1,...,6*) и температурный отклик КА RHESSI (для *i ≥ 7*). Данная матрица построена с использованием спектральной матрицы отклика RHESSI (*Spectral Response Matrix*, см. [120]) и функции теплового тормозного излучения (*f_vth.pro*, доступна в среде *OSPEX*) для температур плазмы, которые содержатся в температурном отклике КА SDO/AIA. На рис. 2.2 показаны функции теплового отклика КА SDO/AIA (сплошные линии) и тепловой отклик КА RHESSI для 7, 12, 15 и 24 кэВ (пунктирные линии). Матрица **R** покрывает диапазон температур от 0.043 кэВ (0.5 МК) до ~86 кэВ (1000 МК). Продление до 86 кэВ необходимо для аппроксимации модельными функциями высокоэнергетической части спектра должным образом. Для этого файлы базы данных, содержащие табличные спектры теплового РИ были пересчитаны с использованием

*CHIANTI 7.1* и теперь являются частью программного обеспечения *SSW* и доступны через *OSPEX*.

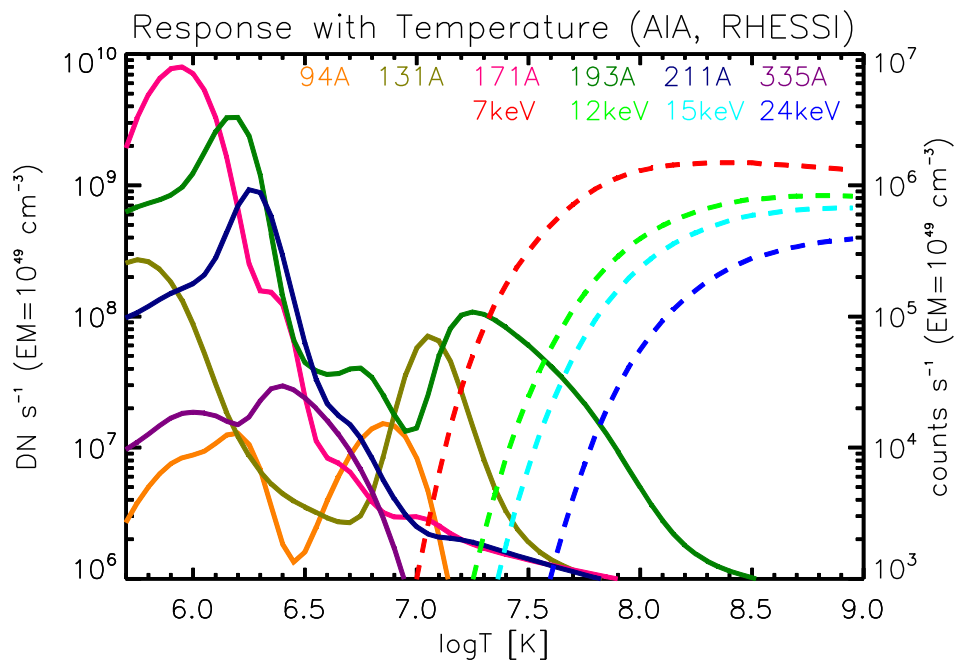

Рис. 2.2. Температурный отклик КА SDO/AIA для шести КУФ каналов (сплошные линии, левая ось) и температурный отклик КА RHESSI для 7, 12, 15 и 24 кэВ (пунктирные линии, правая ось), для меры эмиссии *EM*=$10^{49}$ см$^{-3}$.

Следует также отметить, что поскольку КА RHESSI не чувствителен к низким температурам, температурный отклик КА RHESSI в матрице **R** установили равным нулю для всех энергий, которые представляют температуры <0.1 кэВ.

## 2.4 Анализ событий 14.08.2010 и 08.05.2015

Вспышка 14 августа 2010 г. началась в 09:25:40 UT и относится к рентгеновскому классу C4.1 по GOES [138], вспышка 8 мая 2015 г. – к рентгеновскому классу C1.5, началась в 8:00:40 UT. Обе вспышки были хорошо видны как с КА RHESSI, так и с КА SDO/AIA. На примере вспышки 14 августа 2010 г. в работах [100, 33] подробно рассмотрена разработанная методика аппроксимации модельными функциями SDO/AIA и RHESSI данных одновременно, описанная в разделе 2.3. Для лимбовой вспышки 8 мая 2015 г. была произведена временная эволюция параметров плазмы – меры эмиссии, температуры и концентрации [17].

Методика исследования и обработки данных: а) поиск солнечной вспышки, которая наблюдалась обоими космическими аппаратами; б) подготовка наблюдательных данных для работы с ними (калибровка); в) сравнение существующих методов: метода регуляризации Тихонова для SDO/AIA данных, метода аппроксимации ДМЭ функцией RHESSI данных, с разработанным методом аппроксимации ДМЭ функцией SDO/AIA и RHESSI данных одновременно; г) используя преобразование Лапласа через найденную ДМЭ функцию определение энергетического распределения электронов, а также счет меры эмиссии, температуры и концентрации.

Солнечная вспышка 14 августа 2010 года (SOL2010-08-14T10:05) [32] имела максимум жесткого РИ в 09:46 UT. Она имеет простую однопетлевую морфологию с ярко выраженным источником в вершине петли, наблюдавшийся как с помощью КА RHESSI, так и КА SDO/AIA, и слабое излучение в основаниях петли при энергиях выше ~18 кэВ. Таким образом, данное событие идеально подходит для изучения распределения электронов в источнике в вершине петли, без необходимости использования пространственной спектроскопии. Интервал времени 09:42–09:43 UT, для которого производилась аппроксимация данных модельными функциями, находится на фазе роста, перед пиком вспышки (09:46 UT), где эффект (*pileup*) [120], при котором прилетающие практически одновременно фотоны регистрируются как один, а их энергии суммируются, был незначительным, и наблюдаемое жесткое РИ не превышало ~24 кэВ.

Данные КА RHESSI для момента времени 09:42-09:43 UT были аппроксимированы с помощью много-температурной модели (*f_multi_therm_2pow.pro*), доступной в среде *OSPEX*

$$\xi(T) = \frac{EM}{T_0 \mathrm{B}(\alpha+1, \beta-\alpha-1)} \left(\frac{T}{T_0}\right)^{\alpha} \left(1 + \frac{T}{T_0}\right)^{-\beta} \qquad (2.4.1),$$

где $\mathrm{B}(x,y) = \int_0^{\infty} t^{x-1}/(1+t)^{x+y}\, dt$ - бета-функция, и моделью тонкой мишени (*f_thin2.pro,* реализована в *OSPEX*) (см., напр., [128]). При критерии $\chi^2$=0.84 были получены следующие параметры: $EM$=5×10$^{47}$ см$^{-3}$, $T_0$=0.75 кэВ, $T_{max}=T_0\alpha/(\beta-\alpha)$=0.25 кэВ, $\alpha$=3, $\beta$=12, спектральный индекс $\delta$=3.2, низкоэнергетическая граница $E_c$=7.75 кэВ. Температура и мера эмиссии, полученные по данным спутника GOES, для того же самого интервала составили $T_{GOES}$=0.8 кэВ и $EM_{GOES}$=5×10$^{47}$ см$^{-3}$.

В ходе анализа данных РИ, зарегистрированных на КА RHESSI, был предварительно вычтен фон, соответствующий предвспышечной фазе. Поскольку возможность аппроксимации одновременно RHESSI и SDO/AIA данных не является встроенной в среде *OSPEX*, была использована стандартная процедура *mpfit.pro* из пакета IDL. Ошибки RHESSI данных рассчитывались, как $C_{err} = \sqrt{(C+B)/L + B_{err}^2 + (0.02C)^2}$, где $C$ – скорость счета, $B$ – фоновая скорость счета, $B_{err}$ – статистические ошибки фоновой скорости счета и $L$ – время регистрации сигнала детектором [120]. Предполагалось, что систематические ошибки скорости счета равны 2%.

Данные в диапазоне крайнего ультрафиолетового излучения, полученные на КА SDO/AIA в шести КУФ каналах 94, 131, 171, 193, 211, 335 Å, были дополнительно откалиброваны с учетом перемещения, вращения, масштабирования с помощью программы *aia_prep.pro* (доступна в *SSW*), а также нормированы на время выдержки (*exposure time*). Ошибки на данные КА SDO/AIA ($DN_{err}$) включали систематические ошибки в размере 20% и вычислялись по формуле $DN_{err} = \sqrt{DN + (0.2DN)^2}$ (напр., [86]). Следует отметить, что SDO/AIA карты были взяты практически в один и тот же момент времени, и временной интервал между ними не превышал 12с. При этом предполагалось, что одна и та же излучающая плазма наблюдается на всех длинах волн, а основной процесс энерговыделения происходит из области, равной 50% от максимума интенсивности RHESSI карты. На рис. 2.3 представлена SDO/AIA 131 Å карта (09:42:57.62 UT) с RHESSI 20, 30, 50% контурами для энергетического диапазона 8-10 кэВ при использовании алгоритма CLEAN [69] для временного интервала 09:42-09:43 UT.

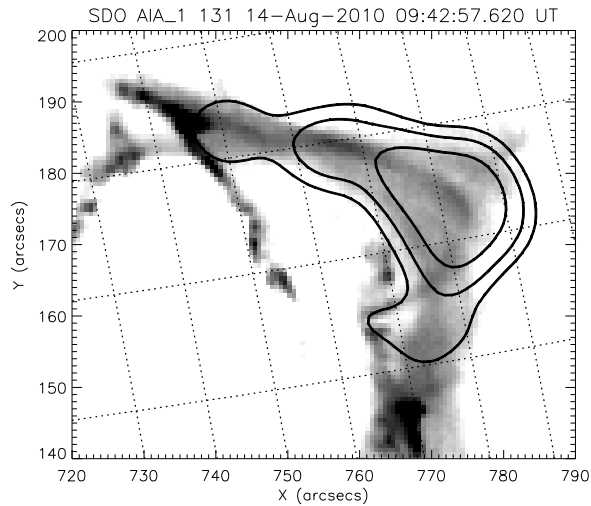

Рис. 2.3. SDO/AIA 131 Å карта с 20, 30, 50% RHESSI контурами для временного интервала 09:42–09:43 UT.

Из рис. 2.3 видно, что источники в рентгеновском и крайнем ультрафиолетовом диапазонах пространственно совместимы. Таким образом, для поиска $\xi(T)$ были использованы данные КА SDO/AIA из вершины петли, из области, соответствующей 50% RHESSI контуру.

### 2.4.1 Вычисление дифференциальной меры эмиссии одновременно для SDO/AIA и RHESSI данных

В настоящем разделе представлено вычисление ДМЭ в виде непрерывной функции, из небольшого числа наблюдаемых интенсивностей для КА SDO/AIA и числа отсчетов для КА RHESSI, которые являются результатом свертки $\xi(T)$ с комбинированной матрицей отклика температуры **R** (см. раздел 2.3).

Для данной области (рис. 2.3) были найдены $\xi(T)$ тремя различными способами:

1) С помощью метода регуляризации (*data2dem_reg.pro*) [25, 75] для SDO/AIA данных, разработанного авторами [63];
2) В результате аппроксимации много-температурной функции (*f_multi_therm_2pow.pro*), имеющей $\xi(T)$ в виде уравнения (2.4.1), RHESSI данных;
3) В результате аппроксимации тепловой моделью $\xi_T(T)$, где ДМЭ представлена выражением (2.2.7), и нетепловой моделью (модель тонкой мишени, *f_thin2.pro*, доступна в *OSPEX*) одновременно RHESSI и SDO/AIA данных.

На рис. 2.4 (левая панель) представлена ДМЭ, найденная тремя вышеперечисленными способами, а также данные КА SDO/AIA, поделенные на функции температурного отклика КА SDO/AIA (*loci-curves*), ограничивающие $\xi(T)$ сверху.

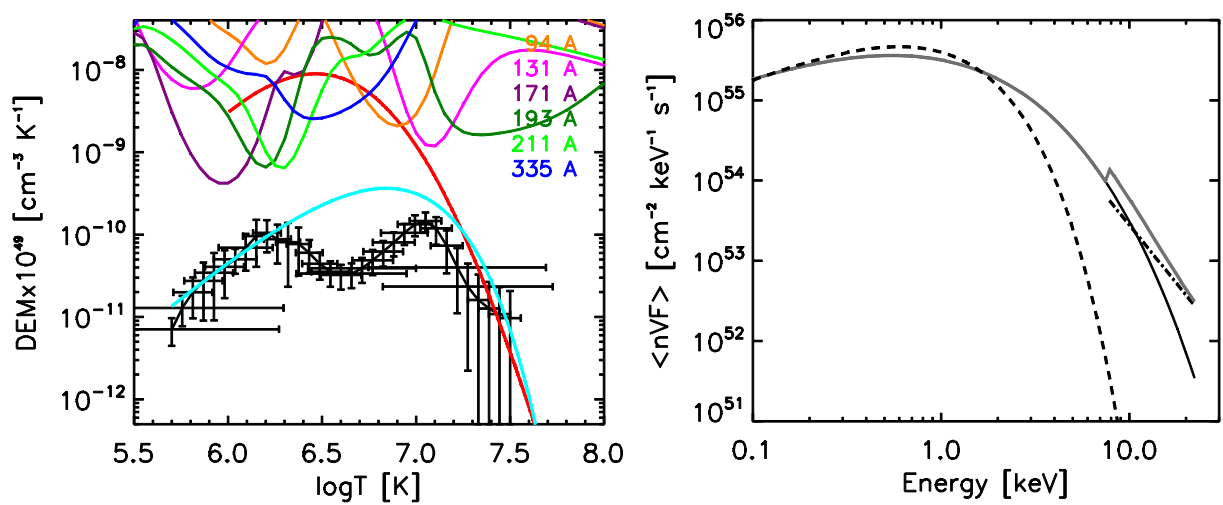

Рис. 2.4. Слева: регуляризированная $\xi(T)$, вычисленная по данным КА SDO/AIA (черная линия), $\xi(T)$ – по данным КА RHESSI (красная линия), $\xi_T(T)$ – по данным КА SDO/AIA и RHESSI одновременно (голубая линия), остальные линии – данные SDO/AIA, поделенные на функции температурного отклика SDO/AIA. Справа: энергетическое распределение электронов, найденное с помощью аппроксимации одновременно КА SDO/AIA и RHESSI данных функцией $\xi_T(T)$ (черная сплошная линия) и моделью тонкой мишени (черная пунктир-точка линия) и их сумма (серая линия), а также распределение Максвелла, соответствующее $EM=4.64\times10^{46}$ см$^{-3}$ и $T=T_{max}=0.58$ кэВ (черная пунктирная линия).

Видно, что $\xi(T)$ имеет сложную форму, поэтому довольно трудно подобрать функцию, имеющую подобный вид. Отметим также, что попытки аппроксимирования два набора данных исключительно функцией $\xi(T)$ без добавления нетепловой модели оказались безуспешными для данной вспышки (критерий $\chi^2>2$), это может свидетельствовать либо о неудачном подборе самой модельной функции, либо о наличии нетепловой компоненты.

В результате одновременной аппроксимации RHESSI и SDO/AIA данных функцией $\xi_\Gamma(T)$ и моделью тонкой мишени (критерий $\chi^2=0.83$) были получены следующие параметры: $EM=4.64\times10^{46}$ см$^{-3}$, $T_0=0.3$ кэВ, $T_{max}=0.58$ кэВ, $<T>=0.9$ кэВ, $\alpha=1.96$, спектральный индекс $\delta=2.9$, низкоэнергетическая граница $E_c=7.78$ кэВ. Энергетическое распределение электронов $\langle nVF(E) \rangle_K$ для найденных параметров одновременной аппроксимации RHESSI и SDO/AIA данных, а также распределение Максвелла, соответствующее значениям найденной меры эмиссии и максимальной температуры, представлены на рис. 2.4 (правая панель). Видно, что отклонение $\langle nVF(E) \rangle_K$ от распределения Максвелла присутствует не только на высоких, но и на низких энергиях, то есть распределение частиц имеет более сложную структуру.

На рис. 2.5 представлены результаты одновременной аппроксимации SDO/AIA и RHESSI данных. Порядок КУФ каналов КА SDO/AIA был выбран таким образом, чтобы показать рост температур, к которым они наиболее чувствительны. Из рис.2.5 видно, что результаты аппроксимации, полученные комбинированием RHESSI и SDO/AIA наблюдений, согласуются с исходными данными, хотя можно заметить, что для фильтра с длиной волны 171 Å, который ответственен за низкие температуры, данные КА SDO/AIA аппроксимируются недостаточно хорошо. Таким образом, разница между данными КА SDO/AIA и результатами аппроксимации для фильтра 171 Å порядка одной сигма может быть объяснена тем, что луч зрения пересекает плазму с температурой ~$10^6$ K, расположенную выше или ниже корональной петли, для которой мы ищем дифференциальную меру эмиссии.

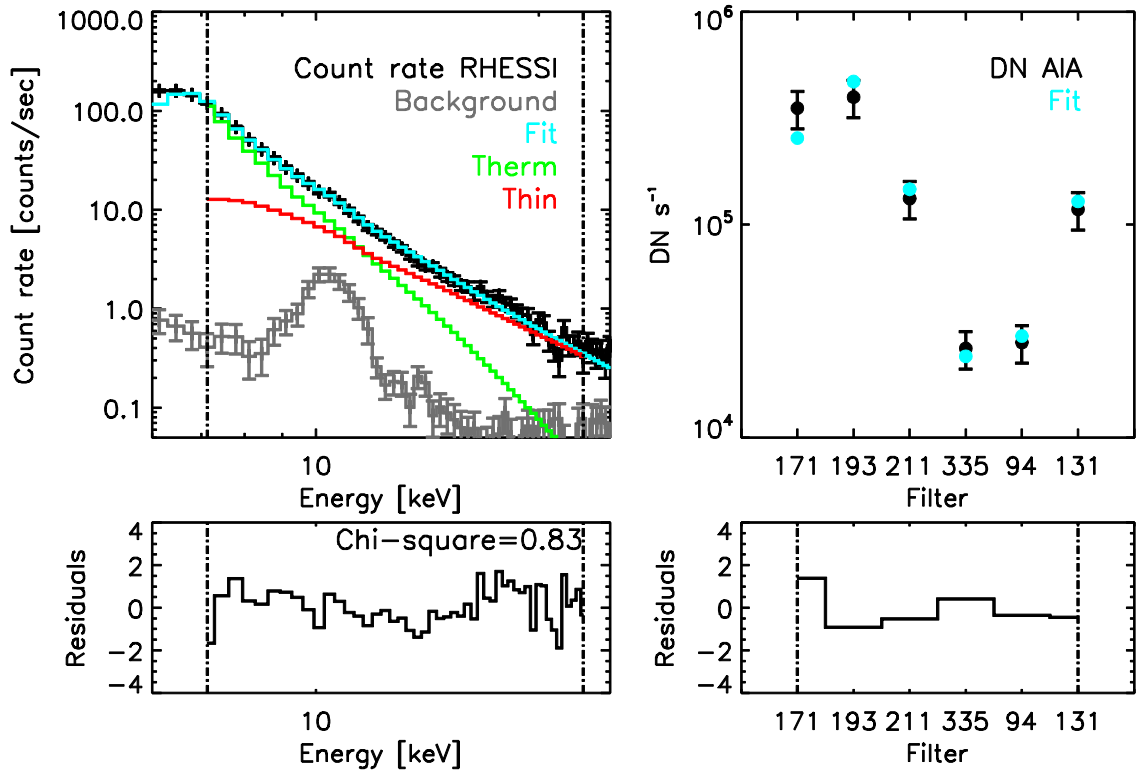

Рис. 2.5. Результаты аппроксимации одновременно RHESSI (слева) и SDO/AIA (справа) данных модельными функциями. Сверху слева: RHESSI данные (*Count rate RHESSI*), результаты аппроксимации (*Fit*), тепловая (*Therm*) и нетепловая (*Thin*) модели. Сверху справа: SDO/AIA данные (*DN AIA*), результаты аппроксимации (*Fit*); внизу показано отношение разности между наблюдательными данными и результатами аппроксимации к соответствующим ошибкам для RHESSI и SDO/AIA измерений (*Residuals*).

Следует отметить, что с помощью комбинированного анализа SDO/AIA и RHESSI данных впервые был найдено энергетическое распределение электронов для широкого диапазона энергий: 0.1 – 20 кэВ.

### 2.4.2 Временная эволюция параметров плазмы в солнечных вспышках на основе RHESSI и SDO/AIA наблюдений

Для лимбового события 8 мая 2015 г., которое началось в 8:00:40 UT, имеющего источник в вершине вспышечной петли, была проанализирована временная эволюция параметров вспышечной плазмы (мера эмиссии, температура, концентрация) с использованием разработанной методики (см. раздел 2.3) одновременной аппроксимации SDO/AIA и RHESSI данных для десяти временных интервалов на фазе роста, максимума и спада интенсивности излучения (рис.2.6). Методика обработки данных аналогична рассмотренной ранее в разделе 2.4.1. На рис. 2.7 показаны карты КА SDO/AIA в шести КУФ фильтрах с наложенными 20, 30, 50% RHESSI контурами (CLEAN алгоритм [69]).

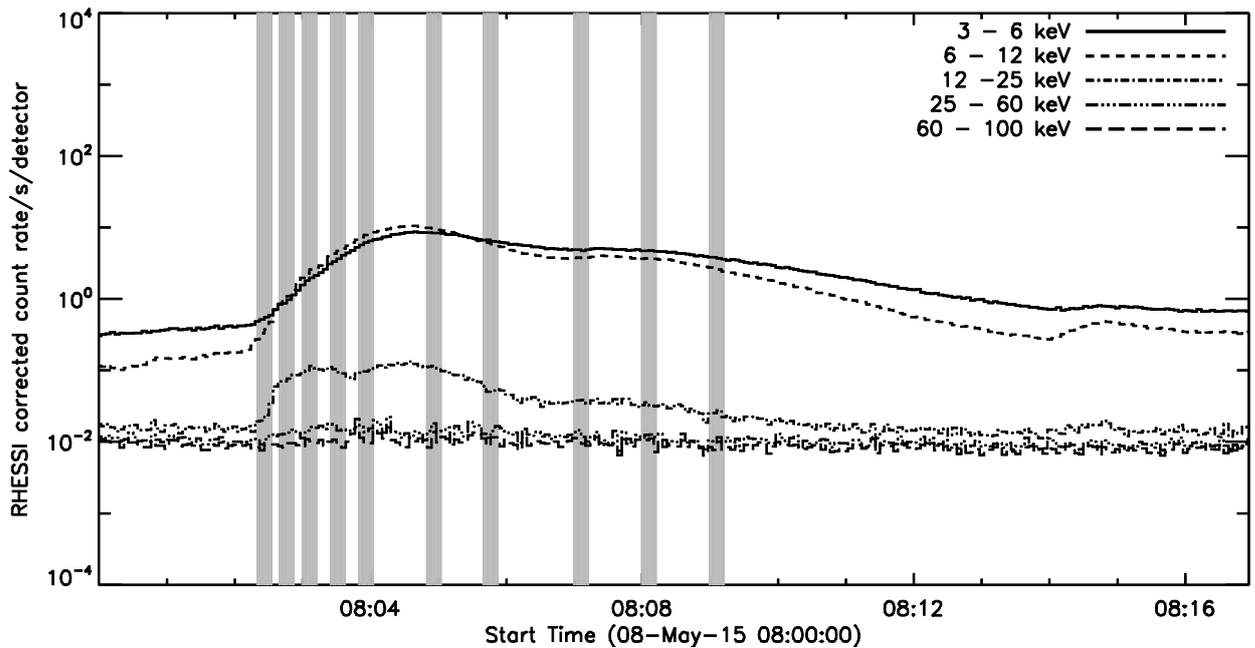

Рис. 2.6. Кривые блеска РИ по данным КА RHESSI для вспышки 8 мая 2015 года. Серыми областями показаны рассматриваемые в работе моменты времени.

Так как аппроксимация данных одной функцией $\xi(T)$ оказалась неудовлетворяющей критерию $\chi^2$, то в данном случае использовались две функции $\xi(T)$, соответствующие двум компонентам плазмы: холодная, соответствующая фоновой плазме, $\xi_Г(T)$ (см. раздел 2.2.1), и горячая, соответствующая вспышечной плазме, $\xi_к(T)$ (см. раздел 2.2.2). Результаты аппроксимации двумя функциями $\xi(T)$ одновременно SDO/AIA и RHESSI данных дали удовлетворительные результаты, и можно сказать, что они взаимно дополняют друг друга.

На рис. 2.8 и рис. 2.9 показана временная эволюция физических параметров для горячей и холодной компонент плазмы: максимальная температура (рис. 2.8, левая панель), средняя температура (рис. 2.8, правая панель), мера эмиссии (рис. 2.9, левая панель), концентрация (рис. 2.9, правая панель), где концентрация находилась из выражения $n=(EM/V)^{1/2}$, в котором объем излучающей области $V=A^{3/2}$ [см$^3$], а площадь $A$ бралась для 50% максимума интенсивности RHESSI карты для энергии 6-10 кэВ (CLEAN алгоритм).

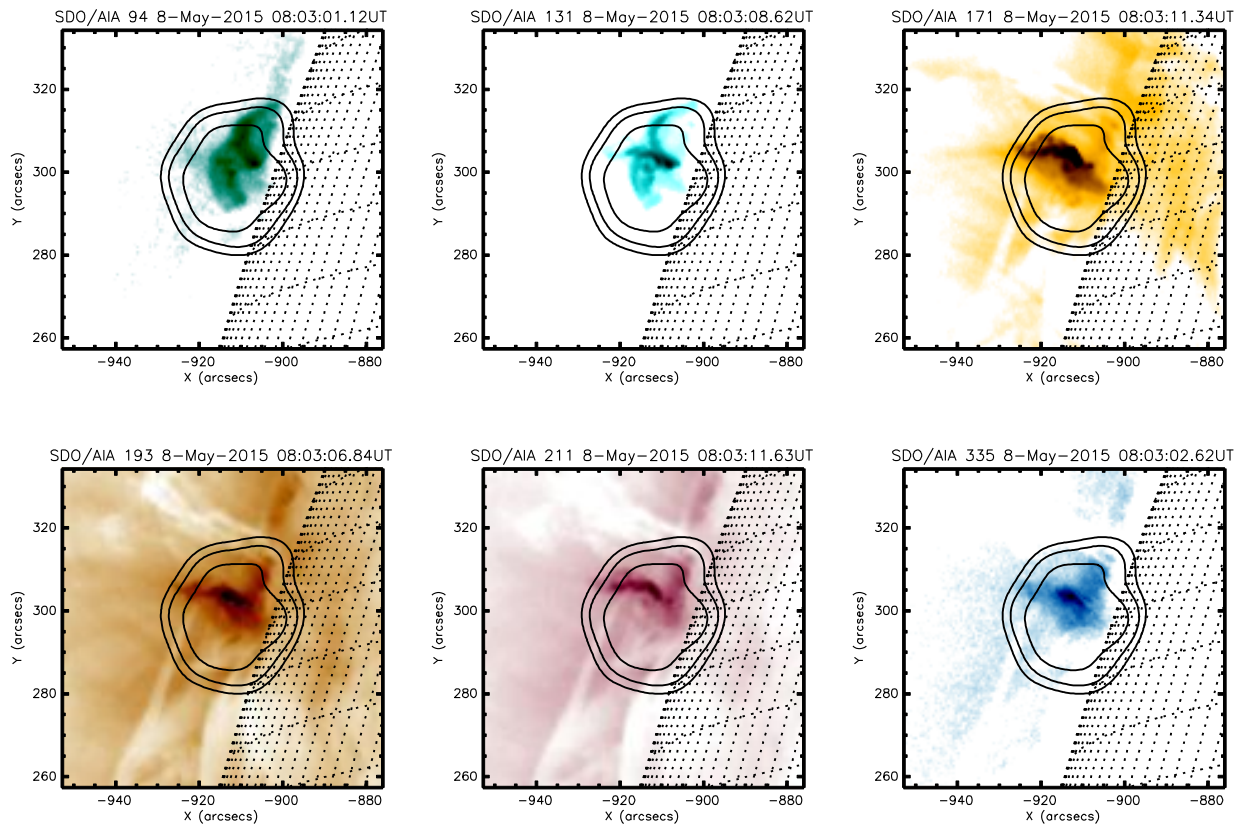

Рис. 2.7. AIA 94, 131, 171, 193, 211, 335 Å карты с 20, 30, 50% RHESSI контурами для временного интервала 08:03:00-08:03:12 UT.

Как видно из рис. 2.8 и рис. 2.9 параметры холодной компоненты плазмы (фоновая плазма) не сильно меняются со временем, в то время как параметры горячей компоненты плазмы изменяются каждые 12-20 секунд. Средняя и максимальная температура горячей компоненты повторяют временной ход РИ, а пики меры эмиссии и концентрации плазмы запаздывают примерно на 2 минуты.

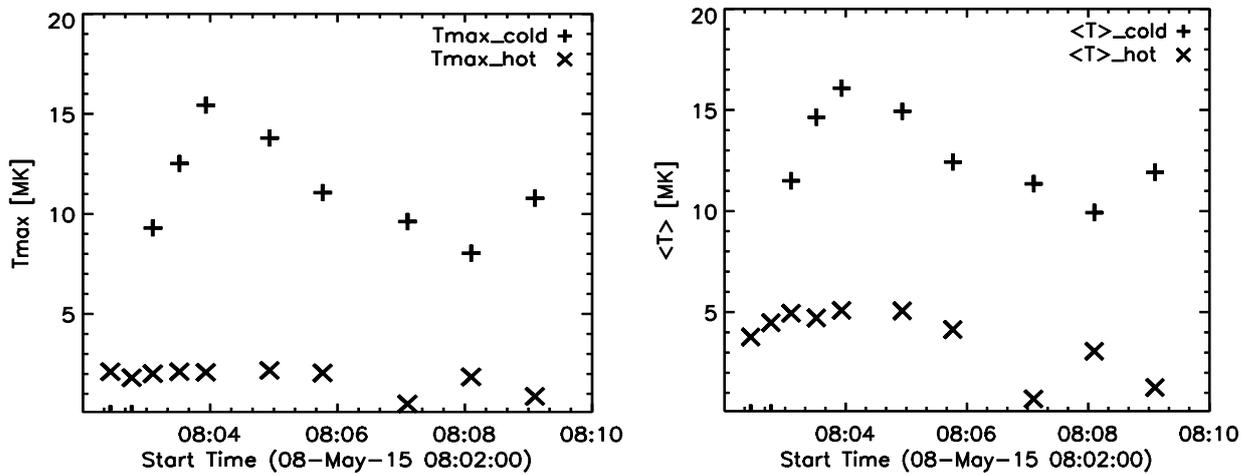

Рис. 2.8. Максимальная (слева) и средняя температура (справа) для холодной (*cold*) и горячей (*hot*) компонент плазмы, найденные с помощью одновременной аппроксимации двумя функциями ДМЭ RHESSI и SDO/AIA данных для десяти временных интервалов.

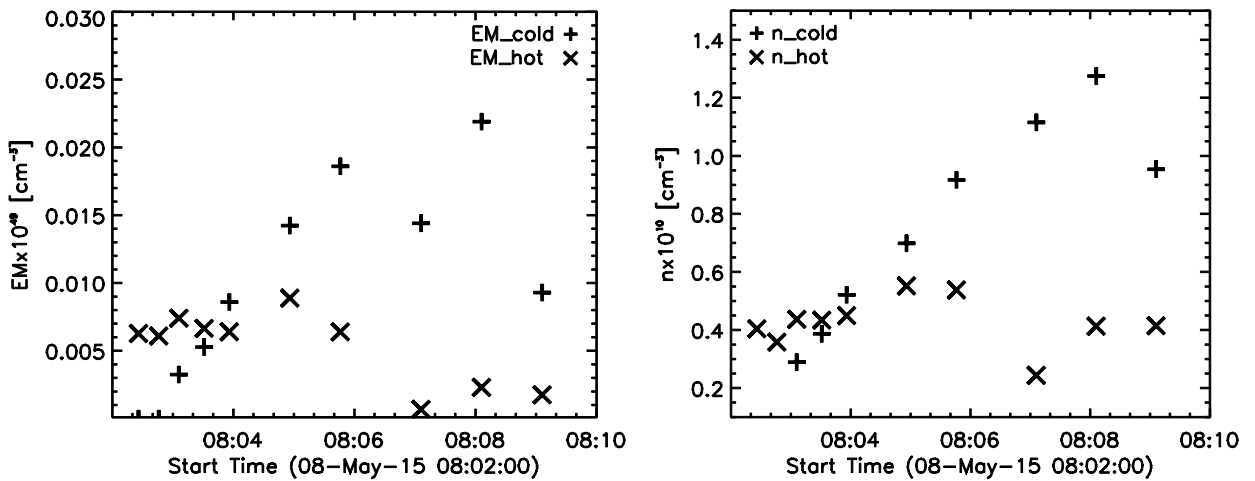

Рис. 2.9. Мера эмиссии (слева), концентрация (справа) для холодной (*cold*) и горячей (*hot*) компонент плазмы, найденные с помощью одновременной аппроксимации двумя функциями ДМЭ RHESSI и SDO/AIA данных для десяти временных интервалов.

Таким образом, полученные результаты показывают подробную эволюцию параметров плазмы в солнечной вспышке, где запаздывание меры эмиссии по отношению к временным профилям мягкого РИ свидетельствует об энерговыделении из вспышечной петли с более высокой плотностью, в то время как температура является результатом первичного энерговыделения, и запаздывания не происходит. Данный эффект является модификацией эффекта Нойперта [101, 53]. Эффект Нойперта заключается в линейной зависимости между потоком жесткого РИ и производной по времени от потока мягкого РИ, который в свою очередь является одним из основных критериев применимости «стандартной» модели солнечной вспышки.

## 2.5 Энергетическое распределение электронов в предположении каппа-распределения

В данном разделе произведена аппроксимация двумя различными функциями ($\xi_\kappa(T)$ в сравнении с функцией *f_thin_kappa.pro*, далее именуемой как *thin_kappa*, доступна в *OSPEX*) на примере события 14 августа 2010 года, рассмотренного в работе [32], а также ранее в разделе 2.4. Временные профили по данным КА RHESSI в трех энергетических диапазонах показаны на рис. 2.10 вместе с кривой блеска по данным КА GOES в канале 1-8 Å [138], серой областью указан момент времени 09:42:00–09:42:32 UT, для которого производилась аппроксимация модельными функциями.

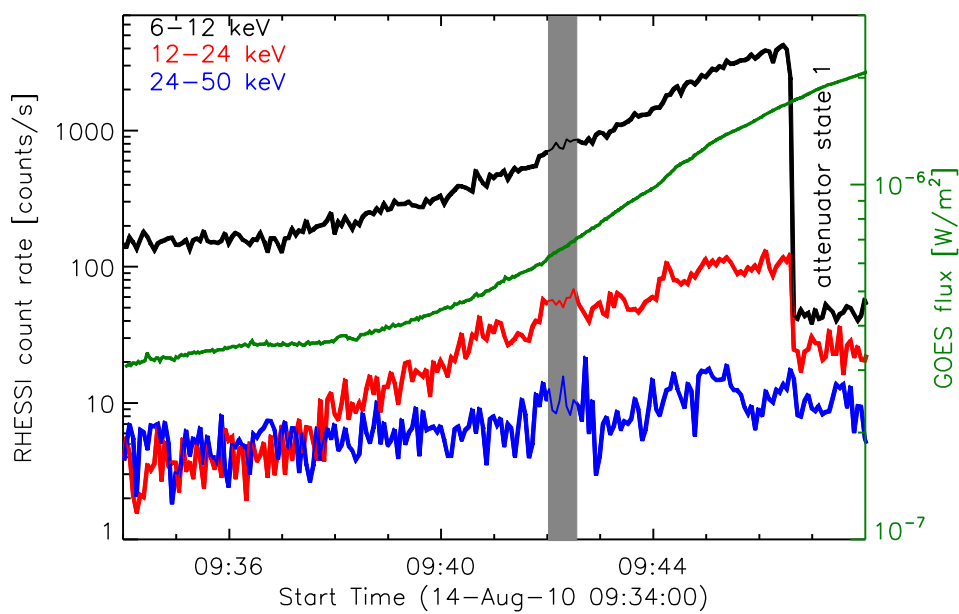

Рис. 2.10. Временные профили РИ по данным космических аппаратов RHESSI (слева) и GOES (справа) для вспышки 14 августа 2010 года. Серой областью показан момент времени (09:42:00–09:42:32 UT), для которого производилась аппроксимация модельными функциями.

Данный временной промежуток приблизительно соответствует рассмотренному в разделе 2.4 (09:42-09:43 UT), однако для лучшего согласования RHESSI и SDO/AIA данных, был взят 32-х секундный интервал.

На рис. 2.11 показаны AIA карты в четырех различных каналах с наложенными 30%, 50%, 70% RHESSI контурами, для получения которых использовался CLEAN алгоритм для энергий 6-12 кэВ [69].

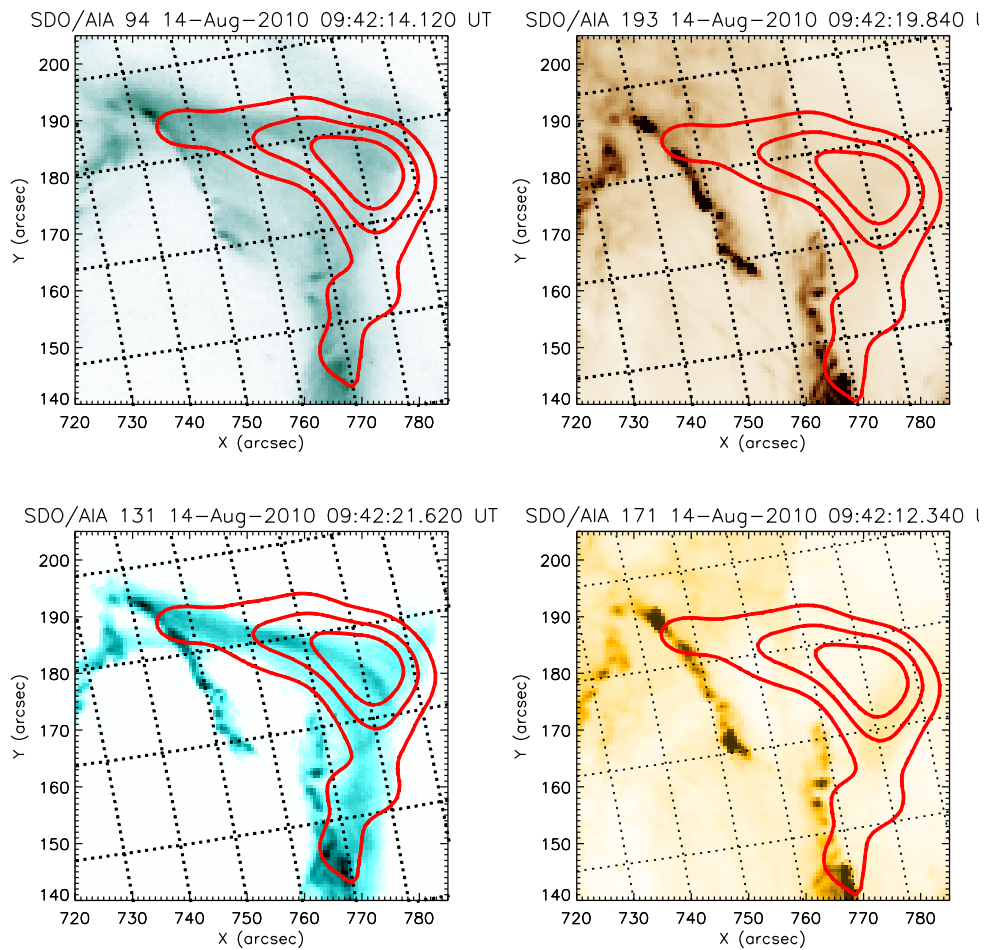

Рис. 2.11. SDO/AIA 94, 193, 131, 171 Å карты и 30%, 50%, 70% RHESSI контуры (красные линии) для момента времени 09:42:00–09:42:32 UT.

SDO/AIA 193 Å и 171 Å карты отображают два вспышечных волокна (правая панель, рис. 2.11), соединенные петлей, видимой на 131 Å и 94 Å картах (левая панель, рис. 2.11), и являются пространственно совместимыми с RHESSI источником в диапазоне энергий 6-12 кэВ. Данные КА RHESSI с предварительно вычтенным фоном, соответствующим предвспышечной фазе, были аппроксимированы функциями $\xi_\kappa(T)$ (*f_multi_therm_pow_exp.pro*, доступна в *OSPEX*) и *thin_kappa* для сравнения в диапазоне энергий 7-24 кэВ. На рис. 2.12 показаны спектры мягкого РИ и результаты аппроксимации двумя различными моделями. Сравнение полученных параметров аппроксимации приведено в Таблице 2.1. В случае аппроксимации моделью *thin_kappa* была добавлена дополнительная гауссиана, чтобы учесть комплекс линий высокоионизованного железа (*Fe-line complex*) около 6.7 кэВ [107].

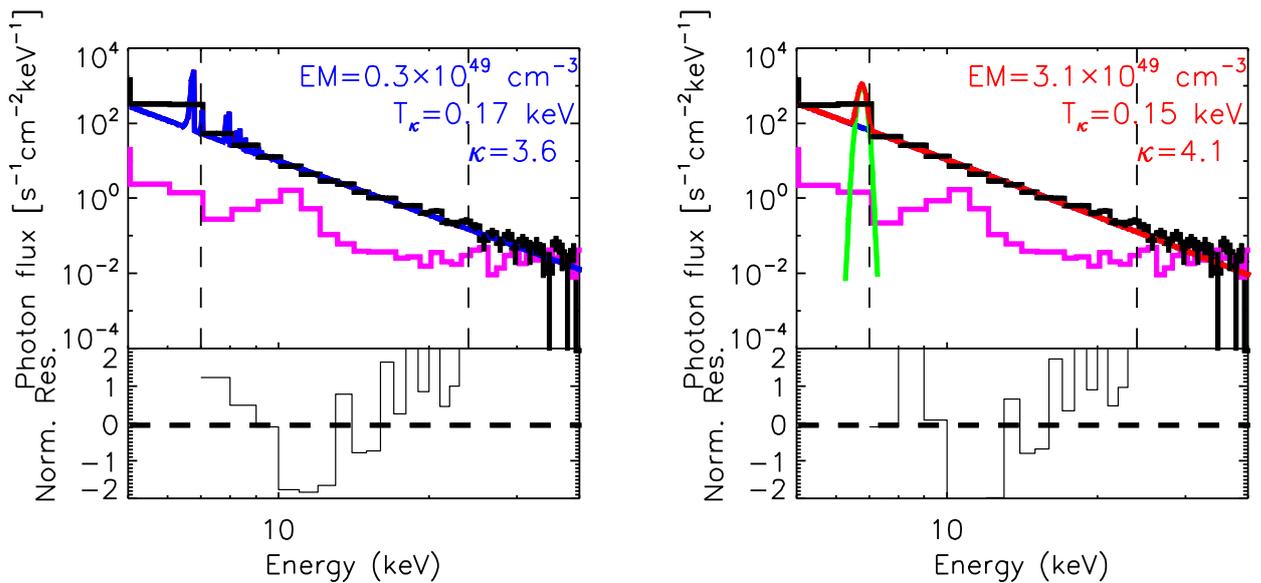

Рис. 2.12. Справа: Спектр фотонов по данным КА RHESSI (черная гистограмма), фоновое излучение (фиолетовая гистограмма), модель $\xi_\kappa(T)$ (синяя линия). Внизу показано отношение разности между наблюдательными данными и результатами аппроксимации к соответствующим ошибкам для RHESSI измерений. Справа: то же самое, что и слева для модели *thin_kappa* (красная линия) плюс гауссиана (зеленая линия). Вертикальные пунктирные линии указывают анализируемый диапазон энергий (7-24 кэВ).

В случае $\xi_\kappa(T)$ добавление гауссианы не требуется, так как излучение в линиях входит в данную модель. В то время как высокоэнергетическая часть спектра хорошо ограничена при аппроксимации, целый ряд значений для меры эмиссии и температуры могут быть найдены данным методом с аналогичными значениями критерия $\chi^2$ для всех моделей. По этой причине аппроксимация модельными функциями была произведена несколько раз с различными начальными параметрами, и используя незначительный сдвиг нижнего и верхнего пределов энергий аппроксимирования (±1 кэВ), чтобы получить оценку неопределенностей. Эти неопределенности приведены в Таблице 2.1 для обозначения диапазона возможных параметров аппроксимации. Для обеих моделей было рассчитано энергетическое распределение электронов $\langle nVF(E) \rangle$ (рис. 2.13).

Таблица 2.1. Параметры, полученные из результатов аппроксимации различными моделями, электронная концентрация $n$, плотность энергии вспышки $U_к$ и полная энергия вспышки $E_{tot}$

| Модель | $EM$ ($10^{48}$ см$^{-3}$) | $K$ | $T_к$ (кэВ) | $n$ ($10^{10}$ см$^{-3}$) | $U_к$ (эрг см$^{-3}$) | $E_{tot}$ ($10^{28}$ эрг) |
|---|---|---|---|---|---|---|
| Аппроксимация модельными функциями данных КА RHESSI ||||||| 
| *thin_kappa* | $31^{+39}_{-9}$ | $4.1^{+0.1}_{-0.1}$ | $0.15^{+0.01}_{-0.02}$ | $14^{+7}_{-2}$ | $52^{+31}_{-14}$ | $7.8^{+4.7}_{-2.1}$ |
| $\xi_к(T)$ | $3^{+11}_{-0.9}$ | $3.6^{+0.1}_{-0.1}$ | $0.17^{+0.03}_{-0.07}$ | $4.5^{+5}_{-0.7}$ | $18^{+28}_{-9}$ | $2.7^{+4.2}_{-1.4}$ |
| Аппроксимация модельными функциями данных КА RHESSI и SDO/AIA ||||||| 
| а) $\xi_к^{hot}(T)$ | $0.03^{+0.01}_{-0.01}$ | $3.9^{+0.2}_{-0.1}$ | $0.97^{+0.04}_{-0.17}$ | $0.45^{+0.02}_{-0.02}$ | $10^{+1}_{-2}$ | $1.5^{+0.2}_{-0.3}$ |
| б) $\xi_к^{cold}(T)$ | $0.002^{+0.001}_{-0.001}$ | $4.5^{+4}_{-1}$ | $0.21^{+0.01}_{-0.08}$ | | | |

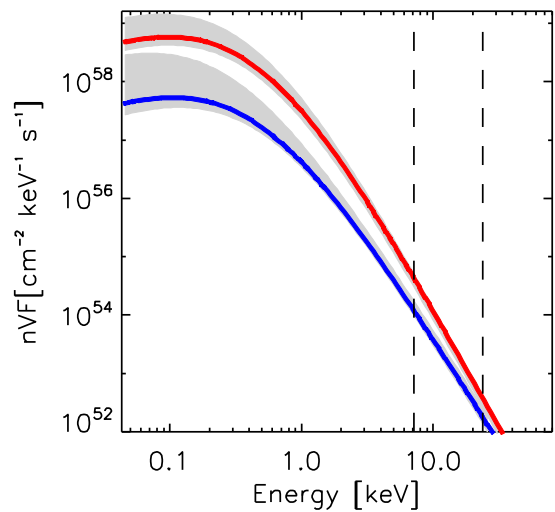

Рис. 2.13. Энергетическое распределение электронов $\langle nVF(E) \rangle$, полученное из результатов аппроксимации моделями *thin_kappa* (красная линия) и $\xi_к(T)$ (синяя линия). Серые области обозначают доверительный интервал. Вертикальные пунктирные линии указывают на аппроксимируемый диапазон энергий (7-24 кэВ).

Из рис. 2.13 видно, что энергетические распределения электронов на низких энергиях существенно отличаются друг от друга вследствие недостаточного ограничения меры эмиссии и температуры. Это указывает на необходимость улучшения ограничений низкоэнергетической компоненты, которую можно получить из данных КА SDO/AIA.

## 2.5.1 Применение одновременной аппроксимации модельными функциями с использованием дифференциальной меры эмиссии $\xi_\kappa(T)$ одновременно для RHESSI и SDO/AIA данных

Применим метод, описанный в разделе 2.3, к солнечной вспышке SOL2010-08-14T10:05 для того же момента времени, который был рассмотрен выше в разделе 2.5 (09:42:00-09:42:32 UT) и проверим, аппроксимирует ли одно много-температурное каппа-распределение энергетический диапазон, расширенный до 0.043 кэВ. В данном разделе была применена методика обработки данных, описанная в разделе 2.4. Данные КА RHESSI и аппроксимируемый диапазон энергий, составлявший 7-24 кэВ, были взяты аналогичными, как и в анализе, представленном выше в разделе 2.5.

Для КА SDO/AIA были использованы незасвеченные карты для шести КУФ каналов (94 Å, 131 Å, 171 Å, 193 Å, 335 Å), взятые в момент времени около 09:42:15 UT. SDO/AIA 131 Å и 94 Å карты отображают пространственную совместимость вспышечной петли в КУФ диапазоне с петлей в мягком рентгеновском диапазоне, изображенной 30%, 50%, 70% контурами от максимума интенсивности RHESSI карты (рис. 2.11). Таким образом, как и в разделе 2.4, предполагаем, что оба инструмента наблюдают одну и ту же излучающую плазму.

Зная из предыдущего раздела, что аппроксимация всего температурного диапазона одной ДМЭ не приводит к удовлетворительному критерию $\chi^2$, к функции $\xi_\kappa(T)$ была добавлена вторая $\xi_\kappa(T)$ и произведена аппроксимация соответственно функциями $\xi_\kappa^{hot}(T)$ и $\xi_\kappa^{cold}(T)$. На рис. 2.14 показаны результаты аппроксимации данными функциями одновременно SDO/AIA (левая панель) и RHESSI (правая панель) данных ($\chi^2$=1.6).

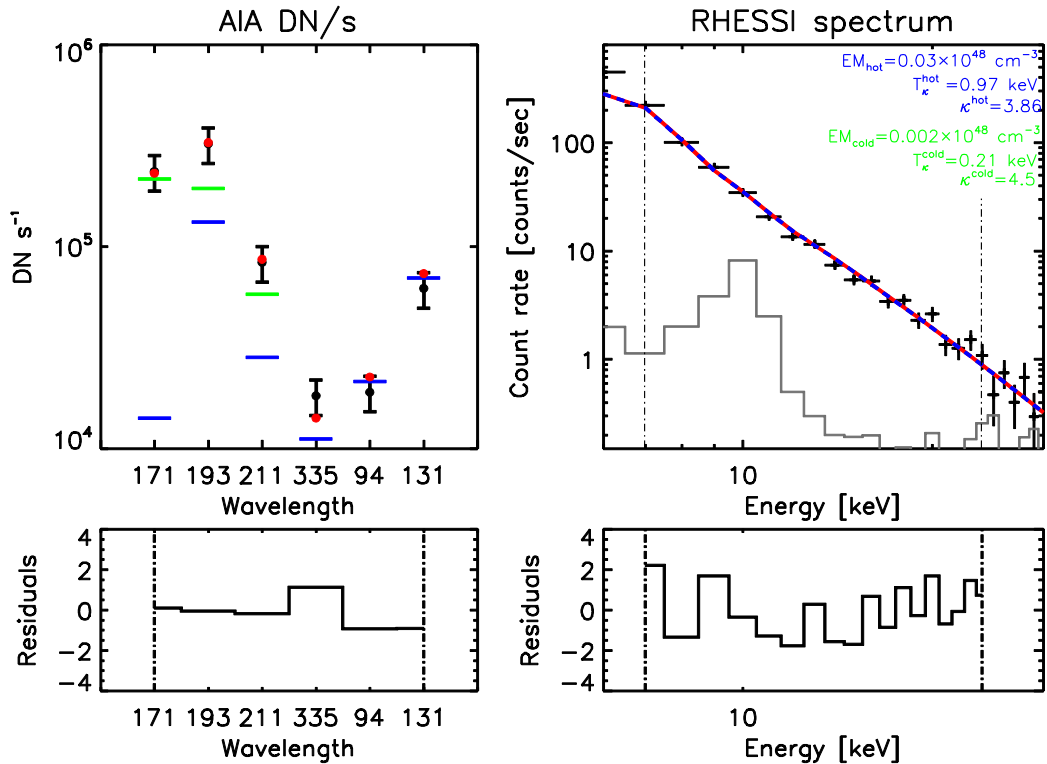

Рис. 2.14. Результаты аппроксимации модельными функциями одновременно SDO/AIA и RHESSI данных. Слева: данные КА SDO/AIA (*DN/s*) в виде зависимости от длины волны канала (черные кружки) и *DN/s*, полученные в результате аппроксимации, которая состоит из двух компонент $\xi_\kappa(T)$ ($\xi_\kappa^{hot}(T)$ - синие пунктиры, $\xi_\kappa^{cold}(T)$ - зеленые пунктиры). Итоговый результат показан красными кружками. Справа: спектр скорости счета, зарегистрированный КА RHESSI и результат аппроксимации, аналогично с левой панелью. Внизу показано отношение разности между наблюдательными данными и результатами аппроксимации к соответствующим ошибкам для КА RHESSI и SDO/AIA измерений (*Residuals*).

Отметим, что $\xi_\kappa^{cold}(T)$ влияет на результат только на очень низких температурах, поэтому эта компонента не видна в спектре КА RHESSI (правая панель, рис. 2.14), но вносит значительный вклад в излучение на длинах волн 171 Å, 193 Å и 211 Å в SDO/AIA данные (левая панель, рис. 2.14). Видно, что $\xi_\kappa^{hot}(T)$ аппроксимирует оба набора данных, в то время как холодную компоненту $\xi_\kappa^{cold}(T)$ можно обнаружить только по данным КА SDO/AIA, но не КА RHESSI. Кроме того, для этой компоненты параметр $\kappa$ слабо ограничен, и другие параметры аппроксимации (*EM*, *T*, $\chi^2$) малочувствительны к его значению. Это позволяет предположить, что для компоненты $\xi_\kappa^{cold}(T)$ преобладает тепловая фоновая корональная плазма вдоль луча зрения с температурой ~1-2 МК, аналогичное предположение было сделано в работах [31, 84] для различных событий.

На рис. 2.15 представлены ДМЭ, найденные путем аппроксимации различными моделями только RHESSI данных, одновременно SDO/AIA и RHESSI данных, и с помощью метода регуляризации только SDO/AIA данных [63]. В дополнение, на рис. 2.15 (левая панель) показаны кривые КА SDO/AIA *loci-curves* (т.е. наблюдаемые данные, разделенные на функции отклика температуры), которые указывают, для каких температур инструмент SDO/AIA наиболее чувствителен. Следует отметить, что ДМЭ, найденная только по данным КА SDO/AIA, резко убывает на краях диапазона температурной чувствительности КА SDO/AIA ($\log_{10} T = 5.7$ и $\log_{10} T = 7.5$).

Кроме того, ДМЭ, полученная путем аппроксимации одновременно RHESSI и SDO/AIA данных не имеет "провала" в районе $\log_{10} T = 6.6$, как ДМЭ, найденная методом регуляризации SDO/AIA данных. Этот "провал" является следствием реконструкции ДМЭ, так как ни один из каналов КА SDO/AIA, которые используются для расчета ДМЭ, не имеет четкого пика вблизи этих температур в функциях температурного отклика [32]. Таким образом, аппроксимация совместно SDO/AIA и RHESSI данных в целом улучшает реконструкцию ДМЭ. Вместо непрерывно возрастающей ДМЭ, полученной в результате аппроксимации модельной зависимостью только RHESSI данных, полученная путем аппроксимации одновременно SDO/AIA и RHESSI данных, ДМЭ имеет ярко выраженный максимум около $\log_{10} T = 6.8$ и пик поменьше около $\log_{10} T = 6.1$.

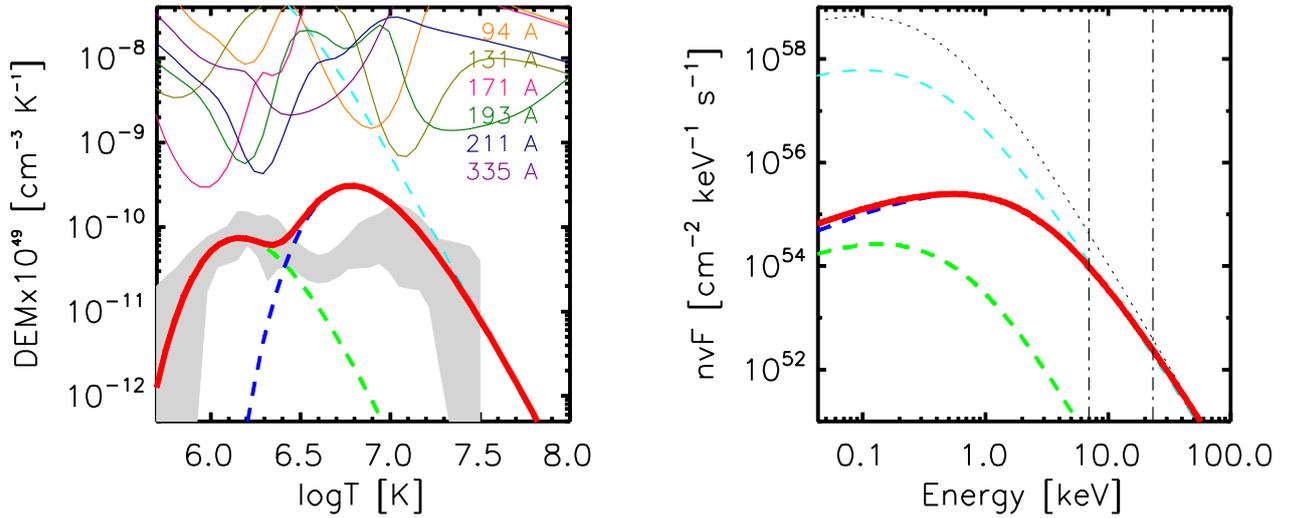

Рис. 2.15. Слева: сравнение ДМЭ, полученных разными методами: ДМЭ – путем аппроксимации одной $\xi_\kappa(T)$ RHESSI данных (пунктирная светло-голубая линия); ДМЭ – путем аппроксимации одновременно RHESSI и SDO/AIA данных двумя $\xi_\kappa(T)$ ($\xi_\kappa^{hot}(T)$ - пунктирная синяя линия, $\xi_\kappa^{cold}(T)$ - пунктирная зеленая линия, аналогично рис. 2.14). Итоговая суммарная ДМЭ показана красной линией, остальные линии – данные КА SDO/AIA, поделенные на функции температурного отклика КА SDO/AIA (*loci-curves*). Серой областью показана ДМЭ с доверительным интервалом, найденная с помощью метода регуляризации только из данных КА SDO/AIA. Справа: энергетическое распределение электронов: $\langle nVF(E) \rangle$, полученное в результате аппроксимации одновременно SDO/AIA и RHESSI данных (красная линия) двумя компонентами (зеленая и синяя линии); $\langle nVF(E) \rangle$, полученное в результате аппроксимации RHESSI данных моделью тонкой мишени *thin_kappa* (черная пунктирная линия), одной $\xi_\kappa(T)$ функцией (пунктирная светло-голубая линия). Две вертикальные пунктир-точка линии соответствуют энергетическому диапазону, для которого производилась аппроксимация RHESSI данных.

Результирующие спектры электронов $\langle nVF(E) \rangle$ показаны на правой панели рис. 2.15. Видно, что при добавлении SDO/AIA данных, т.е. путем введения дополнительных ограничений на низких температурах, распределение электронов в области низких энергий (до ~1 кэВ) уменьшается примерно на порядок по сравнению с $\langle nVF(E) \rangle$, найденное только из RHESSI данных. Полученный результат в дальнейшем повлияет на оценку полной энергии вспышки, в которой преобладает спектр в области около 1 кэВ (см. раздел 2.5.2).

### 2.5.2 Обсуждение результатов

Рассмотрен новый подход к каппа-распределению посредством дифференциальной меры эмиссии, и представлен метод, позволяющий аппроксимировать новой функцией как RHESSI данные отдельно, так и одновременно с данными КА SDO/AIA. В принципе, любая функция, для которой выражение $\xi(T(t))/t^{1/2}$ (см. раздел 2.2) имеет аналитический вид преобразования Лапласа, может быть использована для аппроксимации модельной зависимостью с помощью этого метода [100], но каппа-распределение имеет четкую физическую интерпретацию. В то время как спектр КА RHESSI может быть хорошо аппроксимирован одним каппа-распределением, присоединение данных КА SDO/AIA требует добавления второй, низкотемпературной составляющей. Весь спектр анализируемой вспышки согласуется с моделью суммы двух каппа-распределений: низкотемпературной (~2 МК) "ядро" и высокотемпературной (~11 МК) компонент. Это напоминает спектры электронов, наблюдаемые *in situ* в солнечном ветре (напр., [91, 93, 94]), которые часто могут быть аппроксимированы более холодным "ядром" и каппа-распределением. Вместе с тем, следует отметить, что низкотемпературная компонента КА SDO/AIA ДМЭ (вносящая преобладающий вклад в излучение в 171 Å и 211 Å), вероятно, относится к холодной фоновой корональной плазме вдоль луча зрения, как было отмечено в работах [31, 84] на основе временной эволюции ДМЭ до и во время вспышек. В то время как данная компонента влияет на интерпретацию спектра (ядро-плюс-каппа-распределение) ускоренных во вспышке электронов, она не оказывает воздействия на полную энергию, поскольку ее вклад в полную электронную концентрацию и плотность энергии минимален. Расчеты показывают, что ранее использованная аппроксимация моделью традиционной тонкой мишени только RHESSI спектров переоценивает количество электронов, необходимое для объяснения рентгеновского и КУФ излучения более чем на порядок (см. рис. 2.15, правая панель).

### 2.5.3 Полная электронная концентрация и энергия вспышки

Электронная концентрация $n$ может быть найдена из наблюдаемой меры эмиссии $EM$ и объема излучающей области $V$ через $n=\sqrt{EM/V}$, где $V=A^{3/2}=1.5\times10^{27}$ см$^3$, а площадь $A$ взята для 50% контура RHESSI CLEAN карты для диапазона энергий 6–12 кэВ. Для события 14.08.2010 полная электронная концентрация, выведенная из данных КА RHESSI, $n=1.4\times10^{11}$ см$^{-3}$ для случая, когда спектр КА RHESSI аппроксимирован моделью *thin_kappa*, и $n=4.5\times10^{10}$ см$^{-3}$, когда моделью $\xi_\kappa(T)$. Сравнение электронных концентраций, найденных двумя разными способами, показало, что модель $\xi_\kappa(T)$ дает значение в 3 раза меньшее. При добавлении SDO/AIA данных для большего ограничения самых низких энергий $n=4.5\times10^{9}$ см$^{-3}$ для случая, где учитывались обе компоненты $\xi_\kappa(T)$. Данный результат меньше $n$, полученной с помощью *thin_kappa*, примерно в 30 раз (см. также Таблицу 2.1). Верхний и нижний пределы рассчитаны с использованием верхнего и нижнего пределов параметров аппроксимации.

В дополнение к электронной концентрации, также была рассчитана полная плотность энергии $U_\kappa$ [эрг см$^{-3}$]. Температура $T_\kappa$ может быть определена через характеристическую скорость $\theta^2 = (2k_B T_\kappa/m_e)(\kappa-1.5)/\kappa$ (см. раздел 2.2.2). Следует отметить, что в данной интерпретации каппа-распределения как суммы распределений Максвелла температура $T_\kappa$ имеет значение средней температуры $\xi_\kappa(T)$. Исходя из этого определения, полная плотность энергии $U_\kappa$ и средняя энергия $\langle E \rangle$ могут быть выражены через $U_\kappa = \frac{3}{2} n k_B T_\kappa$ и $\langle E \rangle = \frac{3}{2} k_B T_\kappa$ соответственно, как в случае распределения Максвелла. Это может быть показано прямым интегрированием энергетического распределения электронов

$$\langle nVF(E) \rangle = C \frac{\dfrac{E}{k_B T_\kappa (\kappa - 1.5)}}{\left(1 + \dfrac{E}{k_B T_\kappa (\kappa - 1.5)}\right)^{(\kappa+1)}} \qquad (2.5.1),$$

где $C = \dfrac{n^2 V 2^{3/2} \Gamma(\kappa+1)}{(\pi m_e)^{1/2} (k_B T_\kappa)^{1/2} (\kappa - 1.5)^{3/2} \Gamma(\kappa - 0.5)}$ (2.5.2).

Тогда

$$U_\kappa = \frac{1}{nV} \int_0^\infty \langle nVF(E) \rangle \frac{E}{v} dE \qquad (2.5.3),$$

где скорость $v = \sqrt{2E/m_e}$. Подставляя выражение (2.5.1) в (2.5.3)

$$U_\kappa = \frac{C}{nV} \int_0^\infty \frac{\dfrac{E}{k_B T_\kappa (\kappa - 1.5)}}{\left(1 + \dfrac{E}{k_B T_\kappa (\kappa - 1.5)}\right)^{(\kappa+1)}} \frac{E}{\sqrt{2E/m_e}} dE \qquad (2.5.4).$$

Используя замену переменных $x = E/(k_B T_\kappa (\kappa - 1.5))$ и решая полученный интеграл, применяя бета-функцию $\mathrm{B}(x, y)$

$$\int_0^\infty \frac{x^{3/2}}{(1+x)^{\kappa+1}} dx = \mathrm{B}\left(\frac{5}{2}, \kappa - \frac{3}{2}\right) = \frac{\Gamma(5/2)\Gamma(\kappa - 3/2)}{\Gamma(\kappa + 1)} \qquad (2.5.5),$$

что приводит к

$$U_\kappa = \frac{C}{nV} \sqrt{\frac{m_e}{2}} \left(k_B T_\kappa (\kappa - 1.5)\right)^{3/2} \frac{\Gamma(5/2)\Gamma(\kappa - 3/2)}{\Gamma(\kappa + 1)} \qquad (2.5.6).$$

Подставляя выражение (2.5.2) в (2.5.6) и используя свойство гамма-функции $\Gamma(\kappa - 1/2) = \Gamma(\kappa - 3/2 + 1) = (\kappa - 3/2)\Gamma(\kappa - 3/2)$, находим

$$U_\kappa = \frac{2 n k_B T_\kappa \Gamma(5/2)}{\pi^{1/2}} = \frac{3}{2} n k_B T_\kappa \qquad (2.5.7).$$

Таким образом, полная плотность энергии $U_к$ может быть вычислена с помощью выражения (2.5.7), а полная энергия $E_{tot}$ – путем умножения $U_к$ на объем $V$ (см. Таблицу 2.1). Аппроксимация RHESSI спектров моделью $\xi_к(T)$ уменьшает полную энергию примерно в 2.9 раза по сравнению с *thin_kappa*. Аппроксимация одновременно RHESSI и SDO/AIA данных двумя $\xi_к(T)$ дает значение в ~5 раз меньшее, чем *thin_kappa*. Это говорит о том, что, в то время как аппроксимация RHESSI спектров моделью $\xi_к(T)$ уже приводит к значительному уменьшению полного числа электронов, самые низкие энергии электронов могут быть восстановлены только используя аппроксимацию одновременно RHESSI и SDO/AIA данных (см. рис. 2.15).

## 2.6 Заключение к главе 2

Сформулируем основные результаты второй главы диссертации:

1. Разработаны две модели нахождения ДМЭ, первая из которых является функциональной зависимостью и представляет собой гамма-распределение, а вторая модель – каппа-распределение, которые аппроксимируют как SDO/AIA, так и RHESSI данные. Кроме того, обе модели автоматически дают распределение электронов и ключевые параметры плазмы: меру эмиссии и температуру.

2. Впервые, с помощью комбинированного анализа SDO/AIA и RHESSI данных были реконструированы ДМЭ и энергетическое распределение электронов $\langle nVF(E) \rangle$ на примере вспышечного события 14.08.2010 в энергетическом диапазоне от ~0.1 до 20-24 кэВ. Показано, что отклонение $\langle nVF(E) \rangle$ от распределения Максвелла присутствует не только на высоких, но и на низких энергиях, то есть распределение частиц имеет более сложную структуру.

3. Для вспышки 14 августа 2010 г. показано, что $\xi_Г(T)$, полученная из RHESSI и SDO/AIA наблюдений с добавлением нетепловой

компоненты, находится в хорошем соответствии с регуляризированной ДМЭ, полученной из данных КА SDO/AIA, а также ДМЭ, следующей из данных КА RHESSI. На примере события 8 марта 2015г. рассмотрена временная эволюция полученных с помощью представленного метода параметров: температуры, меры эмиссии и концентрации.

4. Показано на примере события 14 августа 2010 г., что RHESSI данные отдельно могут быть аппроксимированы с помощью одного каппа-распределения, однако добавление SDO/AIA данных обуславливает необходимость добавления второго, соответствующего низкотемпературной компоненте, которое отвечает за холодную, не относящуюся к вспышке, корональную плазму вдоль луча зрения, вклад в излучение которой доминирует на некоторых длинах волн для SDO/AIA данных. Это говорит о том, что одно много-температурное каппа-распределение аппроксимирует относительно широкий диапазон энергий электронов, кроме диапазона энергий около ~0.1 кэВ, что свидетельствует о дополнительном вкладе корональной 1-2 МК плазмы.

5. Аппроксимация моделью тонкой мишени с каппа-распределением RHESSI данных приводит к переоценке полного числа высокоэнергичных электронов, необходимых для генерации наблюдаемого РИ примерно в 3 раза по сравнению с результатами аппроксимации моделью $\xi_\kappa(T)$.

6. Для того, чтобы должным образом ограничить область низких энергий популяции электронов, необходима аппроксимация модельной зависимостью одновременно RHESSI и SDO/AIA данных, что приводит в целом к уменьшению полного числа электронов примерно в 30 раз. Этот вывод имеет принципиально важное значение для оценки как полного числа электронов, так и полной энергии, которая выделяется во время вспышки.

# Глава 3
# Нагрев вспышечных корональных петель и жесткое рентгеновское излучение солнечных вспышек

## 3.1 Введение к главе 3

Как было сказано ранее, в последнее время широкое распространение получила CSHKP модель или «стандартная» модель солнечной вспышки, которая описывает процессы ускорения заряженных частиц и нагрева окружающей плазмы. Поскольку укоренные электроны должны оказывать определяющее влияние на нагрев горячей рентгеновской плазмы, то одним из основных критериев применимости «стандартной» модели является эффект Нойперта [101] (см. раздел 2.4.2). Сравнительно простым критерием является также задержка наступления максимума меры эмиссии относительно максимума температуры [26]. Оценки и результаты численного моделирования показывают, что для некоторых событий описанный сценарий хорошо согласуется с наблюдениями [136, 117]. Вместе с тем часто наблюдаются существенные отклонения от соотношения Нойперта [135]. При этом иногда максимумы температуры горячей корональной плазмы вплоть до нескольких минут опережают пики жесткого РИ [126], что свидетельствует о важной роли тепловых механизмов энерговыделения.

Заметим, что механизм генерации жесткого РИ в хромосфере (столкновительная модель толстой мишени, см. раздел 1.1.2) иногда является малоэффективным, поскольку скорость кулоновских потерь энергии ускоренными электронами в $10^5$ раз превышает мощность тормозного излучения. При этом практически все тепловые электроны, содержащиеся в корональной части вспышечной петли, должны быть ускорены [97], что представляется маловероятным. Браун и др. [39] предложили модель дополнительного ускорения электронов в хромосфере Солнца (*local re-acceleration thick target model*), которая позволяет повысить эффективность генерации жесткого РИ, поскольку один и тот же излучающий электрон может ускоряться несколько раз в области излучения (*re-acceleration*). Кроме того, эта модель позволяет избежать трудностей, связанных с ускорением и распространением большого числа заряженных частиц в разреженной короне. Однако до сих пор каких-либо веских аргументов в пользу ее применимости получено не было.

**Целью главы 3** является анализ теплового баланса горячих вспышечных петель на основе размерностных соотношений и особенностей РИ в рамках «стандартной» модели солнечной вспышки.

Данная глава направлена на решение следующих задач:

(а) Оценить степень эффективности нагрева ускоренными электронами корональной вспышечной плазмы и порогового значения энергии, при которой электроны могут термализоваться в короне.

(б) С помощью полученных оценок из пункта (а) проанализировать зависимость отношения потоков жесткого РИ в корональной части и основаниях петли от относительной энергии фотонов при различных значениях показателя спектра $\delta$.

(в) Рассмотреть два вспышечных события и на основе данных РИ сделать оценку эффективности нагрева корональной плазмы ускоренными электронами из различных частей петли.

Для решения поставленной задачи были рассмотрены солнечные вспышки 23.08.2005 г. и 09.11.2013 г. Выбор данных событий обусловлен тем, что они наблюдались на современных инструментах, а именно на КА RHESSI и SDO/AIA, и имели ярко выраженные источники РИ. Кроме того, авторы работ [71, 118, 119] определили ряд важных параметров петли, которые были учтены при получении соответствующих оценок. Результаты главы 3 опубликованы в работе [131].

## 3.2 Нагрев плазмы и рентгеновское излучение вспышечных петель

Нагрев корональной вспышечной плазмы ускоренными электронами может происходить как вследствие «испарения» горячего хромосферного вещества, так и в результате термализации ускоренных электронов непосредственно в короне. Чтобы оценить степень эффективности нетеплового нагрева, рассмотрим скорость энергетических потерь петли, которая, в первую очередь, определяется электронной теплопроводностью и зависит от температуры.

Мощность теплопроводных потерь столкновительной горячей корональной плазмы можно представить следующим образом [22]

$$q = \kappa \frac{d}{ds}\left(T^{5/2}\frac{dT}{ds}\right) = \kappa \frac{2}{7}\frac{d^2 T^{7/2}}{ds^2} = \kappa \frac{8}{7}\frac{T^{7/2}}{L^2} \qquad (3.2.1),$$

где $T$ – температура плазмы, $L$ – характерный размер области охлаждения (изменения температуры), которую мы считаем равной длине корональной петли, $\kappa = 10^{-6}$ эрг см$^{-1}$ с$^{-1}$ K$^{-1}$ – коэффициент электронной теплопроводности Спитцера [122]. Откуда, полагая тепловую энергию $E_{th} = 3nk_B T$, где $n$ – концентрация электронов и $k_B$ – постоянная Больцмана, характерное время теплопроводных потерь оценим следующим образом

$$\tau_{cond} = \frac{E_{th}}{q} = \frac{21 n k_B T}{8\kappa T^{7/2}/L^2} \qquad (3.2.2).$$

Результаты численных расчетов $\tau_{cond}$ при различных значениях $n$ представлены на рис. 3.1 (левая панель). Из рис. 3.1 (левая панель) видно, что величина $\tau_{cond}$ меняется в широких пределах и в случае достаточно горячей плазмы превосходит несколько секунд. Отметим, что радиационными потерями при температуре горячей плазмы $T \geq 10^7$ K можно пренебречь [105, 45].

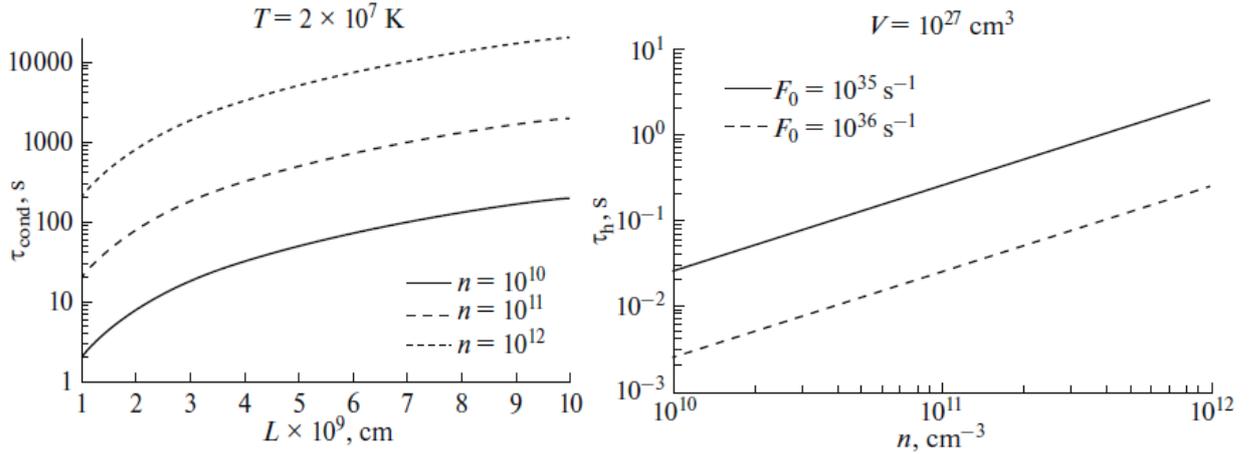

Рис. 3.1. Слева: зависимость характерного времени потерь тепловой энергии $\tau_{cond}$ от длины корональной петли $L$ при различных значениях концентрации электронов $n$. Справа: характерное время нагрева плазмы $\tau_h$ объемом $V$ интегральным потоком ускоренных электронов $F_0$.

Максимальную мощность энерговыделения пучка электронов $P$ [эрг /с], выделяемую в объеме $V$ из-за столкновений с частицами фоновой плазмы можно оценить с помощью соотношения [137]:

$$P = \frac{\delta - 1}{\delta - 2} F_0 E_0 \qquad (3.2.3),$$

где $\delta$, $F_0$ и $E_0$ – показатель спектра, интегральный поток и нижний предел энергии ускоренных электронов соответственно. На рис. 3.1 (правая панель) представлены результаты численного расчета характерного времени нагрева вспышечной плазмы

$$\tau_h = 3 n k_B T V / P \qquad (3.2.4),$$

где для простоты для $\delta > 4$ значение $P$ было взято как $P \approx F_0 E_0$, так как в случае $\delta \to \infty$ выражение $(\delta-1)/(\delta-2)$ стремится к 1. Из рис. 3.1 (правая панель) видно, что для принятых значений ($E_0 = 20$ кэВ, $V = 10^{27}$ см³, $T = 2 \times 10^7$ K) даже в случае достаточно плотной плазмы ($n = 10^{11}$ см⁻³) нагрев может происходить чрезвычайно быстро – за доли секунды, несмотря на значительные теплопроводные потери.

Ускоренные в вершине вспышечной петли электроны даже в корональной части должны испытывать сильное торможение. Электроны термализуются в короне, если их энергия меньше порогового значения [136, 118]

$$E_{loop} \approx 10\sqrt{N_{19}/\mu} \qquad (3.2.5),$$

где $N_{19} = 10^{-19} nL/2$, $\mu$ - косинус питч-угла ускоренного электрона. В частности, для $nL = 6\times10^{19}$ см$^{-2}$ и $\mu = 0.5$ получим $E_{loop} = 25$ кэВ. Найденная оценка свидетельствует о возможности эффективного нагрева ускоренными электронами вспышечной плазмы в ходе их распространения из вершины арки к основаниям. Это также предполагает, что для фотонов с энергией ε ≤ $E_{loop}$ основная доля потока жесткого РИ вспышки должна приходиться на корональную часть петли. Действительно, используя соотношение для потоков жесткого РИ от оснований $I_{fp}$ петли и всего источника $I_{tot}$ [136]

$$\frac{I_{fp}}{I_{tot}} = \frac{\delta-2}{2} B\left(\frac{1}{1+(\varepsilon/E_{loop})^2}, \frac{\delta-2}{2}, \frac{1}{2}\right)\left(\frac{\varepsilon}{E_{loop}}\right)^{\delta-2} \qquad (3.2.6),$$

где неполная бета-функция $B\left(y, \frac{\delta-2}{2}, \frac{1}{2}\right) = \int_0^y x^{\delta/2-2}(1-x)^{-1/2} dx$, с учетом того, что полное излучение из корональной части вспышечной петли равно $I_{lp} \approx I_{tot} - I_{fp}$, из выражения (3.2.6) находим

$$\frac{I_{lp}}{I_{fp}} = \frac{2}{\delta-2} \frac{1}{B\left(\frac{1}{1+(\varepsilon/E_{loop})^2}, \frac{\delta-2}{2}, \frac{1}{2}\right)} \left(\frac{\varepsilon}{E_{loop}}\right)^{2-\delta} - 1 \qquad (3.2.7).$$

Зависимость $I_{lp}/I_{fp}$ от $\varepsilon/E_{loop}$ для различных значений $\delta$ представлена на рис. 3.2. Из рис. 3.2 видно, что величина $I_{lp}/I_{fp}$ сильно зависит от показателя спектра $\delta$, а при $\delta \geq 5$ и ε ≤ $E_{loop}$ излучение в корональной части петли становится доминирующим.

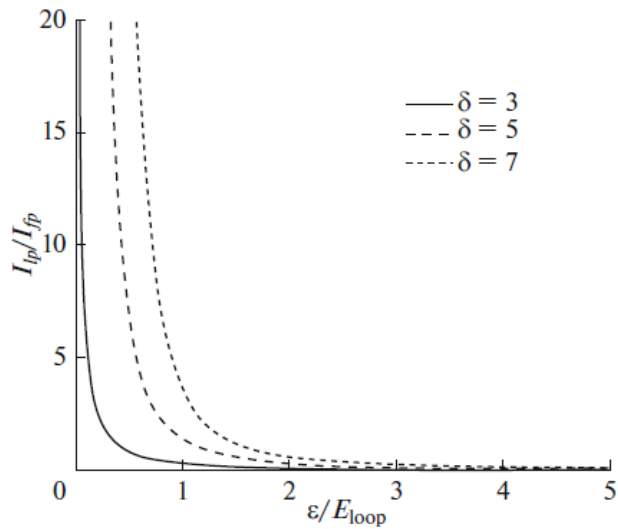

Рис. 3.2. Зависимость отношения потоков жесткого РИ в корональной части и основаниях петли $I_{lp}/I_{fp}$ от относительной энергии фотонов $\varepsilon/E_{loop}$ при различных значениях показателя спектра $\delta$.

## 3.3 Вспышечные события 23.08.2005 г. и 09.11.2013 г.

Рассмотрим динамику тепловой плазмы и нетепловых электронов на примере двух вспышечных событий 23.08.2005 г. и 09.11.2013 г., которые были хорошо изучены ранее благодаря спутниковым наблюдения в широком диапазоне длин волн.

*Событие 23.08.2005 г.*

Лимбовая вспышка 23.08.2005 г. рентгеновского класса М3.0 была подробно описана в работе [71]. Она интересна тем, что, несмотря на слабые изменения потока ускоренных электронов в основной фазе вспышки (импульсная фаза выражена крайне слабо), температура вспышечной плазмы непрерывно уменьшалась со временем (см. рис. 3.3, левая панель). Из рис. 3.3 видно, что задержка между соответствующими пиками температуры и меры эмиссии достигает 10 минут. Кроме того, РИ в основаниях петель оказалось крайне слабым, корональный источник был виден на энергиях вплоть до ~25 кэВ, тогда как основания петли с трудом можно было отождествить в диапазоне 30–40 кэВ [71].

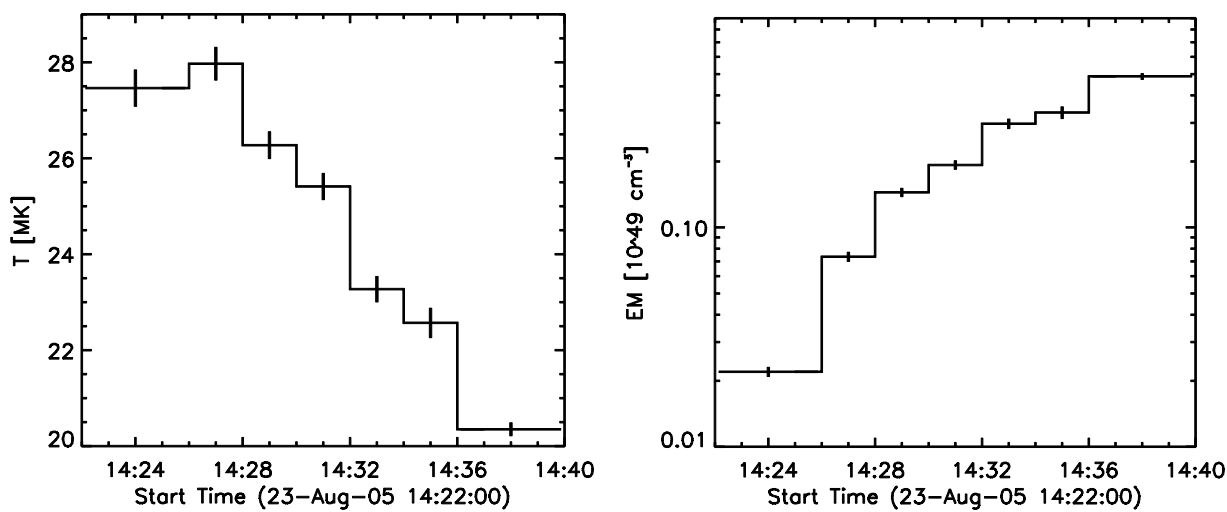

Рис. 3.3. Временные профили температуры (слева) и меры эмиссии (справа) тепловой плазмы вспышечного события 23.08.2005 г., следующие из результатов аппроксимации моделями однородной изотермической плазмы и толстой мишени рентгеновских данных, полученных на КА RHESSI.

Если исходить из результатов работы [71], а также оценок, полученных из результатов аппроксимации спектров жесткого РИ, которое проводилось с помощью пакета программ *OSPEX* в предположении изотермической тепловой модели и модели толстой мишени (см. рис. 3.4, левая панель), то, используя значения, представленные в Табл. 3.1, из соотношений (3.2.2) и (3.2.3) следует, что характерное время теплопроводных потерь $\tau_{cond}$ значительно меньше $\tau_h$ в течение длительного (>15 мин.) времени (рис. 3.4, правая панель).

Следовательно, ускоренные электроны не могут обеспечить нагрев корональной плазмы. Обращает также на себя внимание, что в области энергий 10–12 кэВ в период импульсной фазы вспышки основной вклад в РИ давала тепловая плазма (рис. 3.4, левая панель). Кроме того, несмотря на относительное уменьшение теплопроводных потерь (рис. 3.4, правая панель), температура плазмы продолжала падать (см. рис. 3.3, левая панель). Это указывает на важную роль тепловых механизмов энерговыделения, а также необходимость модификации «стандартной» вспышечной модели.

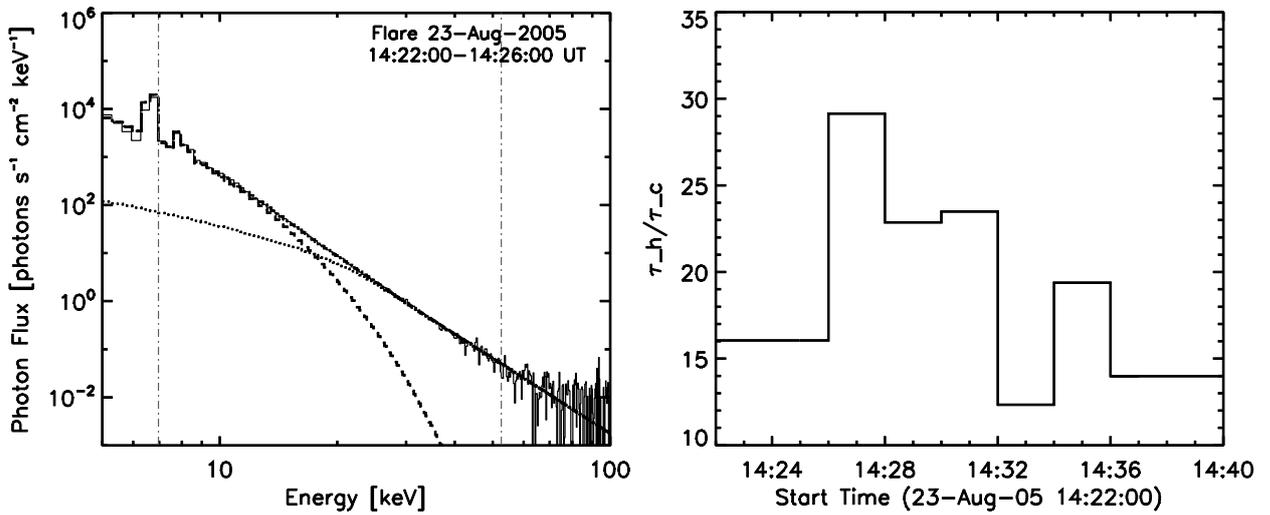

Рис. 3.4. Слева: пример аппроксимации спектра РИ (сплошная линия) вспышки 23.08.2005 г. (14:22-14:26 UT) в рамках тепловой однородной изотермической плазмы (штриховая линия) и модели толстой мишени (пунктирная линия): $EM = 2.22\times10^{47}$ см$^{-3}$, $T = 27.46$ МК, $F_0 = 0.81\times10^{35}$ с$^{-1}$, $\delta = 5.96$, $E_0 = 25.1$ кэВ. Справа: зависимость отношения характерных времен нагрева $\tau_h$ и охлаждения $\tau_{cond}$ плазмы, полученные из рентгеновских наблюдений (КА RHESSI) вспышечного события 23.08.2005 г. и размерностных соотношений (3.2.2) и (3.2.4).

Таблица 3.1. Параметры источника излучения (10–12 кэВ) и ускоренных электронов в событии 23.08.2005 г., полученные из результатов аппроксимации рентгеновских спектров и работы [71]. Оценки, взятые из работы [71], обозначены[*]

| UT | $EM$ ($10^{49}$см$^{-3}$) | $T$ (МК) | $V^*$ ($10^{27}$см$^3$) | $L^*$ ($10^7$ см) | $E_0$ (кэВ) | $F_0$ ($10^{35}$ с$^{-1}$) | $\delta$ |
|---|---|---|---|---|---|---|---|
| 14:22-14:26 | 0.022 | 27.46 | 2.4 | 398.75 | 25.1 | 0.81 | 5.96 |
| 14:26-14:28 | 0.074 | 27.97 | 1.4 | 311.75 | 29.9 | 0.39 | 6.46 |
| 14:28-14:30 | 0.145 | 26.27 | 1.35 | 275.5 | 27.52 | 0.54 | 6.84 |
| 14:30-14:32 | 0.193 | 25.41 | 1.3 | 275.5 | 28.36 | 0.44 | 7.01 |
| 14:32-14:34 | 0.297 | 23.27 | 1.7 | 261.0 | 25.66 | 0.98 | 6.63 |
| 14:34-14:36 | 0.334 | 22.57 | 1.8 | 257.38 | 25.1 | 0.61 | 6.13 |
| 14:36-14:40 | 0.489 | 20.35 | 2.5 | 261.0 | 21.55 | 0.85 | 4.79 |

*Событие 09.11.2013 г.*

Данное событие рентгеновского класса С2.7, произошедшее около центра диска Солнца (S11W03) в активной области AR 11890, одновременно наблюдалось на нескольких космических инструментах [118, 119], включая RHESSI. Оно характеризовалось постепенным ростом потоков КУФ, мягкого и жесткого РИ, начавшимся в 06:24 UT, и ярко выраженным пиком жесткого РИ в 06:25:46 UT. На картах (см. рис. 3.5) во всех длинах волн четко выделяются три источника, один из которых был расположен в короне [118]. Этот источник отличало мощное жесткое РИ, доминировавшее в отдельные моменты времени.

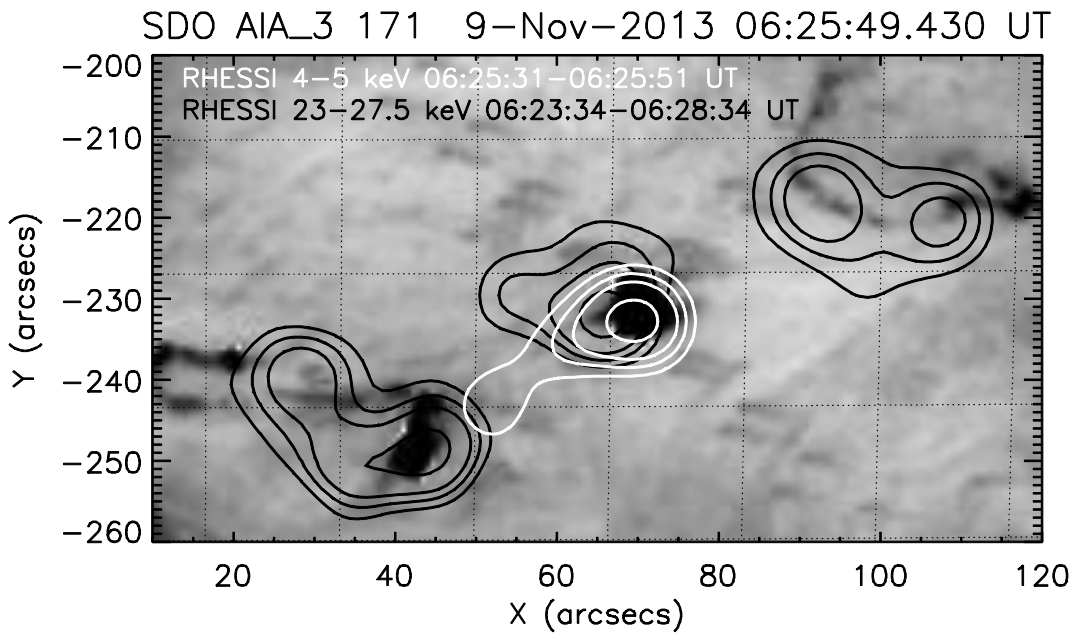

Рис. 3.5. КУФ карта вспышки 09.11.2013 г. на длине волны 171 Å (КА SDO/AIA), совмещенное с рентгеновскими контурами (50, 60, 70, 90% от пика интенсивности), полученными на КА RHESSI в диапазоне энергий 4–5 кэВ (белые контуры) и 23–27.5 кэВ (черные контуры) около максимума импульсной фазы (06:25:49 UT) (см. также [118]). Для создания изображений КА RHESSI был применен алгоритм CLEAN [69].

Симоес и др. [118] оценили электронную концентрацию $n = 3 \times 10^{11}$ см$^{-3}$ для коронального источника и характерный продольный размер коронального источника 9–18 угл. сек. в момент пика жесткого РИ (~06:25:41 UT). Высокая плотность предполагает, что ускоренные в вершине петли электроны должны быстро терять свою энергию, эффективно генерируя жесткое РИ. Положив $nL/2 = (1-2) \times 10^{20}$ см$^{-2}$, при $\mu = 0.5$ находим $E_{loop} \approx 10\sqrt{N_{19}/\mu} = 45\text{-}63$ кэВ. Откуда, считая показатель спектра ускоренных электронов $\delta = 5.6$ [119] и $\varepsilon/E_{loop}= 0.5$, из (3.2.7) следует $I_{lp}/I_{fp} \approx 11$ (см. рис. 3.2). Найденное значение противоречит изображениям источников жесткого РИ в канале 23–27 кэВ, полученных на КА RHESSI в указанный момент времени (рис. 3.5), и указывает на необходимость модификации «стандартной» модели. Учет тормозящих ускоренные электроны электрических полей обратных токов в рассматриваемом событии едва ли существенным образом скажется на полученных результатах, поскольку из-за условия квазинейтральности плазмы они должны быть сравнимы между собой. К тому же, предложенная модель не учитывает действие диамагнитных сил, вызывающих отражение ускоренных электронов от «магнитных пробок» в области оснований [130].

### 3.4 Заключение к главе 3

Таким образом, основными результатами третьей главы диссертации являются:

1. Для события 23.08.2005 г. было показано, что во вспышечной петле теплопроводные потери в течение нескольких минут превышали мощность энергии, содержащейся в ускоренных частицах, хотя температура корональной плазмы увеличивалась. Это указывает на существование дополнительного источника тепловой энергии, связанного, например, с джоулевой диссипацией электрических токов. В рассматриваемой вспышечной петле с концентрацией тепловых электронов $n \approx 3\times10^{10}$ см$^{-3}$ [71] и температурой плазмы $T \approx 2\times10^7$ K (Табл. 3.1), длина свободного пробега электрона $\lambda=5.21\times10^3 T^2/n \approx 10^8$ см. Полученное значение $\lambda$ гораздо меньше характерного масштаба изменения температуры $L \approx 10^9$ см (Табл. 3.1). Поэтому возможность подавления электронной теплопроводности из-за отклонений от максвелловского распределения плазмы или возбуждения плазменной турбулентности [104] во внимание не принималась.

2. Для события 09.11.2013 г. исходя из сценария, следующего из «стандартной» модели, согласно которому ускорение частиц происходит в области вершины петли, было показано, что из-за больших кулоновских потерь энергии в короне электроны не могут генерировать наблюдаемое РИ в основаниях петель. Вероятно, полученный результат является следствием действия "ускорительной машины" не только в вершине, но и в основаниях корональной петли. Данный вывод согласуется с идеей работы [39] о дополнительном хромосферном ускорении, а также моделями солнечной вспышки [62, 140]. В первом случае вспышечное энерговыделение связывается со слиянием магнитных жгутов в частично ионизованной плазме, а во втором, – с развитием желобковой неустойчивости в хромосфере Солнца. Отметим, что гипотеза о распределенном источнике ускорения в корональной части петли (см., напр., [139]) выглядит маловероятной, поскольку в этом случае вместе с увеличением потока ускоренных электронов в основаниях петли должна возрастать эффективность генерации жесткого РИ в корональной части арки.

# Заключение

Диссертационная работа основана на анализе данных в крайнем ультрафиолетовом, мягком и жестком рентгеновском диапазонах, полученных на современных космических инструментах, и посвящена восстановлению энергетических распределений электронов в солнечных вспышках, а также анализу теплового баланса вспышечных петель и интенсивности РИ из различных областей вспышки.

Основные результаты диссертации заключаются в следующем:

1) Произведена реконструкция спектров жесткого РИ на основе данных регистрации спектрометра ИРИС, установленного на борту КА КОРОНАС-Ф, с учетом приборной функции методом случайного поиска. На основе восстановленных спектров жесткого РИ сделана реконструкция энергетических спектров излучающих электронов с помощью метода регуляризации Тихонова и рассмотрена их динамика для солнечной вспышки 15 апреля 2002 г. Особенность в спектрах электронов, связанная с наличием локального минимума в диапазоне энергий 40−60 кэВ, не может быть выявлена прямыми методами.

2) Разработана методика восстановления энергетических распределений электронов в солнечных вспышках одновременно по данным КА SDO/AIA и RHESSI, соответствующим КУФ и РИ, с помощью аппроксимации модельными функциями ДМЭ. С помощью данной методики рассмотрены события 14 августа 2010 г. и 8 мая 2015 г., для которых построены ДМЭ и энергетическое распределение электронов. Для события 8 мая 2015 г. рассмотрена динамика температуры, меры эмиссии и концентрации для десяти временных интервалов.

3) Предложена функциональная форма ДМЭ, для которой найдено аналитическое выражение энергетического распределения электронов в солнечной вспышке. С помощью предложенной ДМЭ и разработанной методики (пункт 2) автоматически можно получать основные параметры вспышечной плазмы (температуру и меру эмиссии). Данный вид ДМЭ рассмотрен на примере вспышечных событий 14 августа 2010 г. и 8 мая 2015г., получены параметры вспышечной плазмы.

4) Рассмотрено каппа-распределение электронов во вспышечной области в виде ДМЭ с использованием метода из пункта 2. Данный подход помог сделать оценку полного количества электронов, которое необходимо для генерации наблюдаемого рентгеновского и крайнего ультрафиолетового излучения.

5) Произведен анализ теплового баланса и сравнение потоков жесткого РИ из различных частей вспышечной петли в рамках «стандартной» модели солнечной вспышки. Для события 23 августа 2005 г. показано, что теплопроводные потери в течение нескольких минут превышали мощность энергии ускоренных частиц при возрастании температуры корональной плазмы, что указывает на существование дополнительного источника тепловой энергии, связанного, например, с джоулевой диссипацией электрических токов. Для события 9 ноября 2013 г. в рамках «стандартной» модели, согласно которой ускорение частиц происходит в области вершины петли, показано, что из-за больших кулоновских потерь энергии в короне электроны не могут генерировать наблюдаемое в основаниях РИ, что указывает на необходимость модификации данной модели.

# *Литература*

[1]  Алтынцев А.Т., Банин В.Г., Куклин Г.В. Солнечные вспышки. – М.: Наука, 1982. – 248 с.

[2]  Богачев С.А., Сомов Б.В. Сравнение эффективности ускорения Ферми и бетатронного ускорения в коллапсирующих магнитных ловушках // Письма в Астрономический журнал. – 2005. – Т.31. – №8. – С.601-610.

[3]  Верлань А.Ф., Сизиков В.С. Методы решения интегральных уравнений с программами для ЭВМ. – Киев: Наукова думка, 1978. – 292 с.

[4]  Верлань А.Ф., Сизиков В.С. Интегральные уравнения: методы, алгоритмы, программы. – Киев: Наукова думка, 1986. – 544 с.

[5]  Гинзбург В.Л., Сыроватский С.И. Происхождение космических лучей. – М.: АН СССР, 1963. – 147 с.

[6]  Грим Г. Спектроскопия плазмы. – М.: Атомиздат, 1969. – 452 с.

[7]  Дмитриев П.Б., Кудрявцев И.В., Лазутков В.П. и др. Особенности рентгеновского излучения солнечных вспышек, зарегистрированных спектрометром ИРИС во время полета ИСЗ КОРОНАС-Ф // Астрономический вестник. – 2006. – Т.40. – №2. – С.160-170.

[8]  Кельнер С.Р., Скрынников Ю.И. Поляризация и направленность жесткого рентгеновского тормозного излучения в солнечных вспышках // Астрономический журнал. – 1985. – Т.62. – №4. – С.760-767.

[9]  Корчак А.А. О модельных представлениях источника рентгеновского излучения вспышек // Астрономический журнал. – 1976. – Т.53. – №2. – С.370-376.

[10] Кузнецов В.Д. Солнечно-земная физика: Результаты экспериментов на спутнике КОРОНАС-Ф. – М.: Физматлит, 2009. – 488 с.

[11] Москаленко Е.И. Методы Внеатмосферной астрономии. – М.: Наука, 1984. – 280 с.

[12] Моторина Г.Г., Кудрявцев И.В., Лазутков В.П. и др. Восстановление энергетического распределения электронов, ускоренных во время солнечной вспышки 26 июля 2002 года, по данным жесткого рентгеновского излучения // Сборник трудов XV ежегодной конференции по физике Солнца «Солнечная и солнечно-земная физика-2011». – 2011. – ГАО РАН, Санкт-Петербург. – С.171-174.

[13] Моторина Г.Г., Кудрявцев И.В., Лазутков В.П. и др. К вопросу о реконструкции энергетического распределения электронов, ускоренных во время солнечных вспышек // Журнал технической физики. – 2012. – Т.82. – №12. – С.11-15.

[14] Моторина Г.Г., Кудрявцев И.В., Лазутков В.П. и др. Реконструкция энергетического спектра электронов, ускоренных в солнечной вспышке 15 апреля 2002 года // Сборник трудов XVI ежегодной конференции по физике Солнца «Солнечная и солнечно-земная физика-2012». – 2012. – ГАО РАН, Санкт-Петербург. – С.301-304.

[15] Моторина Г.Г., Кудрявцев И.В., Лазутков В.П. и др. Эволюция энергетических спектров жесткого рентгеновского излучения солнечной вспышки 15 апреля 2002 года // Сборник трудов XVII ежегодной конференции по физике Солнца «Солнечная и солнечно-земная физика-2013». – 2013. – ГАО РАН, Санкт-Петербург. – С.161-164.

[16] Моторина Г.Г., Контарь Э.П. Дифференциальная мера эмиссии, полученная в результате комбинирования RHESSI, SDO/AIA наблюдений // Сборник трудов XVIII ежегодной конференции по физике Солнца «Солнечная и солнечно-земная физика-2014». – ГАО РАН, Санкт-Петербург. – С.307-310.

[17] Моторина Г.Г., Контарь Э.П. Временная эволюция энергетического распределения электронов в солнечных вспышках на основе RHESSI и SDO/AIA наблюдений // Сборник трудов XIX ежегодной конференции по физике Солнца «Солнечная и солнечно-земная физика-2015». – 2015. – ГАО РАН, Санкт-Петербург. – С.289-292.

[18] Моторина Г.Г., Кудрявцев И.В., Лазутков В.П. и др. Реконструкция энергетического спектра электронов, ускоренных во время солнечной вспышки 15 апреля 2002 года, на основе измерений рентгеновским спектрометром ИРИС // Журнал технической физики. – 2016. – Т.86. – №4. – С.47-52.

[19] Нахатова Г.Г., Кудрявцев И.В. К вопросу о реконструкции энергетических спектров ускоренных во время солнечных вспышек электронов, на основе данных по тормозному рентгеновскому излучению // Сборник трудов XIV ежегодной конференции по физике Солнца «Солнечная и солнечно-земная физика-2010». – 2010. – ГАО РАН, Санкт-Петербург. – С.287-290.

[20] Нахатова Г.Г., Кудрявцев И.В. К вопросу о реконструкции энергетических спектров ускоренных во время солнечных вспышек электронов // Труды XII конференции молодых ученых «Взаимодействие полей и излучения с веществом». – 2011. – Иркутск. – С.90-92.

[21] Постнов К.А., Засов А.В. Курс общей астрофизики. – М.: Физический факультет МГУ, 2005. –192 с.

[22] Прист Э.Р. Солнечная магнитная гидродинамика. – М.: Мир, 1985. – 589 с.

[23] Сомов Б.В., Сыроватский С.И. Физические процессы в атмосфере Солнца, вызываемые вспышками // УФН. – 1976. – Т.120. – С.217–257.

[24] Сыроватский С.И., Шмелева О.П. Нагрев плазмы быстрыми электронами и нетепловое рентгеновское излучение при солнечных вспышках // Астрон. журн. – 1972. – Т.49. – С.334-347.

[25] Тихонов А.Н., Арсенин В.Я. Методы решения некорректных задач. – М.: Наука, 1979. – 286 с.

[26] Цап Ю.Т., Гольдварг Т.Б., Копылова Ю.Г., Степанов А.В. О природе пульсаций нетеплового излучения солнечной вспышки 5 ноября 1992 года // Геомагнетизм и Аэрономия. – 2013. – Т.53. – №7. – С.827–830.

[27] Цыпкин Я.З. Адаптация и обучение в автоматических системах. – М.: Наука, 1968. – 400 с.

[28] Aschwanden M.J., Alexander D. Flare plasma cooling from 30 MK down to 1 MK modeled from Yohkoh, GOES, and TRACE observations during the Bastille day event (14 July 2000) // Solar Phys. – 2001. – V.204. – №1/2. – P.91-120.

[29] Aschwanden M.J. Particle acceleration and kinematics in solar flares – a synthesis of recent observations and theoretical concepts (Invited Review) // Space Science Reviews. – 2002. – V.101. – №1. – P.1–227.

[30] Aschwanden M.J. Physics of the solar corona. An introduction with problems and solutions (2nd edition) // Springer. – 2005. – 892 p.

[31] Battaglia M., Kontar E.P. RHESSI and SDO/AIA Observations of the chromospheric and coronal plasma parameters during a solar flare // Astrophys. J. – 2012. – V.760. – №2. – 9 p.

[32] Battaglia M., Kontar E.P. Electron distribution functions in solar flares from combined X-ray and Extreme-ultraviolet observations // Astrophys. J. – 2013. – V.779. – №2. – 9 p.

[33] Battaglia M., Motorina G., Kontar E.P. Multi-thermal representation of the kappa-distribution of solar flare electrons and application to simultaneous X-ray and EUV observations // Astrophys. J. – 2015. – V.815. – №1. – 8 p.

[34] Bian N.H., Emslie A.G., Stackhouse D.J., Kontar E.P. The formation of kappa-distribution accelerated electron populations in solar flares // Astrophys. J. – 2014. – V.796. – №2. – 11 p.

[35] Bornmann P.L. Limits to derived flare properties using estimates for the background fluxes - examples from GOES // Astrophys. J. – 1990. – V.356. – P.733-742.

[36] Brown J.C. The Deduction of energy spectra of non-thermal electrons in flares from the observed dynamic spectra of hard X-ray bursts // Solar Phys. – 1971. – V.18. – №3. – P.489-502.

[37] Brown J.C., Emslie A.G. Analytic limits on the forms of spectra possible from optically thin collisional bremsstrahlung source models // Astrophys. J. – 1988. – V.331. – P.554-564.

[38] Brown J.C., Emslie A.G., Holman G.D. et al. Evaluation of algorithms for reconstructing electron spectra from their bremsstrahlung hard X-ray spectra // Astrophys. J. – 2006. – V.643. – №1. – P.523-531.

[39] Brown J.C., Turkmani R., Kontar E.P. et al. Local re-acceleration and a modified thick target model of solar flare electrons // Astron. Astrophys. – 2009. – V.508. – №2. – P.993–1000.

[40] Brown J. C., Mallik P.C.V., Badnell N.R. Non-thermal recombination - a neglected source of flare hard X-rays and fast electron diagnostics (Corrigendum) // Astron. and Astrophys. – 2010. – V.515. – 3 p.

[41] Carmichael H. A process for flares // NASA Special Publication. – 1964. – V.50. – P.451-456.

[42] Charikov Yu.E., Dmitrijev P.B., Koudriavtsev I.V. et al. Solar flare hard X-rays measured by spectrometer "IRIS": spectral and temporal characteristics // IAU Symposium. – 2004. – V.223. – P.429- 432.

[43] Charikov Yu.E., Melnikov V.F., Kudryavtsev I.V. Intensity and polarization of the hard X-ray radiation of solar flares at the top and footpoints of a magnetic loop // Geomagnetism and Aeronomy. – 2012. – V.52. – №8. – P.1021-1031.

[44] Cody W. J. SPECFUN - a portable FORTRAN package of special functions and test drivers // ACM Transactions on Mathematical Software. – 1993. – V.19. – № 1. – 22-32.

[45] Colgan J., Abdallah Jr., Sherrill M.E., Foster M. Radiative losses of solar coronal plasmas // Astrophys. J. – 2008. – V.689. – №1. – P.585-592.

[46] Crannell C.J., Frost K.J., Saba J.L. et al. Impulsive solar X-ray bursts // Astrophys. J. – 1978. – V.223. – P.620-637.

[47] Culhane J.L. Thermal continuum radiation from coronal plasmas at soft X-ray wavelengths // Mon. Not. R. Astr. Soc. – 1969. – V.144. – P. 375-389.

[48] Culhane J.L., Acton L.W. A simplified thermal continuum functionfor the X-ray emission from coronal plasmas // Mon. Not. R. Astr. Soc. – 1970. – V.151. – P.141-147.

[49] Culhane J.L., Harra L.K., James A.M. et al. The EUV Imaging Spectrometer for Hinode // Sol. Phys. – 2007. – V.243. – №1. – P.19-61.

[50] Datlowe D.W., Lin R.P. Evidence for thin-target X-ray emission in a small solar flare on 26 February 1972 // Solar Phys. – 1973. – V.32. – №2. – P.459-468.

[51] Del Zanna G., Landini M., Mason H.E. Spectroscopic diagnostics of stellar transition regions and coronae in the XUV: AU Mic in quiescence // Astron. Astrophys. – 2002. – V.385. – P.968-985.

[52] Del Zanna G., Dere K.P., Young P.R. et al. CHIANTI - An atomic database for emission lines. Version 8 // Astron. Astrophys. – 2015. – V.582. – 12 p.

[53] Dennis B.R., Zarro D.M. The Neupert effect - What can it tell us about the impulsive and gradual phases of solar flares? // Solar Phys. – 1993. – V.146. – №1. – P.177-190.

[54] Dere K.P., Landi E., Mason H.E. et al. CHIANTI - an atomic database for emission lines // A & A Supplement series. – 1997. – V.125. – P.149-173.

[55] Donnelly R.F., Kane S.R. Impulsive extreme-ultraviolet and hard X-ray emission during solar flares // Astrophys. J. – 1978. – V.222. – P.1043-1053.

[56] Dudík J., Kašparová J., Dzifčáková E. et al. The non-Maxwellian continuum in the X-ray, UV, and radio range // Astron. and Astrophys. – 2012. – V.539. – 12 p.

[57] Emslie A.G., Brown J.C., Donnelly R.F. The inter-relationship of hard X-ray and EUV bursts during solar flares // Solar Phys. – 1978. – V.57. – P.175-190.

[58] Emslie A.G., Smith D.F. Microwave signature of thick-target electron beams in solar flares // Astrophys. J. – 1984. – V.279. – P.882-895.

[59] Falewicz R. Plasma heating in solar flares and their soft and hard X-ray emissions // Astrophys. J. – 2014. – V.789. – №1. – 15 p.

[60] Garcia H.A. Forecasting methods for occurrence and magnitude of proton storms with solar hard X rays // Space Weather. – 2004. – V.2. – №6. – 10 p.

[61] Goff C.P., van Driel-Gesztelyi L., Harra L.K. et al. A slow coronal mass ejection with rising X-ray source // Astron. Astroph. – 2005. – V.434. – №2. – P.761-771.

[62] Gold T., Hoyle F. On the origin of solar flares // Mon. Not. Roy. Astron. Soc. – 1960. – V.120. – №2. – P.89–105.

[63] Hannah I.G., Kontar E.P. Differential emission measures from the regularized inversion of Hinode and SDO data // Astron. Astrophys. – V.539. – 2012. – 14 p.

[64] Hirayama T. Theoretical model of flares and prominences. I: Evaporating flare model // Solar Phys. – 1974. – V.34. – №2. – P.323-338.

[65] Holman G. D., Sui, L., Schwartz R.A., Emslie A.G. Electron bremsstrahlung hard X-ray spectra, electron distributions, and energetics in the 2002 July 23 solar flare // Astrophys. J. – 2003. – V.595. – №2. – P.L97-L101.

[66] Holman G.D., Aschwanden M.J., Auras H. et al. Implications of X-ray observations for electron acceleration and propagation in solar flares // Space Sci. Rev. – 2011. – V.159. – №1-4. – P.107-166.

[67] Hudson H.S. Thick-target processes and white-light flares // Solar Phys. – 1972. – V.24. – №2. –P.414-428.

[68] Hudson H.S., Canfield R.C., Kane S.R. Indirect estimation of energy deposition by non-thermal electrons in solar flares // Solar Phys. – 1978. – V.60. – P.137-142.

[69] Hurford G.J., Schmahl E.J., Schwartz R.A. et al. The RHESSI Imaging Concept // Solar Phys. – 2002. – V.210. – №1. – P.61-86.

[70] Inglis A.R., Christe S. Investigating the differential emission measure and energetics of microflares with combined SDO/AIA and RHESSI observation // Astrophys. J. – 2014. – V.789. – №2. – 12 p.

[71] Jeffrey N.L.S., Kontar E.P. Temporal variations of X-ray solar flare loops: length, corpulence, position, temperature, plasma pressure, and spectra // Astrophys. J. – 2013. – V.766. – №2. – 12 p.

[72] Kašparová J., Karlický M. Kappa distribution and hard X-ray emission of solar flares // Astron. Astrophys. – 2009. – V.497. – №3. – P.L13-L16.

[73] Kattenberg A. Solar radio bursts and their relation to coronal magnetic structures // Diss. Rijksuniversiteit te Utrecht. – 1981. – 168 p.

[74] Koch H.W., Motz J.W. Bremsstrahlung cross-section formulas and related data // Reviews of modern phys. – 1959. – V.31. – №4. – P.920-955.

[75] Kontar E.P., Piana M., Massone A.M. et al. Generalized regularization techniques with constraints for the analysis of solar bremsstrahlung X-ray spectra // Solar Phys. – 2004. – V.225. – №2. – P.293-309.

[76] Kontar E.P., Emslie A.G., Massone A.M. et al. Electron-electron bremsstrahlung emission and the inference of electron flux spectra in solar flares // Astrophys. J. – 2007. – V.670. – №1. – P.857-861.

[77] Kontar E.P., Dickson E., Kašparová J. Low-energy cutoffs in electron spectra of solar flares: statistical survey // Solar Phys. – 2008. – V.252. – №1. – P.139–147.

[78] Kontar E.P., Brown J.C., Emslie A.G. et al. Deducing electron properties from hard X-ray observations // Space Sci. Rev. – 2011. – V.159. – №1-4. – P.301-355.

[79] Kontar E.P., Jeffrey N.L.S, Emslie A.G., Bian N.H. Collisional relaxation of electrons in a warm plasma and accelerated nonthermal electron spectra in solar flares // Astrophys. J. – 2015. – V.809. – №1. – 11 p.

[80] Kopp R.A., Pneuman G.W. Magnetic reconnection in the corona and the loop prominence phenomenon // Solar Phys. – 1976. – V.50. – P.85-98.

[81] Korchak A.A. Possible mechanisms for generating hard X-rays in solar flares // Soviet Astronomy. – 1967. – V.11. – №2. – P.258-263.

[82] Korchak A.A. On the origin of solar flare X-rays // Solar Phys. – 1971. – V.18. – №2. – P.284-304.

[83] Krucker S., Battaglia M., Cargill P.J. et al. Hard X-ray emission from the solar corona // Astron. & Astrophys. Rev. – 2008. – V.16. – P.155-208.

[84] Krucker S., Battaglia M. Particle densities within the acceleration region of a solar flare // Astrophys. J. – 2014. – V.780. – №1. – Id.107. – 6 p.

[85] Landi E., Feldman U., Dere K.P. CHIANTI - an atomic database for emission lines. V. Comparison with an isothermal spectrum observed with SUMER // Astrophys. J. Suppl. Ser. – 2002. – V.139. – №1. – P.281-296.

[86] Landi E., Young P.R. The relative intensity calibration of Hinode/EIS and SOHO/SUMER // Astrophys. J. – 2010. – V.714. – №1. – P.636-643.

[87] Landi E., Young P.R., Dere K.P. et al. CHIANTI – An atomic database for emission lines. XIII. Soft X-ray improvements and other changes // Astrophys. J. – 2013. – V.763. – №2. – 9 p.

[88] Lemen J.R., Title A.M., Akin D.J. et al. The Atmospheric Imaging Assembly (AIA) on the Solar Dynamics Observatory (SDO) // Solar Phys. – 2012. – V.275. – №17. – P.17-40.

[89] Lin R.P. Non-relativistic solar electrons // Space Science Reviews. – 1974. – V.16. – №1-2. – P.189-256.

[90] Lin R.P., Hudson H.S. Non-thermal processes in large solar flares // Solar Phys. – 1976. – V.50. – P.153-178.

[91] Lin R.P., Larson D.E., Ergun R.E. et al. Observations of the solar wind, the bow shock and upstream particles with the WIND 3D Plasma instrument // Advances in Space Research. – 1997. – V.20. – P.645-654.

[92] Lin R.P., Dennis B.R., Hurford G.J. et al. The Reuven Ramaty High-Energy Solar Spectroscopic Imager (RHESSI) // Solar Phys. – 2002. – V.210. – № 1. – P.3-32.

[93] Maksimovic M., Pierrard V., Lemaire, J. F. A kinetic model of the solar wind with Kappa distribution functions in the corona // Astron. Astrophys. – 1997. – V.324. – P.725-734.

[94] Marsch E. Kinetic physics of the solar corona and solar wind // Living Reviews in Solar Phys. – 2006. – V.3. – №1. – 100 p.

[95] McTiernan J.M., Fisher G.H., Li P. The solar flare soft X-ray differential emission measure and the Neupert effect at different temperatures // Astrophys. J. – 1999. – V.514. – №1. – P.472-483.

[96] Melnikov V.F., Charikov Yu.E., Kudryavtsev I.V. Spatial brightness distribution of hard X-Ray emission along flare loops // Geomagnetism and Aeronomy. – 2013. – V.53. – №7. – P.863-866.

[97] Miller J.A., Cargill P.J., Emslie A.G. et al. Critical issues for understanding particle acceleration in impulsive solar flares // J. Gephys. Res. – 1997. – V.102. – №A7. – P.14631-14660.

[98] Milligan R.O., Chamberlin P.C., Hudson H.S. et al., Observations of enhanced extreme ultraviolet continua during an X-class solar flare using SDO/EVE // Astrophys. J. Lett. – 2012. – V.748. – №1. – Id.L14. – 6 p.

[99] Milligan R.O. EUV irradiance observations from SDO/EVE as a diagnostic of solar flares // IAU Symposium. – 2016. – V.320. – P.41-50.

[100] Motorina G.G., Kontar E.P. Differential emission measure and electron distribution function reconstructed from RHESSI and SDO observations // Geomagnetism and Aeronomy. – 2015. – V.55. – №7. – P.995-999.

[101] Neupert W.M. Comparison of solar X-ray line emission with microwave emission during flares // Astrophys. J. – 1968. – V.153. – P.59L-65L.

[102] Oka M., Ishikawa S., Saint-Hilaire P. et al. Kappa distribution model for hard X-ray coronal sources of solar flares // Astrophys. J. – 2013. – V.764. – №1. – 8 p.

[103] Oka M., Krucker S., Hudson H.S., Saint-Hilaire P. Electron energy partition in the above-the-looptop solar hard X-ray sources // Astrophys. J. – 2015. – V.799. – №2. – 14 p.

[104] Oreshina A.V., Somov B.V. On the heat conduction in a high-temperature plasma in solar flares // Astronomy Letters. – 2011. – V.37. – №10. – P.726-736.

[105] Peres G., Rosner R., Serio S., Vaiana G.S. Coronal closed structures. IV. Hydrodynamical stability and response to the heating perturbations // Astrophys. J. – 1982. – V.252. – P.791-799.

[106] Phillips K.J.H., Feldman U. Properties of cool flare with GOES class B5 to C2 // Astron. Astrophys. – 1995. – V.304. – P. 563-575.

[107] Phillips K.J.H. The solar flare 3.8-10 keV X-ray spectrum // Astrophys. J. – 2004. – V.605. – №2. – P.921-930.

[108] Phillips K.J.H., Feldman U., Landi E. Ultraviolet and X-ray spectroscopy of the Solar atmosphere // Cambridge University Press. – 2012. – 388 p.

[109] Piana M., Massone A.M., Kontar E.P. et al. Regularized electron flux spectra in the 2002 July 23 solar flare // Astrophys. J. – 2003. – V.595. – №2. – L127-L130.

[110] Press W.H., Teukolsky S.A., Vetterling W.T., Flannery B.P. Numerical recipes in FORTRAN. The art of scientific computing // Cambridge: University Press. – 1992. – 992 p.

[111] Priest E.R., Forbes T. Magnetic reconnection: MHD theory and applications // Cambridge University Press. – 2000. – 616 p.

[112] Priest E.R., Forbes T. The magnetic nature of solar flares // Astron. Astrophys. Rev. – 2002. – V.10. – №4. – P.313-377.

[113] Saint-Hilaire P., Benz A.O. Thermal and non-thermal energies of solar flares // Astron. Astrophys. – 2005. – V.435. – №2. –P.743-752.

[114] Schwartz R.A., Csillaghy A., Tolbert A.K. et al. RHESSI data analysis software: rationale and methods // Solar Phys. – 2002. – V.210. – №1. – P165-191.

[115] Shibata K., Masuda S., Shimojo M. et al. Hot-plasma ejections associated with compact-loop solar flares // Astrophys. J. Lett. – 1995. – V.451. – P.L83-L85.

[116] Shibata K. Evidence of magnetic reconnection in solar flares and a unified model of flares // Astrophys. and Space Sci. – 1999. – V.264. – №1/4. – P.129-144.

[117] Siarkowski M., Falewicz R., Rudawy P. Plasma heating in the very early phase of solar flares // Astrophys. J. Lett. – 2009. – V.705. – №2. – P.L143-L147.

[118] Simões P.J.A., Graham D.R., Fletcher L. Direct observation of the energy release site in a solar flare by SDO/AIA, Hinode/EIS, and RHESSI // Astrophys. J. – 2015. – V.577. – 9 p.

[119] Simões P.J.A., Graham D.R., Fletcher L. Impulsive heating of solar flare ribbons above 10 MK // Solar Phys. – 2015. – V. 290. – № 12. – P.3573-3591.

[120] Smith D.M., Lin R.P., Turin P. et al. The RHESSI Spectrometer // Solar Phys. – 2002. – V.210. – № 1. – P.33-60.

[121] Somov B.V. Classical and anomalous heat conduction in solar flares // Pisma v Astronomicheskii Zhurnal. –1979. – V.5. – P.50-53.

[122] Spitzer L. Physics of fully ionized gases // New York: Interscience. – 1962. – 192 p.

[123] Stepanov A.V., Zaitsev V.V. The challenges of the models of solar flares // Geomagnetism and Aeronomy. – 2016. – V.56. –№8. – P.952–971.

[124] Sturrock P.A. Model of the high-energy phase of solar flares // Nature. – 1966. – V.211. – №5050. – P.695–697.

[125] Sui L., Holman G.D., Dennis B.R. Evidence for magnetic reconnection in three homologous solar flares observed by RHESSI // Astrophys. J. – 2004. – V.612. – №1. – P.546-556.

[126] Sui L., Holman G.D., Dennis B.R. Determination of low-energy cutoffs and total energy of nonthermal electrons in a solar flare on 2002 April 15 // Astrophys. J. – 2005. – V.626. – №2. – P.1102-1109.

[127] Svestka Z. Solar flares // Springer. – 1976. – 415 p.

[128] Tandberg-Hanssen E., Emslie A.G. The physics of solar flares // Cambridge University Press. – 1988. – 286 p.

[129] Trottet G., Barat C., Ramaty R. et al. Thin target γ-ray line production during the 1991 June 1 flare // AIP Conf. Proc. – 1996. – V.374. – P.153-161.

[130] Tsap Y.T. The stochastic acceleration of upper chromospheric electrons // Astron. Rep. – 1998. – V.42. – №2. – P. 275-281.

[131] Tsap Yu.T., Motorina G.G., Kopylova Yu.G. Flare coronal loop heating and hard X-ray emission from solar flares of August 23, 2005, and November 9, 2013 // Geomagnetism and Aeronomy. – 2016. – V.56. – №8. – P.1104-1109.

[132] Tsuneta S. Moving plasmoid and formation of the neutral sheet in a solar flare // Astrophys. J. – 1997. – V.483. – №1. – P.507-514.

[133] Tucker W. Radiation processes in astrophysics // Cambridge, Mass., MIT Press. – 1975. – 320 p.

[134] Vasyliunas V.M. A survey of low-energy electrons in the evening sector of the magnetosphere with OGO 1 and OGO 3 // JRG. – 1968. – V.73. – P.2839-2884.

[135] Veronig A., Vršnak B. Dennis B.R. et al. Investigation of the Neupert effect in solar flares. I. Statistical properties and the evaporation model // Astron. Astrophys. – 2002. – V.392. – №1. – P.699–712.

[136] Veronig A.M., Brown J.C. A coronal thick-target interpretation of two hard X-ray loop events // Astrophys. J. – 2004. – V.603. – №2. – P.L117-L120.

[137] Veronig A.M., Brown J.C., Dennis B.R. et al. Physics of the Neupert effect: estimates of the effects of source energy, mass transport, and geometry using RHESSI and GOES data // Astrophys. J. – 2005. – V.621. – №1. – P.482-497.

[138] White S.M., Thomas R.J., Schwartz R.A. Updated expressions for determining temperatures and emission measures from GOES soft X-ray measurements // Solar Phys. – 2005. – V.227. – №2. – P.231-248.

[139] Xu Y., Emslie A.G., Hurford G.J. RHESSI hard X-ray imaging spectroscopy of extended sources and the physical properties of electron acceleration regions in solar flares // Astrophys. J. – 2008. – V.673. – №1. – P.576–585.

[140] Zaitsev V.V., Urpo S., Stepanov A.V. Temporal dynamics of Joule heating and DC-electric field acceleration in single flare loop // Astron. Astrophys. – 2000. – V.357. – №3. – P.1105–1114.

[141] Zaitsev V.V., Stepanov A.V. Particle acceleration and plasma heating in the chromosphere // Solar Phys. – 2015. – V.290. – №12. – P.3559–3572.